\documentclass[journal,10pt]{IEEEtran}
\usepackage[paperwidth=8.5in, paperheight=11in,
            left=0.75in,
            right=0.75in,
            top=0.75in,
            bottom=0.98in,          
            headheight=12pt,
            headsep=18pt,
            footskip=0.5in       
            ]{geometry}

\makeatletter
\def\ps@headings{%
\def\@oddhead{\mbox{}\scriptsize\rightmark \hfil \thepage}%
\def\@evenhead{\scriptsize\thepage \hfil \leftmark\mbox{}}%
\def\@oddfoot{}%
\def\@evenfoot{}}
\makeatother
\ifCLASSINFOpdf
\else
\fi

\usepackage{caption} 

\let\biconditional\leftrightarrow
\newcounter{problem}
\newcounter{save@equation}
\newcounter{save@problem}

\makeatletter

\usepackage{epsfig}
\usepackage{array}
\usepackage[short]{optidef}
\usepackage{tabularx}
\usepackage{amsthm}
\usepackage{mathrsfs}
\usepackage{flexisym}
\usepackage{graphicx}
\usepackage{algorithmicx}
\usepackage{algorithm}
\usepackage{siunitx}
\usepackage{comment}
\usepackage{pifont}
\DeclareUnicodeCharacter{2717}{\ding{55}}
\usepackage{soul}
\usepackage{algpseudocode}
\usepackage{cite}
\usepackage{url}
\usepackage[utf8]{inputenc}
\usepackage[T1]{fontenc}
\usepackage{amsmath}
\usepackage{amsfonts}
\usepackage{amssymb}
\usepackage[version=4]{mhchem}
\usepackage{stmaryrd}
\usepackage{graphicx}
\usepackage[export]{adjustbox}
\usepackage{breqn}
\usepackage{lipsum}
\usepackage{mathtools}
\usepackage{float}
\usepackage{tikz}
\usepackage{soul}
\usepackage{blindtext, graphicx}
\usepackage{cuted}
\usepackage{color}
\usepackage{multicol}

\newcolumntype{L}[1]{>{\raggedright\let\newline\\\arraybackslash\hspace{0pt}}m{#1}}
\newcolumntype{C}[1]{>{\centering\let\newline\\\arraybackslash\hspace{0pt}}m{#1}}
\newcolumntype{R}[1]{>{\raggedleft\let\newline\\\arraybackslash\hspace{0pt}}m{#1}}

\DeclareUnicodeCharacter{2717}{\ding{55}} 

\newtheoremstyle{case}{}{}{}{}{}{:}{ }{}
\newcommand{\bc}{\begin{center}}
\newcommand{\ec}{\end{center}}
\newcommand{\be}{\begin{equation}}
\newcommand{\ee}{\end{equation}}

\newcommand{\bnu}{\begin{enumerate}}
\newcommand{\enu}{\end{enumerate}}

\usepackage{titlesec}
\titlespacing{\section}{0pt}{*0.5}{*0.5}
\titlespacing{\subsection}{0pt}{*0.5}{*0.5}
\titlespacing{\subsubsection}{0pt}{*0.5}{*0.5}

\begin{document}

\title{OSI Stack Redesign for Quantum Networks: Requirements, Technologies, Challenges, and Future Directions}

\author{ Shakil Ahmed~\IEEEmembership{Memeber, IEEE}, Muhammad Kamran Saeed, and Ashfaq Khokhar~\IEEEmembership{Fellow, IEEE}

\vspace*{-0.8 cm}
\thanks{S. Ahmed, M. Saeed, and A. Khokhar are with the Department of Electrical and Computer Engineering, Iowa State University, Ames, Iowa, USA. (email: \{shakil, kamran, ashfaq\}@iastate.edu)  }}

\maketitle
\thispagestyle{empty}

\begin{abstract}
Quantum communication is poised to become a foundational element of next-generation networking, offering transformative capabilities in security, entanglement-based connectivity, and computational offloading. However, the classical Open Systems Interconnection (OSI) model—designed for deterministic and error-tolerant systems—cannot support quantum-specific phenomena such as coherence fragility, probabilistic entanglement, and the no-cloning theorem. This paper presents a comprehensive survey and architectural redesign of the OSI model tailored for quantum networks, particularly within the context of emerging seventh-generation (7G) requirements.
We propose a Quantum-Converged OSI stack by extending the classical seven-layer model with two additional planes: Layer~0, the \textit{Quantum Substrate Layer}, which manages entanglement, coherence, and teleportation; and Layer~8, the \textit{Cognitive Intent Plane}, enabling semantic orchestration through Large Language Models (LLMs) and Quantum Machine Learning (QML). Each classical layer is reexamined to support quantum mechanisms such as quantum-enhanced medium access control (MAC) protocols, fidelity-aware routing, session tokenization, and twin-based quantum applications.
This survey consolidates over 150 research works from the Institute of Electrical and Electronics Engineers (IEEE), Association for Computing Machinery (ACM), Multidisciplinary Digital Publishing Institute (MDPI), arXiv, and Web of Science (2018–2025), classifying them by OSI layer, enabling technologies such as quantum key distribution (QKD), Quantum Error Correction (QEC), Post-Quantum Cryptography (PQC), and Reconfigurable IntelligentSurfaces (RIS), and use cases such as satellite QKD, unmanned aerial vehicle (UAV) swarms, and quantum IoT. A taxonomy of cross-layer enablers—such as hybrid quantum–classical control, metadata-driven orchestration, and blockchain-integrated quantum trust—is provided, along with simulation tools including Network Simulator for Quantum Information using Discrete events (NetSquid), Quantum Internet Simulation (QuNetSim), and Quantum Internet Simulation Platform (QuISP).
We present several domain-specific applications, including quantum healthcare telemetry, entangled vehicular networks, and satellite mesh overlays. An evaluation framework is proposed using metrics such as entropy throughput, coherence latency, and entanglement fidelity. Finally, the paper outlines critical research challenges and future directions, including programmable quantum stacks, digital twins, and Artificial Intelligence (AI)-defined Quantum Network (QNet) agents.
This work lays the foundation for a scalable, intelligent, and quantum-compliant OSI framework that bridges classical systems with quantum-enabled networks for 7G and beyond.
\end{abstract}


\section{Introduction}
\subsection{Background and Motivation}
\begin{figure}[]
\centering
\includegraphics[width=3.35in]{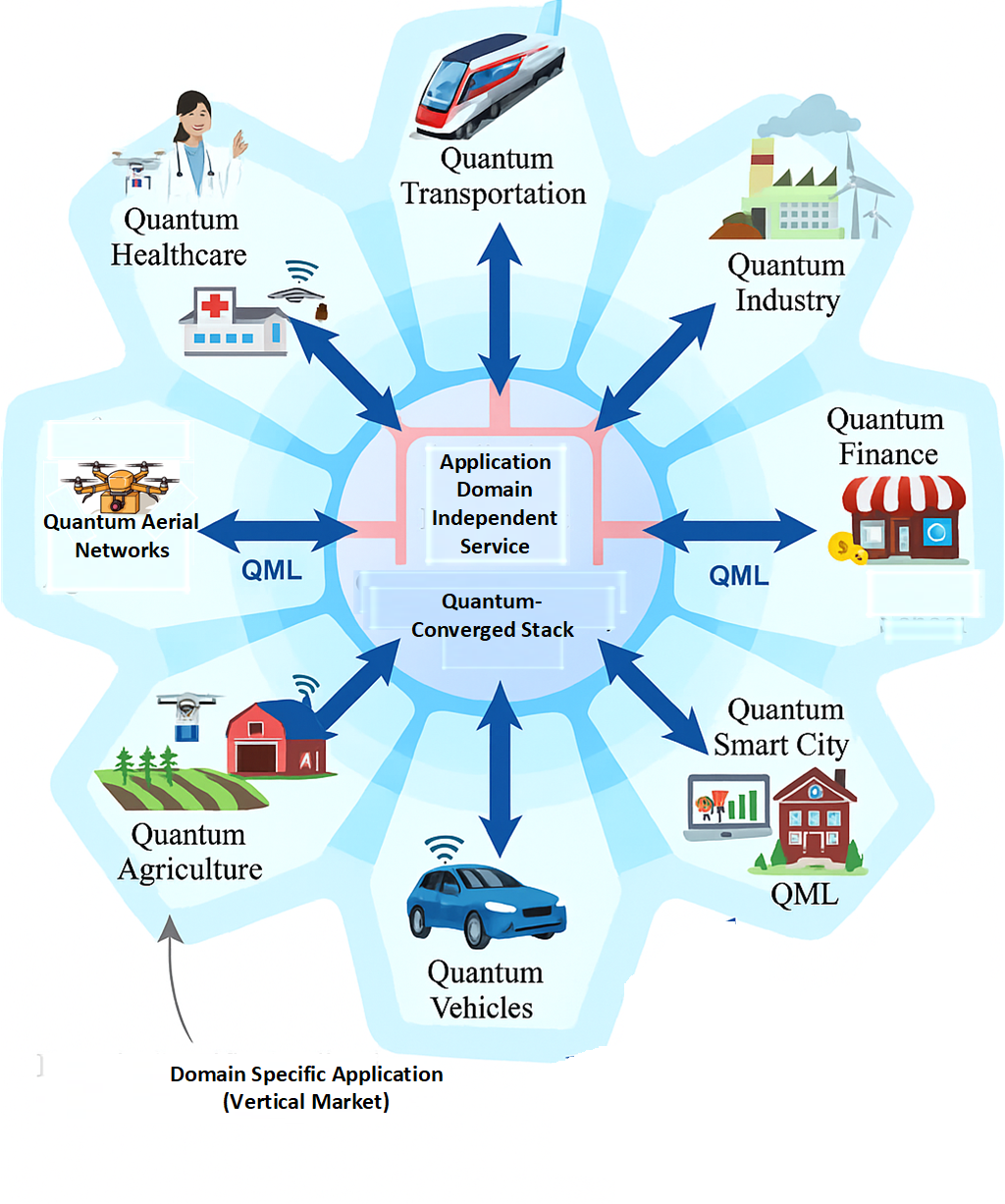}
\caption{ The overall picture of Quantum networks emphasizing the vertical markets and the
horizontal integration}
\label{Fig_application}
\end{figure}

\IEEEPARstart{Q}{uantum} communication is rapidly emerging as a foundational enabler for future wireless systems, particularly in the context of sixth-generation (6G) and seventh-generation (7G) networks \cite{chowdhury}. While classical networks depend on deterministic signal transmission and replication, quantum communication relies on principles such as superposition, entanglement, and the no-cloning theorem, which introduce fundamentally new constraints and capabilities. These include ultra-secure transmission through Quantum Key Distribution (QKD), quantum teleportation for information exchange, and non-local correlation via entanglement routing \cite{Mehic2023}.
The motivation for integrating quantum communication into 7G stems from its alignment with the envisioned properties of 7G systems: autonomous orchestration, ultra-low-latency reliability, global coverage, and end-to-end security. As conventional technologies approach theoretical performance ceilings, quantum systems offer new possibilities in overcoming channel capacity limits, secure satellite communications, and next-generation Internet-of-Things (IoT) deployments \cite{Zhao2024QuantumWireless,ahmed2025quantum}. Fig.~\ref{Fig_application} illustrates the overall concept of the quantum in which every domain-specific application interacts with domain-independent services, whereas, in each domain, sensors and actuators communicate directly with each other.

Despite its exceptional capabilities, quantum networking introduces unique architectural challenges. The foundational Open Systems Interconnection (OSI) model, designed in the classical era, assumes properties incompatible with quantum systems—such as duplicability, deterministic paths, and conventional error recovery. These conflicts necessitate a ground-up reconsideration of protocol design. For example, classical retransmission strategies are invalid in quantum systems due to the no-cloning theorem \cite{Ray2021Vision6GNIB}, and traditional synchronization techniques cannot accommodate coherence lifetimes required by quantum entangled states \cite{Chamola2025FutureConnectivity}.
As a result, integrating quantum communication into the 7G paradigm is not a matter of incremental enhancement. Instead, it requires a layered redesign that spans hardware (e.g., quantum repeaters, metasurfaces) and software (e.g., entanglement routing, quantum session control). Recent surveys have acknowledged the transformative role of quantum in 6G/7G but typically fall short of offering a unified stack-level vision that bridges quantum physics, protocol design, and intelligent orchestration \cite{Glisic2024QuantumNeuroSurvey,Dev2021QuantumDDI}.
This paper responds to that gap by proposing a quantum-converged OSI stack architecture, integrating quantum technologies across all layers and introducing two new conceptual layers: Layer 0, the \textit{Quantum Substrate}, and Layer 8, the \textit{Cognitive Intent Plane}. These additions reflect the physical peculiarities of quantum states and the intelligent, LLM-driven orchestration required for real-time adaptation in 7G networks.

\subsection{Vision of Quantum-Converged Protocol Design}
As 7G networks push toward global-scale intelligence, zero-latency operation, and ubiquitous trust, integrating quantum communication into the protocol architecture is not merely an upgrade but a necessity. Classical networking protocols, which are based on assumptions of replicability, deterministic delivery, and layered abstraction, fail to support the non-cloning, coherence-sensitive, and probabilistic nature of quantum information. Thus, a new vision is required: one that unifies quantum mechanical principles, Artificial Intelligence (AI)-driven cognition, and cross-layer programmability in a coherent, future-ready protocol stack.
Therefore, we propose a \textit{Quantum-Converged OSI Stack}, extending the traditional seven-layer model in two directions. First, we introduce \textit{Layer 0 – the Quantum Substrate} to support physical-level quantum operations such as entanglement generation, teleportation, and decoherence management. Second, we define \textit{Layer 8 – the Cognitive Intent Plane}, responsible for orchestrating cross-layer behaviors via learning-based agents, such as Large Language Models (LLMs), reinforcement learners, and Quantum Machine Learning (QML) modules. This architectural expansion reflects the dual challenge of 7G networks: incorporating quantum physical primitives while embedding intelligence across the stack. 

Unlike classical layers that isolate functionality (e.g., routing at Layer 3 or reliability at Layer 4), quantum-converged layers must work collaboratively to maintain fidelity and coherence. As Glisic and Lorenzo emphasize, the layering must be reinterpreted to accommodate neuro-inspired models, semantic data flows, and quantum interactions \cite{Glisic2024QuantumNeuroSurvey}.
Moreover, Quy et al. proposed the necessity of layered innovation in 6G and beyond through programmable, intelligent stacks that blend classical and quantum resources dynamically \cite{Quy2023InnovativeTrends}. While rooted in classical-quantum coexistence, their vision stops short of a structural OSI expansion—an omission this paper addresses directly. Arshi et al. similarly emphasize the urgency of incorporating full-stack adaptability and propose a quantum-aware IoT protocol framework, but without defining the orchestration layer required for stack-wide cognition \cite{Arshi2024QuantumIoT}.
Our vision thus synthesizes these streams of research into a comprehensive architectural redesign. The proposed quantum-converged stack is not only backward-compatible with classical systems but also future-ready for AI-native operations, post-quantum security primitives, and programmable networks. It enables 7G systems to operate over heterogeneous substrates—including entangled Unmanned Aerial Vehicles (UAVs), satellite QKD constellations, and federated edge quantum devices—while maintaining Service-Level Agreement (SLA) guarantees through intent-based, autonomous control.

\subsection{Limitations of Classical OSI Stack in Quantum Context}
The classical OSI model, a cornerstone of modern network protocol design, was developed to support digital communication's deterministic, linear, and layered abstraction. However, its assumptions fundamentally contradict the properties of quantum information systems. This section outlines the architectural misalignments and their implications for quantum networking in 7G environments.

\textit{1) No-Cloning Theorem:} Classical networking relies on the duplicability of data packets, enabling retransmissions, caching, and redundancy. In quantum communication, the \textit{no-cloning theorem} prohibits copying arbitrary quantum states, making classical packet duplication, backup, and loss recovery infeasible \cite{Khurana2024ICMLA}.

\textit{2) Decoherence and Fidelity Constraints:} Quantum states are inherently fragile and degrade due to interactions with the environment—a process known as \textit{decoherence}. This challenges session persistence, routing stability, and transport reliability. However, the classical OSI stack lacks inherent mechanisms to handle time-sensitive quantum coherence requirements or to model entanglement lifespan within protocol flows \cite{Li2023EntanglementAssisted, Khan2024Heliyon}.

\textit{3) Inadequate Error Correction Models:} Classical error correction assumes bit errors and noise can be corrected via parity checks or redundancy schemes. However, Quantum Error Correction (QEC)  involves maintaining state superpositions and requires entirely new constructs such as surface codes, logical qubits, and entanglement-assisted recovery, whereas OSI layers lack interfaces or representations for QEC codes \cite{Picchi2023NTNQuantum, Luo2024QuantumNetworks}.

\textit{4) Absence of Entanglement Awareness:} Entanglement introduces non-local correlations that are not addressable in classical routing or addressing models. In the OSI architecture, layers such as Network and Transport operate under addressable, isolated endpoints assumptions. Quantum links, by contrast, are shared probabilistic resources and require coordination protocols that span multiple layers to preserve entanglement fidelity and coherence \cite{Granelli2022HybridStack, Chehimi2022PhysicsInformed}.

\textit{5) Lack of Cross-Layer Coordination:} The strict layering principle of OSI discourages real-time interaction across layers. In contrast, quantum protocols often require cross-layer feedback (e.g., fidelity from the physical layer to influence session control or routing). This necessitates a departure from rigid OSI encapsulation toward a more integrated, dynamically coordinated protocol model \cite{Li2021QuantumOSIModel, Das2024CQStackIoT}.
To address these challenges, a quantum-aware redesign of the OSI model is needed to introduce new abstractions, interoperable quantum-classical interfaces, and protocol primitives designed for entanglement-based communication. This motivates the introduction of a bottom-layer physical extension (Quantum Substrate) and a top-layer orchestration interface (Cognitive Intent Plane), as detailed in later sections.

\subsection{Related Surveys and Literature Gaps}
Several surveys have explored quantum communication and networking components in the context of next-generation systems, including 6G and 7G. These studies have addressed diverse topics such as post-quantum cryptography, QKD, quantum-enhanced physical layer security, and AI-driven networking. However, most existing works are technology-centric or security-focused and do not present an integrated, OSI-aligned architectural framework.
For instance, Mahmud and Abdelhadi \cite{Mahmud2025AIQuantumSurvey} provide a broad review of artificial intelligence in quantum communications, covering AI-enhanced traffic optimization and post-quantum security techniques. While their work discusses layering in a conceptual sense, it does not map technologies to specific OSI layers or address structural stack redesign. Similarly, Wijesekara and Gunawardena \cite{Wijesekara2023BlockchainKDN} examine the use of blockchain in Knowledge-Defined Networking (KDN) and propose an augmented management-cognitive layer, with a primary focus on security and trust in cognitive overlays, rather than on entanglement-based protocol stacks.
Khaloopour et al. \cite{Khaloopour2024Resilience} explore resilience-by-design approaches in 6G, proposing adaptive control planes and post-quantum secure protocols. Yet, their work is limited to threat resilience and lacks stack-wide orchestration logic involving LLMs or QML.
Moreover, Baseri et al. \cite{Baseri2024QuantumSafeSurvey} offer a valuable classification of post-quantum security protocols across network layers but do not propose enhancements to accommodate quantum decoherence, entanglement routing, or non-classical error correction. Similarly, Barros \cite{Barros2025ProofHumanity} introduces a multi-layered framework for AI-proof digital provenance using OSI-inspired layering, yet omits quantum substrate modeling and ignores inter-layer coherence feedback.
These surveys, though insightful, fail to converge on a holistic, forward-compatible model that spans physical quantum fidelity management to top-layer cognitive orchestration. To our knowledge, no current work unifies:
\begin{itemize}
    \item A full 9-layer stack model inclusive of a Quantum Substrate (Layer 0) and Cognitive Intent Plane (Layer 8),
    \item A cross-layer analysis of quantum-specific protocol functions,
    \item Integration of simulation environments (e.g., NetSquid, QuNetSim) for fidelity-aware validation,
    \item Mapping of quantum and classical technologies across OSI layers,
    \item AI/LLM-driven orchestration mechanisms for real-time quantum-classical coordination.
\end{itemize}

This paper aims to bridge this critical gap by delivering a unified survey and architectural framework, grounded in OSI principles but extended for quantum-aware, AI-native 7G networks. A comparative summary of our work against state-of-the-art survey articles in quantum networking is presented in Table~\ref{tab:qnet-surveys}. This comparison highlights key differentiators in terms of QKD integration, entanglement transport protocols, Software-Defined Networking (SDN)-based orchestration, stack-layer abstraction, synchronization strategies, hybrid quantum-classical interoperability, and intent-driven control—areas where our contribution offers a unified and forward-looking perspective.

\subsection{Scope and Contributions}
This paper presents a comprehensive survey and architectural proposal for a quantum-converged protocol stack suitable for 7G communication systems. It is motivated by the convergence of multiple disciplines—quantum communication, AI-based orchestration, post-quantum cryptography, and classical networking—which demand a structural revision of the OSI model.
This work adopts a full-stack perspective, unlike prior surveys that focus on specific technologies (e.g., QKD, Quantum Error Correction (QEC), or Post-Quantum Cryptography (PQC) or narrow protocol layers. We synthesize findings from over 150 academic sources and structure the paper around redesigning the classical OSI model to support quantum networking properties such as entanglement routing, decoherence mitigation, coherence-sensitive session handling, and cognitive cross-layer control.
Our work differentiates itself from notable contributions like Glisic and Lorenzo’s neuro-quantum integration model for 6G/7G systems \cite{Glisic2024QuantumNeuroSurvey}, which discusses cognitive-quantum intersections but lacks a stack-level formalism, and Chakraborty et al.'s general conceptual framework on AI and systems design \cite{Chakraborty2020DeepAnalytics}, which does not engage directly with quantum-layer abstractions.

\textit{The scope of this paper includes:}
\begin{itemize}
    \item A full 9-layer OSI-based stack model that incorporates two new layers:
        \begin{itemize}
            \item \textit{Layer 0 – Quantum Substrate:} For physical quantum fidelity, decoherence tracking, and teleportation channels.
            \item \textit{Layer 8 – Cognitive Intent Plane:} Enabling autonomous, LLM- or QML-based cross-layer orchestration.
        \end{itemize}
    \item A survey of quantum technologies and their integration at each OSI layer: quantum MAC, Q-routing, PQC transport, hybrid handshakes, and quantum metadata formatting.
    \item A classification of enabling technologies, including QEC, Reconfigurable Intelligent Surface (RIS), blockchain, LLMs, and programmable networks.
    \item Use-case mapping across domains: entangled UAVs, QKD satellite swarms, and quantum IoT in sensitive industries.
    \item An evaluation framework with fidelity-aware and coherence-aware performance metrics for quantum networks.
    \item A review of open challenges and future research themes: stackless control, quantum-AI orchestration, simulation testbeds, and standardization barriers.
\end{itemize}

\textit{Key Contributions:}
\begin{enumerate}
    \item Introduction of a 9-layer Quantum-Converged OSI model, enhancing structural compatibility with quantum network properties.
    \item Layer-by-layer mapping of quantum technologies, protocols, and interfaces using over 150 scholarly sources.
    \item Cross-layer analysis of orchestration frameworks powered by AI/LLMs, QML, and programmable substrates.
    \item Real-world use cases and benchmarking criteria grounded in quantum fidelity, coherence, and cross-stack latency.
    \item Identification of key research gaps, including lack of quantum-aware OSI models, hybrid-classical integration standards, and orchestration-layer simulation tools.
\end{enumerate}

\begin{figure*}[ht]
\centering
\includegraphics[width=.80\textwidth]{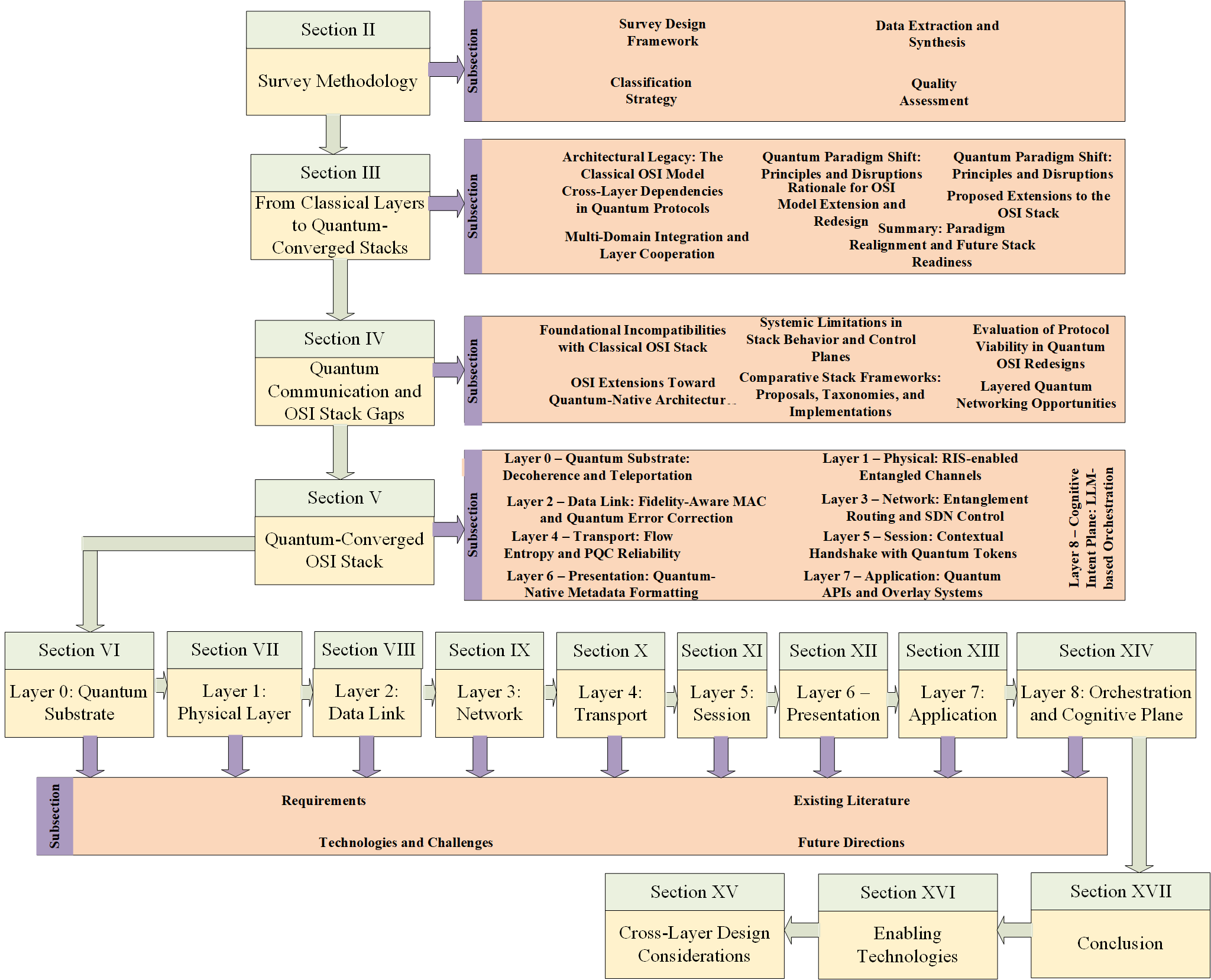}
\caption{Overview of the paper organization.}
\label{fig:paperoverview}
\end{figure*}

By encompassing foundational and futuristic perspectives, this paper provides a reference architecture and knowledge roadmap for researchers and engineers working on post-6G quantum networking systems. The summary of relevant abbreviations in this survey is listed in Table \ref{tab:abbreviations} and  
Fig.~\ref{fig:paperoverview} shows the overview of the paper organization. 

\begin{table}[!ht]
\centering
\caption{Summary of Abbreviations}
\label{tab:abbreviations}
\renewcommand{\arraystretch}{1.2}
\begin{tabular}{|c|l|}
\hline
\textbf{Abbreviation} & \textbf{Definition} \\
\hline
API & Application Programming Interface \\
ARQ & Automatic Repeat reQuest \\
BSM & Bell State Measurement \\
CRC & Cyclic Redundancy Checks \\
DQML & Distributed Quantum Machine Learning \\
ETSI & European Telecommunications Standards Institute \\
EPR & Einstein-Podolsky-Rosen \\
FEC & Forward Error Correction \\
FSO & Free Space Optics \\
GPT & Generative Pre-trained Transformer \\
IDQ & ID Quantique \\
IoT & Internet of Things \\
LLM & Large Language Model \\
MAC & Medium Access Control \\
MDI & Measurement Device Independent \\
NFV & Network Function Virtualization \\
NISQ & Noisy Intermediate-Scale Quantum \\
NV & Nitrogen-Vacancy  \\
OSI & Open Systems Interconnection \\
PHY & Physical Layer \\
PICs & Photonic Integrated Circuits \\
PQC & Post-Quantum Cryptography \\
QaaS & Quantum-as-a-Service \\
QAI & Quantum Artificial Intelligence \\
QASM & Quantum Assembly Language \\
QBER & Quantum Bit Error Rate \\
QEC & Quantum Error Correction \\
QDEs&Quantum Development Environments\\
QFCs & Quantum Frequency Converters \\
QFL & Quantum Federated Learning \\
QIoT & Quantum-enabled IoT \\
QIR & Quantum Intermediate Representation \\
QKD & Quantum Key Distribution \\
QML & Quantum Machine Learning \\
QNID & Quantum Network Identifier \\
QPACE & Quantum Protocol-Aware Cloud Edge \\
QPU & Quantum Processing Unit \\
QRNA & Quantum Repeater Network Architecture \\
QSDs & Quantum Session Descriptors \\
QSMC & Quantum Secure Multiparty Computation \\
QSMs & Quantum Session Managers \\
QSPN & Quantum Source Path Negotiation \\
RIS & Reconfigurable Intelligent Surface \\
RL & Reinforcement Learning \\
SDN & Software-Defined Networking \\
SDQN & Software-Defined Quantum Networking \\
SLA & Service-Level Agreement \\
SLOs & Service Level Objectives \\
SLR & Systematic Literature Review \\
SPDs & Single-Photon Detectors \\
TaaS &Teleportation-as-a-Service\\
TTL & Time-to-Live \\
TCP & Transmission Control Protocol \\
UAV & Unmanned Aerial Vehicle \\
ZKP & Zero-Knowledge Proof \\
ZPL & Zero Phonon Line \\
\hline
\end{tabular}
\end{table}

\section{Survey Methodology}
This section outlines the methodological framework adopted for conducting this layered quantum-converged protocol stack research survey. The methodology is informed by Systematic Literature Review (SLR) practices, combining quantitative and qualitative synthesis from highly curated databases.

\subsection{Survey Design Framework}
To ensure methodological rigor, this survey follows a structured approach inspired by SLR frameworks. The adopted protocol aligns with the methodology that Khan et al.~\cite{Khan2023SysReviewQuantum}, emphasizing four core principles: replicability, thematic coverage, bias mitigation, and traceability. Drawing from established models like PRISMA, our review framework consists of the following sequential phases: identification, screening, eligibility assessment, and inclusion.
The identification phase retrieved literature using advanced keyword queries from leading scientific databases, including IEEE Xplore, SpringerLink, Elsevier’s ScienceDirect, ACM Digital Library, and arXiv. The query design was optimized to capture research at the intersection of quantum communication and protocol engineering. Specific search terms included “quantum networking,” “quantum OSI stack,” “entanglement routing,” “quantum protocol design,” “quantum error correction,” “post-quantum security,” and “AI-driven orchestration in quantum networks.” These terms were selected to reflect the layered nature of the problem space and the multidimensional technologies under consideration, as recommended in thematic review practices for quantum systems~\cite{Caleffi2024DistributedQuantum}.
The time window for literature selection spans from January 2018 to March 2025, reflecting the surge in interest following early demonstrations of QKD networks and the formalization of quantum internet proposals. This window also overlaps with the emergence of quantum protocol stack models and hybrid quantum-classical infrastructure frameworks~\cite{Huang2024LayeredSurvey}.
To qualify for inclusion, publications had to meet specific criteria: (i) be peer-reviewed or published in high-impact conferences or journals; (ii) address at least one OSI or OSI-like layer in the context of quantum communication or classical-quantum hybrid systems; and (iii) provide either empirical results, simulation frameworks, architectural blueprints, or comprehensive theoretical formulations. Articles were excluded if they lacked technical depth, were opinion-based essays or perspectives, or did not appear in reputable indexing systems. Non-peer-reviewed white papers, patents, and vendor-specific documentation were also excluded to maintain academic neutrality.

\subsection{Classification Strategy}
Given the multidisciplinary nature of quantum communication, organizing the surveyed literature required a structured and multidimensional classification scheme. We adopted a three-axis classification strategy to maintain architectural clarity and technological specificity across the protocol stack.
The first axis is \textit{Layer Mapping}, where each paper was assigned to one or more OSI layers. This included the classical seven layers and the two extended layers proposed in this survey: Layer 0 (Quantum Substrate) and Layer 8 (Cognitive Intent Plane). This mapping aimed to ensure that research contributions—whether protocol definitions, hardware innovations, or orchestration mechanisms—were evaluated in relation to their position within the proposed stack. This structural mapping supports stack-wise analysis and highlights research saturation or gaps at specific layers.
The second axis is \textit{Technology Domains}, which classifies each contribution based on its enabling technology or core focus area. For example, works were tagged as focusing on QEC, QKD, RIS, QML, or PQC. This axis allows for a comparison of how foundational quantum technologies are distributed across layers and research communities.
The third axis is \textit{Application Use Cases}, wherein contributions were categorized based on their targeted deployment scenarios. These included quantum-enabled satellite constellations, entangled UAV swarm networks, edge-intelligent quantum IoT systems, privacy-sensitive healthcare telemetry, and programmable quantum internet infrastructure. This dimension contextualizes stack-layer activity with real-world utility.
This classification strategy is inspired by the taxonomies used in Caleffi et al.~\cite{Caleffi2024DistributedQuantum}, who propose layered segmentation in distributed quantum computing, and by Huang et al.~\cite{Huang2024LayeredSurvey}, who map quantum protocols across traditional OSI layers. By integrating architectural and domain-driven perspectives, our framework achieves protocol-level granularity without sacrificing interdisciplinary coverage—an essential requirement for bridging quantum information theory with systems and network engineering.

\subsection{Quality Assessment}
We adopted a structured quality assessment framework to ensure analytical rigor and consistency across the selected literature. This assessment methodology was inspired by systematic review practices employed in emerging quantum-computing and communication domains, where thematic relevance, simulation validity, and architectural alignment are critical for research synthesis. Following best practices outlined by Khurana \cite{Khurana2024ICMLA}, each paper was evaluated against five quality dimensions tailored for protocol-stack-level studies.

\begin{enumerate}
    \item \textit{Relevance to Quantum-Aware Network Design:} The work must focus on quantum networking concepts or contribute to the architecture, protocol design, or orchestration of quantum communication systems.

    \item \textit{Layer or Cross-Layer Specificity:} The study should address one or more layers of the OSI model (including extensions proposed in this paper), or highlight dependencies across layers.

    \item \textit{Clarity in Modeling or Implementation:} Preference was given to articles that presented protocol models, empirical results, or simulation frameworks such as NetSquid, QuNetSim, or QuISP. Textual discussions without architectural grounding were marked low on this metric.

    \item \textit{Innovation in Quantum-Classical Integration:} Works proposing hybrid-stack designs, AI-based control interfaces, quantum SDN, or fidelity-aware orchestration mechanisms received higher scores.

    \item \textit{Reference Density and Scientific Rigor:} Finally, the methodological robustness of each paper was judged based on peer-review status, number of citations, and the granularity of technical explanations.
\end{enumerate}

Each paper was scored on a 5-point scale across these metrics, and only those achieving a minimum cumulative score of 15 out of 25 were retained for deeper analysis. This filtration ensured that only conceptually robust, layer-relevant, and technically mature papers formed the basis of the proposed OSI redesign.

\subsection{Data Extraction and Synthesis}
A systematic data extraction and synthesis methodology was implemented once relevant articles were selected through the screening and quality assessment phases. This phase ensures the structural organization, thematic alignment, and interoperability of findings across diverse layers and technologies of the quantum stack. The strategy is adapted from the systematic protocol synthesis framework discussed by Thakur et al.~\cite{Thakur2023DeepSurvey}, which outlines best practices in capturing structured knowledge from diverse research sources, particularly when dealing with emerging, multidisciplinary domains.
Data were extracted manually and curated in structured matrices using predefined fields: publication metadata (author, venue, year), OSI layer relevance, enabling technology, application context, and evaluation type (analytical, simulation, prototype). Cross-referencing handled overlapping contributions, while contradictions were flagged for consensus-based resolution.
To synthesize insights, thematic coding was used to cluster articles according to technological themes (e.g., QKD, QEC, SDN, LLMs), architectural functions (e.g., routing, synchronization, handshake), and simulation methods (e.g., NetSquid, QuNetSim). From this classification, layer-specific patterns and emerging cross-layer dependencies were extracted. The process also enabled the identification of architectural primitives—such as quantum session handshakes or entropy-coherent switching—that transcend specific layers.
This structured synthesis method facilitates stack-wise comparison, highlights key research gaps (e.g., limited work at Layers 5 and 6), and supports the derivation of architectural recommendations.

\begin{table*}[htbp]
\centering
\caption{Comparison of Quantum Network Survey Articles}
\label{tab:qnet-surveys}
\begin{tabular}{|l|c|c|c|c|c|c|c|}
\hline
\textbf{Survey Article} & \textbf{Quantum} & \textbf{Network} & \textbf{Retransmission \& Routing} & \textbf{Control Plane} & \textbf{Classical/Quantum} & \textbf{Security} & \textbf{QAI} \\
\hline
Caleffi \textit{et al.} (2024) \cite{Caleffi2024DistributedQuantum} & $\checkmark$ & $\checkmark$ & $\checkmark$ & $\checkmark$ & $\checkmark$ & $\checkmark$ & $\times$ \\
Illiano \textit{et al.} (2022) \cite{Illiano2022QuantumStackSurvey} & $\checkmark$ & $\checkmark$ & $\times$ & $\checkmark$ & $\checkmark$ & $\checkmark$ & $\times$ \\
Hasan \textit{et al.} (2023) \cite{hasan2023qnet} & $\times$ & $\checkmark$ & $\checkmark$ & $\times$ & $\checkmark$ & $\checkmark$ & $\times$ \\
Pierucci \textit{et al.} (2024) \cite{Pierucci2024NontTerrQuantum} & $\checkmark$ & $\checkmark$ & $\checkmark$ & $\checkmark$ & $\checkmark$ & $\checkmark$ & $\checkmark$ \\
Sidhu \textit{et al.} (2021) \cite{sidhu2021space} & $\times$ & $\checkmark$ & $\times$ & $\times$ & $\checkmark$ & $\checkmark$ & $\times$ \\

Manzalini (2020) \cite{manzalini2020future} & $\times$ & $\checkmark$ & $\times$ & $\times$ & $\checkmark$ & $\times$ & $\times$ \\
Zhang \textit{et al.} (2022) \cite{zhang2022future} & $\checkmark$ & $\checkmark$ & $\checkmark$ & $\checkmark$ & $\checkmark$ & $\checkmark$ & $\checkmark$ \\
\textbf{Ours} & $\checkmark$ & $\checkmark$ & $\checkmark$ & $\checkmark$ & $\checkmark$ & $\checkmark$ & $\checkmark$ \\
\hline
\end{tabular}
\end{table*}

\section{From Classical Layers to Quantum-Converged Stacks}
OSI model has served as the backbone of communication protocol architecture for over four decades, enabling the modularization of complex networking tasks into discrete functional layers. However, the emergence of quantum communication presents a fundamental shift in how information is encoded, transmitted, and secured, thus challenging the validity of the classical OSI stack assumptions. In classical networks, packet duplication, error correction via redundancy, and deterministic routing rely on the physical ability to clone and retransmit digital signals. Quantum communication systems, in contrast, operate under constraints such as the no-cloning theorem, coherence decay, and probabilistic measurement collapse, which defy the foundational expectations of classical protocol design \cite{Huang2024LayeredSurvey}.
Recent advances in quantum internet research have demonstrated the urgent need for revisiting and reengineering the protocol stack to accommodate entanglement-based routing, QKD, and teleportation protocols \cite{Illiano2022QuantumStackSurvey}. Similarly, conventional protocol simulators like SeQUeNCe \cite{Wu2021SeQUeNCe} and QuNetSim \cite{Das2024CQStackIoT} have highlighted the inadequacy of classical-layered abstractions in handling dynamic entanglement states, fidelity-driven sessions, and distributed quantum orchestration.

Moreover, as pointed out by Bassoli et al., a naive mapping of quantum functionalities onto classical OSI layers not only fails to preserve quantum state integrity but also overlooks the essential role of cognitive agents and machine learning models in managing stack behavior across fluctuating coherence conditions \cite{Bassoli2021DesignQuantumNets}. Unlike the strict layer independence in the classical OSI model, quantum protocols often require tight cross-layer coordination, particularly for real-time updates on entanglement quality, decoherence rates, and cryptographic key freshness \cite{Pirker2018ModularQuantum}.
Therefore, this section is a critical architectural prelude to the quantum-aware protocol stack proposed in this survey. We examine the structural divergence between classical and quantum assumptions, discuss the need for OSI model extension, and motivate the introduction of two new layers: Layer 0 (Quantum Substrate) for physical quantum processes and Layer 8 (Cognitive Intent Plane) for orchestrated, AI-guided stack intelligence. This forms the basis of a quantum-converged model essential for 7G-era networks.

\subsection{Architectural Legacy: The Classical OSI Model}
The classical OSI model was a landmark achievement in formalizing communication systems into an abstracted, modular framework. Its influence persists across every modern protocol stack, including TCP/IP, LTE, and 5G architectures. However, its enduring success is also a source of architectural inertia—many of its foundational assumptions are hardwired into current design practices, even as communication paradigms evolve rapidly.
The model’s strict vertical separation of concerns allowed for innovation in individual layers but at the expense of inter-layer responsiveness. This rigidity is particularly problematic for quantum systems, where phenomena like entanglement, coherence decay, and probabilistic collapse require continual feedback loops between physical, transport, and application layers. Moreover, classical assumptions such as determinism, reversibility, and error tolerance through redundancy do not translate into the quantum domain. Therefore,
the transition to quantum networking demands a fundamental rethinking of these legacy constraints. While the OSI model offers a useful starting point for structural comparison, its lack of mechanisms for handling non-classical behaviors (e.g., no-cloning, entangled routing, real-time fidelity assessment) reveals a conceptual and practical mismatch. Contemporary research notes that layered independence must be traded for cross-layer intelligence and real-time adaptability \cite{Pirker2018ModularQuantum, Huang2024LayeredSurvey}. The role of orchestrators, AI agents, and simulation-aware abstractions must be embedded into the architectural core—not retrofitted post hoc.
Thus, while the OSI model continues to guide protocol logic in legacy systems, its principles are increasingly incompatible with the needs of post-classical communication. The upcoming sections argue for a restructured protocol stack that integrates quantum physics, machine intelligence, and application-aware adaptation as first-class design goals.

\subsubsection{Historical Context and Determinism}
The International Organization for Standardization (ISO) established the OSI model in the late 1970s as a reference framework for building interoperable, modular communication systems. Each of the seven layers was intended to encapsulate a well-defined function, from physical transmission to end-user application services. One of the foundational assumptions of this model was \textit{determinism}—the idea that data packets, once dispatched, would follow predictable paths and behavior, suitable for synchronization and flow control \cite{Popovic1998Mechatronics}. In industrial and safety-critical systems, deterministic delivery further evolved to mean bounded latency and guaranteed order preservation \cite{Decotignie2005RealTimeEthernet}.
This property became essential in domains such as SCADA systems, avionics, and robotics, where communication reliability was engineered via layered abstractions. However, this deterministic behavior fundamentally relies on the ability to duplicate, retransmit, and buffer digital information—operations incompatible with the no-cloning constraint in quantum mechanics.

\subsubsection{Assumptions of Reversibility and Redundancy}
OSI-based protocols assume that transmission errors can be mitigated using redundancy. Techniques such as Cyclic Redundancy Checks (CRC), Forward Error Correction (FEC), and packet retransmissions are standard mechanisms in the data link and transport layers. These methods presume that information can be reconstructed, delayed, or resent without consequence—a practice integral to error recovery in classical stacks \cite{Paulitsch2018IndustrialApplications}.
Quantum systems, by contrast, operate under the irreversible collapse of quantum states upon measurement and prohibit copying of unknown qubit states due to the no-cloning theorem. These principles make classical redundancy ineffective or even destructive in quantum communications. Hence, OSI’s error correction paradigms are inadequate and potentially incompatible with the requirements of emerging quantum communication systems, necessitating the development of an upgraded architectural framework.

\subsubsection{Layer Autonomy vs. Functional Entanglement}
A key strength of the OSI model lies in its vertical independence. Each layer can be developed and maintained without detailed knowledge of the others, as long as interface contracts are honored. This independence is an enabler for protocol modularity and layered troubleshooting \cite{Kim2019IndustrialSensors}. However, in quantum networks, layer isolation breaks down due to phenomena like decoherence and entanglement routing.
Quantum protocols are tightly interlinked across the stack layers, as fidelity measurements from the physical layer influence routing (network layer), session establishment (session layer), and cryptographic key validity (application layer). This cross-layer feedback breaks the OSI's assumption of autonomy, necessitating architectural constructs that support upward and downward state propagation between layers in near real-time \cite{Pirker2018ModularQuantum}.

\subsection{Quantum Paradigm Shift: Principles and Disruptions}
The advent of quantum communication introduces fundamental physical principles that directly contradict the assumptions underpinning classical network architectures. These principles—no-cloning, entanglement, superposition, coherence, and measurement collapse—not only redefine how information is encoded and transmitted but also necessitate fundamental changes to how protocol layers operate and interact. This section outlines four foundational quantum principles and their disruptive implications for stack design.

\subsection{Quantum Paradigm Shift: Principles and Disruptions}
\subsubsection{No-Cloning and Non-Replicability Constraints}
Reliability mechanisms such as packet duplication and retransmission are foundational in classical networks. However, the no-cloning theorem prohibits copying an arbitrary quantum state, making these strategies fundamentally infeasible in quantum systems. Zhou et al.~\cite{Zhou2025Teleportation} provide design strategies for teleportation-based quantum routing to circumvent this constraint. Similarly, Picchi~\cite{Picchi2023NTNQuantum} emphasizes that quantum amplifiers must operate without violating the no-cloning theorem, particularly in space-based quantum links. Iqbal et al.~\cite{Iqbal2023ImperfectCloning} analyze imperfect cloning and its detrimental effects on entanglement fidelity, cementing the no-cloning principle as a disruptive constraint.

\subsubsection{Entanglement, Superposition, and Routing Implications}
Entanglement creates non-local correlations that challenge classical routing models. A qubit’s state is no longer an isolated entity but part of a global, correlated system. Routing decisions must now consider fidelity metrics, entanglement lifetimes, and the possibility of quantum teleportation. Zhou et al.~\cite{Zhou2025Teleportation} and Khan et al.~\cite{Khan2024Heliyon} emphasize the central role of entanglement in designing protocols for space and ground-based networks. Iacovelli et al.~\cite{Iacovelli2024ProbabilityEntanglement} propose probabilistic optimization techniques for entanglement distribution and source placement—mechanisms absent in classical routing.

\subsubsection{Coherence Decay and Temporal Fragility}
Quantum coherence is the lifeblood of all qubit-based operations. However, environmental interaction rapidly degrades this property, limiting usable transmission time. Patel~\cite{Patel2019QuantumLimits} discusses the intrinsic difficulty of preserving superposition under real-world conditions. Hu et al.~\cite{Hu2023QuantumNative} explore THz-band coherence engineering to extend lifetime in mobile quantum scenarios. As quantum layers depend on maintaining coherence, every layer must be designed with temporal sensitivity and fast decision cycles.

\subsubsection{Probabilistic Transmission and Measurement Collapse}
Quantum transmission is fundamentally non-deterministic. A qubit, once measured, collapses into a definitive state, rendering classical-style read-receipt and inspection mechanisms inapplicable. Khan et al.~\cite{Khan2024Heliyon} and van Loock~\cite{vanLoock2002Continuous} underscore how quantum measurement unpredictability necessitates probabilistic design strategies and measurement-free protocol verification. This shift calls for monitoring based on statistical channel tomography and estimation—not direct state inspection.

\subsection{Cross-Layer Dependencies in Quantum Protocols}
\begin{figure}[htbp]
    \centering
    \includegraphics[width=0.55\linewidth]{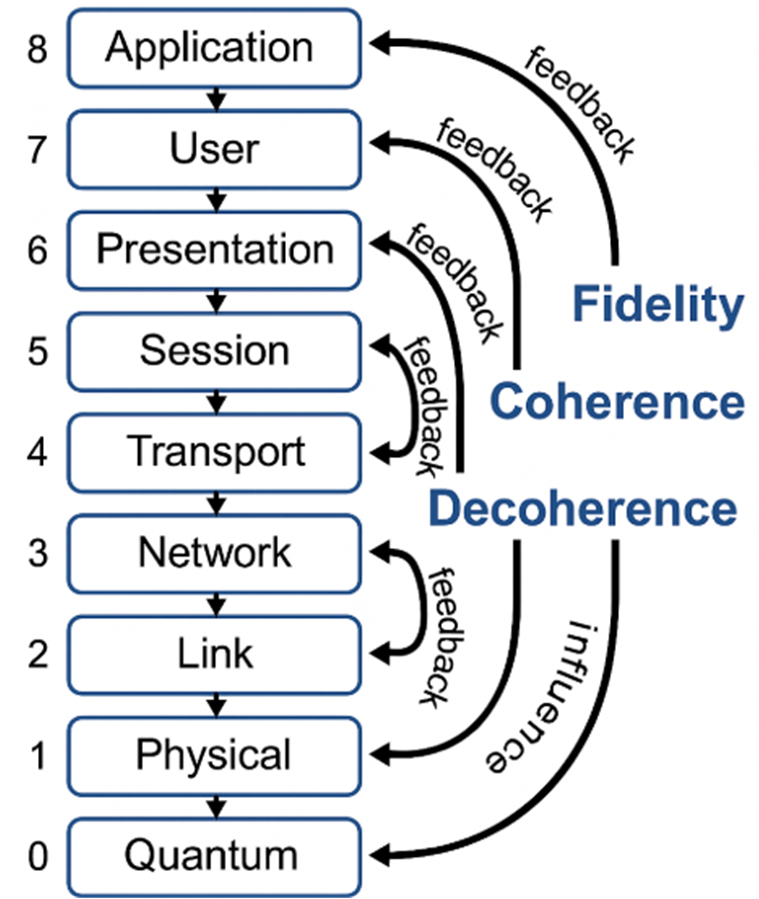}
    \caption{Feedback-coupled Quantum-Converged OSI model. Lower-layer quantum effects such as decoherence, fidelity, and coherence latency provide adaptive feedback to upper layers for dynamic session, application, and semantic orchestration.}
    \label{fig:feedback_stack}
\end{figure}

Unlike classical stacks emphasizing vertical modularity and layer isolation, quantum networking protocols exhibit significant cross-layer dependencies. This cross-layer relationship results in fidelity-aware routing, entanglement management, and decoherence control that require real-time coordination among physical, link, network, and application layers. This section outlines four core dimensions of these dependencies. Fig.~\ref{fig:feedback_stack} illustrates how coherence and fidelity-aware feedback loops across layers redefine classical OSI constraints in quantum-enhanced architectures.

\subsubsection{Feedback-Driven Stack Behavior}
Quantum protocols must account for the dynamic quality of quantum channels, which are susceptible to environmental interference and operational noise. As a result, stack behavior must be continually adapted based on live measurements—such as entanglement fidelity or coherence decay rates. Caleffi et al.~\cite{Caleffi2020Feedback} show that feedback from the physical layer directly influences decisions in entanglement routing and session control. Similarly, the Quantum Internet protocol stack proposed by Dahlberg et al.~\cite{Dahlberg2019QuantumStack} includes explicit upward and downward feedback paths, deviating from the classical OSI philosophy.

\subsubsection{Temporal Correlation Across Quantum Sessions}
Quantum communication sessions, especially those involving entanglement distribution and key generation, often exhibit temporal correlations. Thus, qubits cannot be handled as independent stateless entities across layers. NetSquid-based simulations by de Jong et al.~\cite{deJong2020NetSquid} highlight how session-layer behavior depends on the history and timing of entanglement creation, swap operations, and coherence expiry—all of which are intrinsically tied to the physical layer.

\subsubsection{Quantum Channel Variability and Real-Time Metrics}
Quantum channels are inherently probabilistic, with fidelity, photon loss rate, and Quantum Bit Error Rate (QBER) varying over time. Protocols must be designed to ingest and respond to these metrics across multiple layers. Lu et al.~\cite{Lu2020SecurityRouting} emphasize that real-time QBER evaluation at the physical layer directly affects routing tables in the network layer and session timeout policies. Therefore, the stack requires telemetry agents capable of real-time, multi-layer adaptation.

\subsubsection{Role of Simulation Environments: NetSquid, QuNetSim, QuISP}
Dedicated simulation tools have been developed to design, evaluate, and optimize such feedback-heavy and tightly coupled stacks. \textit{NetSquid} allows fine-grained simulation of quantum devices, channels, and protocol logic, supporting layer-aware fidelity tracking~\cite{deJong2020NetSquid}. \textit{QuNetSim} provides a high-level Python interface for simulating stack behavior across nodes with quantum memory and channel delay~\cite{Loke2022QuNetSim}. \textit{QuISP} focuses on large-scale, distributed quantum networking with realistic entanglement switching and qubit queuing logic~\cite{Fujiwara2020QuISP}. These tools have become essential in modeling how quantum layers co-evolve under real-world variability and noise.

\subsection{Rationale for OSI Model Extension and Redesign}
Though elegant in abstraction and historically foundational, the OSI model falls short when applied to the architectural and physical constraints of quantum communication. Its fundamental assumptions—stateless interfaces, deterministic flows, and signal replicability—are incompatible with quantum mechanics. This inadequacy has led researchers to explore and propose structural extensions integrating quantum principles and cognitive orchestration.

\subsubsection{Quantum-State Aware Abstraction at Sub-Physical Level}
One major limitation of the classical OSI model is the absence of a dedicated abstraction layer for managing the physical nuances of quantum information, such as qubit fidelity, coherence time, and entanglement states. Khan et al.~\cite{Khan2024QuantumFuture} argue for introducing a "Layer 0" beneath the physical layer, responsible for quantum state preparation, purification, teleportation channels, and coherence-aware scheduling. This layer bridges the quantum hardware and classical logic.

\subsubsection{Cognitive-Orchestrated Stack Behavior at Intent Level}
Modern quantum networks also demand high-level orchestration based on user intent, AI-driven control, and policy-based management. The traditional OSI control plane does not serve these requirements, which is largely deterministic and static. Granelli et al.~\cite{Granelli2022HybridStack} and Getu et al.~\cite{Getu2025SemanticComm} propose extending the OSI model upward to include a "Cognitive Intent Plane" (Layer 8), where stack-wide behavior is orchestrated via semantic inference, LLM-based agents, or context-aware orchestration frameworks.

\subsubsection{Disaggregated Intelligence Across Stack Boundaries}
The increasing adoption of SDN principles in classical networks has inspired similar logic for quantum networks. However, quantum systems require even tighter coordination across layers due to temporal fragility and entanglement dependence. Barros~\cite{Barros2025ProofHumanity} suggests that middleware-based disaggregated intelligence modules positioned between OSI layers can enable proactive fidelity control and adaptive entanglement management without violating the layer contracts.

\subsubsection{Hybrid Classical–Quantum Stack Synchronization}
Quantum-classical interoperability is vital, especially since control, synchronization, and most application logic still rely on classical hardware. Khan et al.~\cite{Khan2024QuantumFuture} present a dual-stack approach where classical and quantum protocols coexist and coordinate through a shared substrate interface. This harmonization is essential for managing hybrid sessions such as QKD key delivery alongside classical traffic routing.
These design needs call for an extension to the OSI layered architecture that is downward (Layer 0: Quantum Substrate) and upward (Layer 8: Cognitive Intent Plane), creating a sandwich model where physical coherence and semantic control become bookends of the protocol stack.

\subsection{Proposed Extensions to the OSI Stack}
To support the unique demands of quantum communication, researchers and architects have proposed layered extensions below and above the traditional OSI stack. These proposals address incompatibilities such as non-clonability, semantic intent orchestration, and probabilistic fidelity management. We present the two major proposed extensions: Layer 0 (Quantum Substrate) and Layer 8 (Cognitive Intent Plane).

\subsubsection{Layer 0: Quantum Substrate Layer}
Adding a sub-physical layer—Layer 0—has gained traction to encapsulate hardware-level quantum effects. This layer handles the preparation, preservation, and purification of qubits and entangled pairs, and supports decoherence-aware routing and state monitoring. Tataria et al.~\cite{Tataria2021Layer0} underscore the need for fidelity adaptation and quantum-state telemetry below Layer 1. Similarly, Vista~\cite{Vista2024QuantumSubstrate} proposes quantum substrate controllers interfacing directly with quantum memories, Quantum Processing Units (QPUs), and coherence clocks, providing quantum-context metadata upstream.

\subsubsection{Layer 8: Cognitive Intent Plane}
At the other end of the stack, the Layer 8 extension introduces a plane for cognitive decision-making \cite{hasan2021optimum}, AI orchestration, and policy-driven automation. This cognitive layer enables LLM agents and semantic orchestrators to govern lower-layer behavior, adapting stack functionality to user intent, context, and network conditions in real-time. Zeb et al.~\cite{Zeb2023NextGAI} describe a Layer 8 control plane integrated with microservice orchestration frameworks and edge intelligence. Patwary et al.~\cite{Patwary2023EdgeServices} further illustrate Layer 8 functionality in next-gen edge environments, proposing a tripartite architecture combining orchestration, cognitive control, and quantum-awareness.

\subsubsection{Integration Pathways and Design Patterns}
Recent design efforts advocate for open, programmable interfaces between the classical OSI stack and its proposed extensions. Moriondo~\cite{Moriondo2022Cybernetics} outlines a blueprint for cybernetic stack modularity using intent-driven orchestration. Tataria et al.~\cite{Tataria2021Layer0} also emphasize the value of end-to-end AI-native orchestration that spans all control and user planes, recommending a bifurcated path for classical and quantum intent flow across the stack.

\subsection{Multi-Domain Integration and Layer Cooperation}
The future of quantum communication demands cooperation across traditionally siloed network layers and technology domains. Seamless integration of satellite-based relays, edge-IoT coherence control, and decentralized trust layers mandates a harmonized protocol stack capable of temporal synchronization and cryptographic resilience. We highlight four key integration pillars.

\subsubsection{Satellite Quantum Networks and Inter-Tier Coordination}
Quantum satellite constellations, such as those envisioned by EuroQCI and Chinese Micius satellites, require multi-layer protocol coherence—from quantum-state generation to terrestrial synchronization. Hasanaj et al.~\cite{Hasanaj2024SecureSat} argue that orbital network tiers must synchronize qubit lifetimes and link-switching policies with terrestrial 5G/6G infrastructures, especially for long-distance entanglement distribution. Giannaki~\cite{Giannaki2024EUQuantumSat} further notes the necessity for pooled quantum resources and interoperability of satellite-based secure channels across national borders.

\subsubsection{Edge-Based Quantum IoT and Coherence Synchronization}
Quantum-enabled IoT (QIoT) systems pose unique challenges in maintaining coherence across distributed, edge-located sensors and actuators. Javeed et al.~\cite{Javeed2024QuantumFL} demonstrate the value of federated learning supported by edge-based quantum synchronization using satellite-assisted QKD. The authors advocate for real-time feedback loops to synchronize quantum memory states and perform entropy budgeting across constrained devices.

\subsubsection{Quantum Twin Systems and Overlay Models}
Digital twin systems are being extended to “quantum twins,” replicating entangled system states in cloud-hosted environments. Bishwas and Sen~\cite{Bishwas2024StrategicRoadmap} describe these overlays as policy-governed replicas that can simulate entanglement swap strategies, coherence optimization, and dynamic protocol adaptation. They argue that middleware is essential for aligning classical and quantum twin operations across organizational layers.
\subsubsection{Interoperability with PQC and Blockchain-Driven Identity}
PQC and Decentralized IDentity (DID) systems based on blockchain offer resilience during the quantum transition phase. Idima and Nwaga~\cite{Idima2024PQCBlockchain} explore hybrid identity models where QKD enables forward secrecy and blockchain secures session state and user credentials. This fusion allows session resumption, zero-trust interactions, and seamless stack upgrades without classical session key compromise.
These sub-domains collectively highlight the emergent need for stack-level cooperation between satellite overlays, edge environments, cryptographic substrates, and semantic orchestrators—pushing beyond classical OSI rigidity into a converged quantum-heterogeneous model.

\subsection{Paradigm Realignment and Future Stack Readiness}
The transformation from classical to quantum networking represents a foundational paradigm shift in how communication stacks are conceptualized, implemented, and orchestrated. Unlike the classical OSI model—founded on deterministic processing, independently functioning layers, and replicable digital states—the quantum networking paradigm necessitates architecture that embraces physical constraints such as entanglement, decoherence, and measurement-induced collapse. These phenomena require a stack-wide rethinking of data encoding and transport, session orchestration, and control plane abstraction.
Recent work by Caleffi et al.~\cite{Caleffi2024DistributedQuantum} underscores that quantum protocol design is inherently non-modular in the classical sense, requiring active negotiation between layers based on physical-state feedback. This realignment challenges the modularity dogma of the OSI model, prompting a shift toward context-driven and temporally-aware behavior. Illiano et al.~\cite{Illiano2022QuantumStackSurvey} provide a thorough overview of quantum protocol layering, advocating for cross-layer orchestration mechanisms that leverage software-defined control, post-quantum cryptography, and real-time entropy awareness.

Further, as highlighted in the satellite communication context by de Forges de Parny et al.~\cite{Parny2023SatelliteQuantumNet}, future networks must support the co-existence of classical and quantum layers, enforcing interoperability across entangled satellite constellations, terrestrial fiber infrastructures, and post-quantum secured edge nodes. In response, architectural proposals have introduced Layer 0 (Quantum Substrate) to support entanglement and coherence maintenance, and Layer 8 (Cognitive Intent Plane) to accommodate AI-enabled orchestration, policy adaptation, and semantic control.
Overall, the readiness of the communication stack for quantum networking hinges on its ability to transcend the linear, layered abstraction and embrace a converged, feedback-driven, and policy-aware architecture. It must incorporate semantic intelligence, cross-domain protocol mapping, and resilience against probabilistic and non-reversible quantum operations, which traditional network paradigms are fundamentally ill-equipped to handle.

\section{Quantum Communication and OSI Stack Gaps}
Quantum communication introduces a suite of physical-layer phenomena—such as entanglement, coherence decay, and the no-cloning principle—that fundamentally challenge the design assumptions of classical protocol stacks such as the OSI model. While the OSI stack offers a layered abstraction beneficial for modular design and deterministic behavior, it fails to accommodate quantum constraints that are non-replicable, probabilistic, and time-sensitive.
For instance, the no-cloning theorem directly contradicts the foundational concept of packet retransmission at Layer 2 and 4~\cite{Cobourne2011QKD}. Similarly, coherence loss imposes a strict temporal coupling between protocol layers, as demonstrated in coherence-tracked entanglement routing protocols~\cite{Pompili2022StackEntanglement}. Furthermore, traditional error correction codes are inadequate for quantum noise models, which include phase flip, amplitude damping, and depolarizing errors~\cite{Dutta2024QuantumComms}.

These constraints necessitate a realignment of the stack architecture. Illiano et al.~\cite{Illiano2022QuantumStackSurvey} propose a stack-wide redesign introducing new control and semantic layers tailored to quantum-classical hybrid operations. Similarly, Granelli et al.~\cite{Granelli2022HybridStack} highlight the need for cognitive intent overlays and cross-layer orchestration to interpret quantum session requirements in real-time.
The classical OSI stack must evolve as quantum networking moves closer to deployment—e.g., through satellite QKD systems and fiber-integrated testbeds~\cite{Vedovato2024QBER}. This section investigates these mismatches and outlines emerging solutions, including the formalization of Layer 0 (Quantum Substrate) and Layer 8 (Cognitive Intent Plane), which expand the design space beyond legacy architectural boundaries.

\subsection{Foundational Incompatibilities with Classical OSI Stack}
Quantum networking requires fundamentally reevaluating classical protocol assumptions embedded in the OSI model. Unlike classical digital communication, quantum systems impose physical laws that limit duplication, introduce probabilistic measurements, and demand real-time coherence awareness. These issues permeate every layer of the OSI stack and create profound misalignment between classical expectations and quantum capabilities.

\subsubsection{Violation of the No-Cloning Principle}
One of the most disruptive principles in quantum mechanics is the no-cloning theorem, which asserts that creating an identical copy of an arbitrary unknown quantum state is impossible. This contradicts a foundational concept in classical networking—packet retransmission and duplication, which are standard at Layers 2 and 4 for loss recovery. Dutta and Bhuyan~\cite{Dutta2024QuantumComms} explain how this constraint invalidates Automatic Repeat reQuest (ARQ) and TCP-style recovery mechanisms in quantum contexts, where measurement collapses the quantum state and retransmission becomes infeasible.

\subsubsection{Temporal Fragility and Loss of Coherence}
Quantum states, such as qubits and entangled pairs, are fragile and subject to decoherence—the process by which quantum information is lost due to interaction with the environment. This makes timing and synchronization critical across all stack layers. Picchi~\cite{Picchi2023NTNQuantum} highlights that maintaining coherence across satellite-ground quantum links requires timing precision and continuous cross-layer coordination, especially when switching quantum channels or relaying entangled states.

\subsubsection{Non-Classical Error Models }
Unlike classical channels, which are dominated by bit-flip errors mitigated by cyclic redundancy checks and forward error correction codes, quantum channels suffer from depolarization, amplitude damping, and phase flips. Pintore~\cite{Pintore2021QKDSoftwarised} shows that classical models are not designed to detect or correct these quantum-specific noise models. Instead, quantum error correction schemes such as Shor, Steane, and surface codes must be implemented as cross-layer features, tightly integrated with physical layer feedback.

\subsubsection{Entanglement-Aware Topology vs Stateless Routing}
Classical routing algorithms are typically stateless and rely on abstracted hop counts or link costs. However, routing must consider entanglement fidelity, swapping strategies, and coherence windows in quantum networks. Granelli et al.~\cite{Granelli2022HybridStack} emphasize that quantum routing is stateful and entanglement-aware, where link viability cannot be abstracted statically. This disrupts assumptions in Layers 3 and 4 of the classical model, requiring integration of fidelity and availability metrics in routing logic.

\subsection{Limitations in Stack Behavior and Control Planes}
Beyond foundational mismatches, quantum communication challenges the operational dynamics of the OSI stack, particularly in the behavior of control planes and session orchestration. Classical networks rely on deterministic logic, stateless feedback, and protocol modularity. However, these assumptions falter when applied to quantum systems, which operate under constraints of probabilistic transmission, fidelity-dependent routing, and context-aware orchestration. This subsection outlines three critical systemic limitations that hinder the direct adoption of classical control frameworks in quantum stacks.

\subsubsection{Inadequacy of Deterministic Control Protocols}
Conventional control protocols (e.g., OSPF, BGP, TCP) are built on deterministic state transitions and precomputed rule sets. These approaches assume a stable and repeatable channel environment. In quantum networks, however, probabilistic behavior dominates due to measurement-induced collapse and coherence decay. Minoli and Occhiogrosso~\cite{Minoli2022QuantumComm} emphasize that deterministic decision-making in a probabilistic environment leads to suboptimal control flow, especially for entanglement management. Serrano et al.~\cite{Serrano2022QuantumSoft} argue for goal-oriented orchestration mechanisms that integrate non-deterministic states, highlighting that future control planes must support hybrid probabilistic control primitives.

\subsubsection{Absence of Feedback Loops for Fidelity}
Unlike classical systems, where link state and round-trip time suffice for path computation, quantum systems require real-time fidelity estimation, coherence lifetimes, and QBER. Illiano et al.~\cite{Illiano2022QuantumStackSurvey} present a quantum stack architecture where each layer must provide fidelity-aware feedback to adjacent layers, enabling adaptive switching and entanglement path selection. However, existing network stacks lack telemetry hooks for quantum metrics, limiting their suitability for dynamic quantum environments.

\subsubsection{Semantic Blindness in Traditional Application Layer}
The OSI application layer is traditionally syntax-driven, operating under the assumption that content is independent of physical-layer semantics. In quantum systems, however, content relevance, coherence context, and fidelity budgets are integral to session behavior. Strinati and Barbarossa~\cite{Strinati2021Semantic6G} propose a paradigm shift toward semantic communication layers that adapt to application content and contextual fidelity constraints. Without such semantic awareness, Layer 7 cannot meaningfully coordinate quantum application tasks such as teleportation-based conferencing, QKD session initiation, or cross-layer orchestration.

\subsection{OSI Extensions Toward Quantum-Native Architectures}
As quantum communication reveals the classical OSI stack's critical structural and functional limitations, researchers have advocated for novel architectural augmentations. These extensions aim to natively accommodate the probabilistic, coherence-sensitive, and fidelity-bound operations that define quantum networking. Chief among these enhancements are the proposed additions of Layer 0 and Layer 8, which provide architectural entry and exit points for quantum substrates and semantic-intent orchestration, respectively. Alongside, middleware architectures are being explored to bridge quantum-classical interoperability, enabling hybrid deployment models and transitional adoption pathways.
Layer 0 is introduced to capture the hardware-near abstractions essential for managing entanglement fidelity, coherence times, quantum memory states, and teleportation channels. Gonzalez-Guerrero et al.~\cite{Gonzalez2021AIControlQuantum} discuss this layer as a programmable substrate that interfaces directly with physical quantum devices, tracking quantum metrics in real-time and supplying them to upper stack layers for context-aware routing. This layer also interfaces with qubit reset logic, error correction feedback, and quantum link-layer signaling.
The Cognitive Intent Plane, proposed as Layer 8, translates high-level user goals into quantum-executable protocol flows. Inspired by AI-defined networking and semantic communications, this layer leverages machine reasoning, LLMs, and goal-based orchestration strategies~\cite{Song2022NetworkingAI}. By interpreting quantum session constraints—such as entanglement quality, privacy thresholds, and fidelity margins—Layer 8 performs dynamic policy updates that affect all lower layers in a closed semantic loop.
Another important challenge for near-term deployment is ensuring that hybrid network architectures can coexist seamlessly with classical infrastructures. Middleware sidecars have emerged as critical components to interface quantum protocol logic with classical control and data planes. Granelli et al.~\cite{Granelli2022HybridStack} outline middleware strategies that encapsulate quantum protocol stacks while exposing standardized Application Programming Interfaces (APIs) for SDN controllers, blockchain identity frameworks, and PQC systems. These sidecars provide emulation hooks, security context handlers, and stackless control paths across entangled and classical flows.

\subsection{Comparative Stack Frameworks}
As the limitations of the classical OSI model in quantum networking become increasingly evident, a diverse array of stack architectures has emerged to bridge the conceptual and operational divide. These proposed frameworks vary in granularity, modularity, and compatibility with hybrid deployments. Some aim to minimally adapt existing layers for quantum compatibility, while others advocate for fully restructured quantum-native stacks with new layering principles, APIs, and control abstractions. This section synthesizes major contributions from the literature—ranging from modular quantum stacks and hybrid quantum-classical designs to emerging standards like Quantum Intermediate Representation (QIR) and Software-Defined Quantum Networking (SDQN)—offering a taxonomical view of how researchers envision the future quantum internet protocol architecture.

\subsubsection{Modular Stack Models}
Pirker and Dür were among the first to formalize a modular architecture for quantum networks. Their framework proposed stack decompositions where fidelity tracking, entanglement routing, and teleportation logic were treated as composable layers. Their vision emphasized reliability, protocol independence, and quantum-compatible abstraction boundaries~\cite{Pirker2019Modular}.
Based on this foundation, Caleffi, Cacciapuoti, and colleagues introduced a more integrative stack model incorporating real-time fidelity propagation, cross-layer telemetry, and entropy feedback. Their architecture enables decentralized quantum control by leveraging quantum physical metrics as active triggers in routing and session decisions~\cite{Illiano2022QuantumStackSurvey}.

\subsubsection{Symbiotic Quantum-Classical Stack Designs}
Granelli, Bassoli, and Nötzel introduced a hybrid OSI-compliant architecture that maintains classical control paths while embedding quantum logic in specialized vertical slices of the stack. Their proposal defines standardized APIs between classical and quantum segments, enabling resource mapping, session synchronization, and cross-technology handovers without compromising compatibility with legacy systems~\cite{Granelli2022HybridStack}.
To contextualize our unified OSI redesign within ongoing research efforts, Table~\ref{tab:qnet-challenges} summarizes the current state of quantum networking initiatives across seven key architectural and orchestration challenges. It highlights the need for a cohesive, layered protocol approach, as introduced in this work.
\begin{table*}[htbp]
\centering
\caption{Projects and Research Addressing Quantum Networking Challenges}
\label{tab:qnet-challenges}
\begin{tabular}{|l|p{4.5cm}|p{3cm}|}
\hline
\textbf{Quantum Challenge} & \textbf{Projects/Protocols} & \textbf{Research} \\
\hline
Quantum Stack Architecture & SEQUOIA, Quantum Internet Alliance, EU Quantum Flagship & \cite{Chamola2025FutureConnectivity}, \cite{Khan2023SysReviewQuantum} \\
\hline
Entanglement Distribution & QuNetSim, NetSquid, QRNA, QLoop & \cite{Zhao2024QuantumWireless}, \cite{Glisic2024QuantumNeuroSurvey} \\
\hline
Retransmission and Quantum Routing & QSPN, Q-RoadNet, QStackSim & \cite{Zhao2024QuantumWireless}, \cite{Chamola2025FutureConnectivity} \\
\hline
Control Plane and Stack Orchestration & ETSI ISG QKD, QuantumIRIS, LUMIQ-NET & \cite{Chamola2025FutureConnectivity}, \cite{Mehic2023} \\
\hline
Hybrid Classical/Quantum Integration & QKDnet, Q3P, QPACE & \cite{Mehic2023}, \cite{Zhao2024QuantumWireless} \\
\hline
Quantum Security (QKD, PQC) & SECOQC, OpenQKD, IDQ Network & \cite{Mehic2023}, \cite{Chamola2025FutureConnectivity} \\
\hline
AI/QML-enhanced Network Intelligence & QML-Stack, QuantumGraphNet & \cite{Glisic2024QuantumNeuroSurvey}, \cite{Zhao2024QuantumWireless} \\
\hline
\end{tabular}
\end{table*}

\subsubsection{Emerging Standards: QIR, QCoDeS, and SDQN}
Brito and collaborators have outlined a blueprint for scalable quantum network deployments by leveraging SDN paradigms in quantum systems. Their model includes centralized control loops, intent-aware service orchestration, and programmable interfaces for entanglement allocation. They also examine emerging toolchains such as Microsoft’s QIR and QCoDeS as essential building blocks for hardware-software integration across the quantum stack~\cite{Brito2024BlueprintQuantumNet}.

\subsection{Protocol Viability in Quantum OSI Redesigns}
Designing a robust quantum protocol stack requires architectural innovation and empirical validation through simulation and benchmarking. Unlike classical networks, quantum systems must be evaluated against novel metrics such as coherence cost, fidelity drift, and probabilistic throughput. Additionally, quantum data's non-clonable and non-observable nature presents significant debugging and traceability challenges. This section examines current simulation testbeds, performance metrics, and diagnostic tools developed to assess the viability of quantum OSI stack proposals in realistic conditions.

\subsubsection{Simulators and Testbeds}
A suite of dedicated quantum networking simulators has been developed to evaluate the viability of proposed quantum protocol stacks. These tools support discrete-event simulation, entanglement fidelity tracking, decoherence modeling, and hybrid quantum-classical flows—essential capabilities lacking in classical simulators.
QuNetSim offers a lightweight Python-based framework designed for fast prototyping and educational deployments. DiAdamo et al.~\cite{9465750} emphasize its extensibility for quantum routing and entanglement swapping scenarios, making it ideal for academic validation and early-stage stack logic testing.
NetSquid, developed by QuTech, provides one of the most comprehensive quantum simulation environments. Coopmans et al.~\cite{Coopmans2021NetSquid} highlight its discrete-event architecture and modular component-based modeling, enabling high-fidelity simulation of teleportation, memory decay, and complex quantum topologies. It is widely adopted for research on realistic quantum networks with noise models and decoherence integration.

SimulaQron, as described by Dervisevic et al.~\cite{Dervisevic2024SimulaQron}, supports synchronous and asynchronous quantum operations within a distributed multi-node environment. Its primary strength is abstracting hardware behavior, providing sockets and APIs to simulate qubit exchange and quantum gate invocation across networked nodes.
QuISP takes a more protocol-oriented approach. As shown in Joubert's survey~\cite{joubert2024quisp}, it emphasizes programmable fidelity-driven behaviors and queue-based entanglement scheduling, simulating protocol stacks with fidelity budget enforcement and probabilistic routing decisions. It is particularly well-suited for testing full-stack interactions under stochastic decoherence.
Each simulator thus caters to distinct aspects of the quantum stack validation pipeline—ranging from abstraction-level API testing (QuNetSim) to realistic entanglement fidelity simulations (NetSquid), making them essential tools in quantum OSI redesign evaluation.

\subsubsection{Fidelity Drift, Entropy Rate, Probabilistic Throughput}
Evaluating the viability of quantum network protocols demands the development of performance metrics that account for quantum-specific phenomena—many of which have no classical analogs. Traditional throughput and latency must be reinterpreted in probabilistic and non-deterministic terms.
Fidelity, which measures the closeness between an ideal quantum state and the experimentally produced one, is a central metric. Proctor et al.~\cite{Proctor2022MeasuringQuantumCapabilities} developed a comprehensive framework to characterize quantum capabilities using process fidelity, emphasizing its sensitivity to gate imperfections and decoherence effects.
Fidelity drift, a temporal degradation of quantum gate performance or channel stability, often results from environmental noise or calibration drifts. Nguyen et al.~\cite{Nguyen2024DriftEntropyBenchmark} modeled linear entropy to quantify temporal fidelity variance, highlighting how thermal noise and timing jitter propagate through quantum stacks.

Furthermore, entropy rate is increasingly used to quantify the degree of decoherence and information leakage from a system. Proctor and Baczewski~\cite{Proctor2025QuantumBenchmarking} examined entropy production during quantum evolution and emphasized its role in differentiating gate performance under Noisy Intermediate-Scale Quantum (NISQ) conditions.
Moreover, probabilistic throughput is the likelihood of successfully delivering entangled pairs or executing gate operations under probabilistic noise models. Baum et al.~\cite{Baum2021DeepRLPRX} demonstrated that Reinforcement Learning (RL) models could optimize such throughput by learning to manage entanglement queues and coherence windows adaptively.
Finally, coherence cost refers to the quantum resource expenditure needed to maintain quantum states over time or across hops. Liu et al.~\cite{Liu2010AdvancesSpinControl} provide a thermodynamic perspective, associating coherence loss with entropy dissipation and heat accumulation in spin-based quantum systems.
These metrics collectively form a multidimensional benchmarking strategy for evaluating proposed quantum OSI stacks, helping determine protocol robustness, resource consumption, and cross-layer optimization requirements.

\subsubsection{Cross-Layer Debugging and Temporal Traceability}
One of the most persistent barriers in building robust quantum protocol stacks is the absence of native support for cross-layer debugging and temporal traceability. In classical systems, logging, packet inspection, and state replay mechanisms allow detailed analysis of protocol behavior. However, such strategies are infeasible in quantum contexts due to the destructive nature of measurement and the temporal sensitivity of coherence.
Huang et al.~\cite{Huang2024LayeredSurvey} highlight that decoherence degrades qubit quality and introduces traceability ambiguities when states transition between layers. Their survey indicates the absence of fidelity-preserving diagnostics that can operate non-invasively across protocol boundaries.
Furthermore, protocol simulation environments like NetSquid and SimulaQron have only recently begun to include modular hooks for logging entanglement lifetime, teleportation delays, and measurement noise. These are often non-standardized and layer-specific, hindering stack-wide introspection. Debugging quantum sessions remains fundamentally challenging without structured temporal maps of fidelity, entropy, and routing state.
The need for “quantum state observability,” without collapse, motivates the design of abstract coherence monitors and fidelity beacons—tools capable of offering statistical insight without violating quantum constraints. Their development is essential for practical deployment, particularly in dynamic topologies where entangled paths must be validated end-to-end.

\subsection{Layered Quantum Networking Opportunities}
As quantum technologies mature, they present new opportunities for innovation at every layer of the networking stack. Unlike classical systems, where functionality is often confined to well-defined modular boundaries, quantum networking requires deeply interdependent protocols that manage entanglement, fidelity, and coherence across the entire OSI spectrum. This section explores emerging design opportunities and open research challenges associated with adapting or reimagining each layer—ranging from quantum repeaters and error-resilient MAC protocols at the lower levels to session management, compression, and intent-driven orchestration at the upper layers. By examining each layer through the lens of quantum awareness, this section establishes a roadmap for stack-wide innovation in 7G-era networks and beyond.

\textit{Layer-by-layer overview:}
\begin{itemize}
    \item \textit{Layer 1 – Physical:}
The fundamental challenge at the physical layer of quantum networks is to reliably transmit quantum states—often embodied as photons or spin states—across noisy, lossy, and decoherence-prone media. This layer establishes and preserves quantum entanglement over short and long distances using mechanisms such as optical fibers, free-space optics, satellite links, and emerging quantum RF-based schemes.
Entanglement-based photonic transmission is the cornerstone of quantum communication. Pompili et al.~\cite{Pompili2022StackEntanglement} demonstrated the feasibility of layered entanglement delivery across quantum nodes using optical fiber setups, where fidelity degradation due to physical impairments was addressed via temporal batching and quantum memory buffering.
Quantum RF is an emerging modality to extend entanglement distribution beyond photonic channels. Pierucci et al.~\cite{Pierucci2024NontTerrQuantum} explored radio-based entanglement interfaces for Low-Earth Orbit (LEO) satellite communication, showing that careful electromagnetic shielding and coherent phase alignment can support stable RF-based entangled qubit exchange in non-terrestrial environments.
RIS are now being explored to enhance quantum signal propagation. Zhang et al.~\cite{Zhang2024QuantumRIS} proposed a RIS-assisted entanglement routing scheme to mitigate polarization drift and environmental decoherence during optical transmissions. Their simulations showed significant improvements in end-to-end fidelity and spatial coverage compared to conventional line-of-sight models.
Moreover, decoherence—defined as the loss of quantum coherence due to interaction with the environment—remains a fundamental bottleneck. Koudia et al.~\cite{Koudia2024PhysLayerSurvey} reviewed state-of-the-art physical-layer techniques such as cryogenic cooling, entanglement purification, and photon-number-resolving detectors for coherence preservation.
Finally, quantum repeaters, reviewed extensively by Azuma et al.~\cite{Azuma2023Repeaters}, enable long-distance entanglement distribution by breaking the communication into short entangled segments with intermediate purification and swapping. These repeaters and physical-layer quantum error correction form the backbone of scalable and fault-tolerant quantum networks.
Together, RIS-assisted enhancement, RF entanglement channels, photon-based quantum links, and repeater-assisted entanglement swapping form the key components of the next-generation quantum physical layer.

    \item \textit{Layer 2 – Data Link:}
The data link layer in quantum networks faces novel constraints due to quantum coherence, entanglement volatility, and measurement sensitivity. Unlike classical MAC protocols, which assume deterministic channel access and bit-wise transmission, quantum MAC mechanisms must incorporate probabilistic resource availability and fidelity-aware traffic prioritization.
Granelli et al.~\cite{Granelli2022HybridStack} propose a buffer-prioritized MAC scheme tailored for hybrid classical-quantum communication networks, where qubit queueing policies are adjusted dynamically based on decoherence time and fidelity thresholds. This adaptive MAC design aims to avoid fidelity degradation during channel arbitration by scheduling qubits in descending order of temporal criticality.
Gauthier et al.~\cite{Gauthier2025ModularQuantum} introduced a modular data link framework with layered fidelity guarantees. Their architecture integrates entanglement buffer management with MAC arbitration, enabling programmable delays that account for qubit memory lifetime and purification latency.
In addition, Wang et al.~\cite{Wang2023PPQ} proposed a Prioritized Purification (PP) scheme at the link layer that couples the scheduling of entanglement swapping with fidelity targets. This enables fidelity-constrained routing without compromising entanglement availability.
At the heart of these efforts lies fidelity-aware scheduling, a mechanism wherein packet prioritization is influenced by quantum state integrity and expiry timelines. These policies often require feedback loops from physical-layer fidelity monitors or real-time predictions from coherence estimators~\cite{Avis2025SDQN}.
Collectively, these developments illustrate the emergence of a distinct quantum MAC layer—fundamentally different from classical CSMA/TDMA systems—designed to optimize throughput under probabilistic, temporally-sensitive constraints in quantum communication environments.

    \item \textit{Layer 3 – Network:} The quantum network layer orchestrates the transmission of entangled qubits between distant nodes, necessitating routing mechanisms that differ significantly from IP-based classical networks. The ephemeral nature of entanglement and coherence loss over time and distance means that routing decisions must be made with an awareness of fidelity decay, entanglement rates, and quantum memory lifetimes.
Kumar et al.~\cite{Kumar2023RoutingSurvey} highlighted that quantum entanglement routing requires proactive fidelity estimation, often coupling probabilistic path success rates with purification delays. Unlike deterministic paths, entangled links have temporal constraints that can degrade mid-transit if not refreshed or swapped.
Quantum Software-Defined Networking (Q-SDN) has been proposed as a programmable control paradigm to address these challenges. Chiti et al.~\cite{Pierucci2024NontTerrQuantum} designed an SDN-based architecture for quantum satellite backbones, where entanglement quality, coherence budgets, and memory state were abstracted into controller APIs. This separation of control and data planes enables global optimization of fidelity-aware routing strategies.
Pierucci et al.~\cite{Pierucci2022Metropolitan} extended this concept to drone-based quantum metropolitan networks, implementing a controller-driven approach that dynamically adapts routing paths based on coherence duration and fidelity drift predictions.
Emerging coherence-aware routing schemes also leverage quantum-aware heuristics. Abane et al.~\cite{abane2025entanglementrouting} proposed the Temporal Purification and Entanglement-aware Distribution (TPED) policy, which factors qubit age and remaining coherence time into route computation. This approach achieved higher fidelity stability under mobility and link perturbations in space-air-ground environments.
Finally, Zhang and Liu~\cite{Zhang2023MultipathRouting} introduced multipath segment-based routing for quantum networks, where entangled pairs are distributed concurrently over parallel paths and reconciled through purification, thereby enhancing reliability and resilience.
These innovations lay the foundation for a programmable, coherence-aware network layer with policy-driven fidelity optimization and SDN-augmented control logic.
    \item \textit{Layer 4 – Transport:} 
The transport layer in quantum networks must manage end-to-end session continuity while accounting for entanglement loss, fidelity fluctuations, and the absence of retransmission mechanisms. Unlike TCP in classical networks, which uses acknowledgments and retransmissions to guarantee delivery, quantum transport protocols must operate probabilistically and under irreversible measurement constraints.
Baseri, Chouhan, and Hafid~\cite{Baseri2024QuantumSafeSurvey} conduct a detailed study on quantum-safe protocols at the transport layer, highlighting the shift toward PQC control primitives integrated into TLS-like frameworks. They emphasize that entanglement-driven connections must rely on lightweight session negotiation protocols, including fidelity estimation, quantum key verification, and session token lifetimes. Their proposed architecture supports QTLS—quantum-resilient TLS variants that interoperate with post-quantum public key infrastructure.
In addition to that, a complementary reliability measure is based on entropy modeling. Instead of binary success or failure of message delivery, quantum protocols track entropy gradients to evaluate how much information coherence is preserved during transmission. Entropy-based reliability allows adaptive protocol behaviors, where session strategies are adjusted based on coherence decay rather than fixed thresholds.
Moreover, this layer often incorporates probabilistic congestion control and session multiplexing strategies. For example, fidelity-aware flow control mechanisms, as suggested in QTLS schemes, allow dynamic reallocation of quantum resources to preserve long-distance entanglement under high-variance network conditions.
In short, the quantum transport layer must rethink the classical semantics of reliability, instead adopting entropy-centric, PQC-integrated, and session-aware abstractions for robust operation in probabilistic and non-deterministic environments.

    \item \textit{Layer 5 – Session: }  The quantum session layer plays a pivotal role in orchestrating contextual continuity, identity validation, and metadata synchronization across quantum communication channels. Unlike its classical counterpart that assumes a persistent state and replayable handshakes, the quantum session layer must support ephemeral entanglement, destructive measurements, and probabilistic token exchanges.
Len et al.~\cite{Len2023Interoperability} proposed a hybrid handshake protocol leveraging X3DH (Extended Triple Diffie–Hellman) with post-quantum primitives for interoperable group messaging. Their framework includes quantum-safe metadata fields such as entropy tolerance, coherence lifetime, and verification tokens to initiate session state agreements without collapsing the quantum channel.
Barros~\cite{Barros2025ProofHumanity} introduced a multi-layer quantum network stack where TLS 1.3 handshakes were adapted to include decentralized identity tokens embedded in session metadata. This enables operator-independent identity verification during quantum session establishment, preserving privacy while enabling inter-provider orchestration.
Mehic et al.~\cite{Mehic2022QKDNetworks} outlined session initialization schemes that embed entropy tags and quantum capability descriptors in control plane headers. These descriptors dynamically modulate handshake behavior, allowing entanglement-level capabilities to guide application-layer bindings.
Grubbs et al.~\cite{Grubbs2022Middleboxes} developed session transcript oracles where zero-knowledge token issuance enables metadata-coordinated mutual authentication. Their architecture allows tokens to capture context semantics—such as entanglement fidelity class or memory compatibility—which are then interpreted at the session plane.
These advances push the session layer from a passive boundary handler to an intelligent, decentralized, and metadata-rich orchestrator. Further developments are expected to explore learning-based session resumption strategies and integration with LLM agents that interpret contextual quantum network goals.

    \item \textit{Layer 6 – Presentation:}  The presentation layer in quantum networks is critical in transforming qubit states, encoding semantic metadata, and facilitating format-level interoperability between quantum and classical protocols. With the integration of LLMs, this layer is evolving beyond static encoders to intelligent semantic agents capable of interpreting quantum session goals and dynamically adapting formats for quantum-classical convergence.
Chaoub and Elkotob~\cite{Chaoub2025MobileLLM6G} introduced a hybrid LLM invocation architecture where model agents operate as context processors at the edge of quantum subsystems, enabling format negotiation, encoding optimization, and metadata restructuring in coherence-sensitive channels. Their work highlights that formatting is no longer a fixed function but a learning-based capability responsive to temporal entropy and task semantics.
Veronica~\cite{Veronica2025LLMBioQuantum} emphasized the FAIR (Findable, Accessible, Interoperable, Reusable) compliance of LLM-assisted formatting layers when applied to structured quantum datasets. Their work connects LLM metadata synthesis with entropy-preserving data translation across domain-specific quantum applications, including quantum SAT solvers and QML pipelines.
Ray~\cite{Ray2025ModelContextLLM} explored the use of adaptive encoding layers in LLM-driven presentation systems for post-quantum streaming. Their survey shows how metadata-rich quantum streams can benefit from runtime-optimized tokenization and selective compression guided by language models that interpret coherence limits and fidelity expectations.
Together, these innovations support the emergence of quantum-aware formatting agents—LLM-driven middleware that encodes, translates, and preserves context across layers. These agents are critical for achieving presentation-layer interoperability in applications like federated QKD, distributed QML, and entangled IoT overlays.

    \item \textit{Layer 7 – Application:} The application layer in quantum networks must expose high-level abstractions, APIs, and Software Development Kits (SDKs) that enable quantum-native and hybrid applications to interact with underlying quantum infrastructure. Unlike classical applications, which assume byte-stream interfaces and deterministic message flows, quantum applications must negotiate fidelity, entanglement, and coherence with highly specialized middleware.
Romero-Álvarez et al.~\cite{RomeroAlvarez2024QaaS} propose a comprehensive framework for quantum service-oriented computing, where APIs abstract QKD, entanglement session management, and protocol orchestration into modular services. Their system supports quantum overlay composition, enabling the chaining of distributed QKD, federated quantum learning, and secure quantum messaging.
Arias et al.~\cite{Arias2023QSESurvey} emphasize the growing need for quantum SDKs that integrate seamlessly into DevOps pipelines, especially for dynamic application contexts such as smart cities or healthcare. Their work catalogs popular SDKs (e.g., Qiskit, Cirq, Braket) and proposes formal application-layer patterns for resilience, capability advertisement, and intent negotiation.
Kop~\cite{Kop2020IPQuantum} underscores the legal and regulatory implications of QKD and application-layer APIs, pointing to the necessity of standard-compliant interfaces for interoperability and secure deployment. This is especially critical in sovereign or cross-border environments where overlay QKD services operate across jurisdictional boundaries.
Bakar et al.~\cite{Bakar2024PostQuantumIPSec} present a fully realized software-defined overlay model integrating post-quantum IPsec tunneling with a programmable application interface. Their implementation supports application-aware QKD path selection and overlays for military-grade infrastructure.
Together, these advances establish the application layer as a consumer of quantum resources and a programmable orchestrator of secure, distributed, and intent-driven quantum services.
\end{itemize}

\textit{Classical-Quantum Hybrid Interfaces:} Hybrid stacks must support seamless interaction between quantum and classical layers, particularly in post-quantum cryptography, key distribution, and synchronization scenarios. Cacciapuoti et al.~\cite{Cacciapuoti2020HybridControl} proposed an architecture that integrates classical SDN control with quantum memory state feedback, establishing an interface standard for interlayer consistency.

\textit{Simulation Environments – NetSquid, QuNetSim, QuISP:} Tools like NetSquid~\cite{Coopmans2021NetSquid}, QuNetSim~\cite{9465750}, and QuISP~\cite{joubert2024quisp} serve as critical platforms for validating cross-layer interactions. These tools support fidelity tracking, entanglement queueing, decoherence modeling, and synchronization, allowing researchers to simulate full-stack quantum protocols under realistic noise conditions.

\textit{AI/LLM Agents for Intent-Driven Network Orchestration:} LLMs and AI agents are increasingly leveraged for adaptive orchestration of quantum stacks. Chaoub and Elkotob~\cite{Chaoub2025MobileLLM6G} demonstrated how LLMs can interpret user intent and dynamically reconfigure session policies, routing paths, and fidelity constraints. These cognitive agents operate across layers, mediating between user goals and stack-level reactivity.

\section{Quantum-Converged OSI Stack}
Though pivotal to layered communication design, the classical OSI model lacks the semantic and operational constructs necessary to support quantum phenomena such as entanglement distribution, coherence preservation, and teleportation-based communication. As quantum networking transitions from theoretical proposals to experimentally viable architectures, a converged OSI-like framework must evolve to accommodate quantum-specific operations, metrics, and abstractions across the whole protocol stack. This section introduces a restructured nine-layer model—extending from Layer 0 (Quantum Substrate) to Layer 8 (Cognitive Intent Plane)—to reflect these requirements in quantum-enabled 6G/7G networks.
Layer 0 formalizes the quantum substrate, a foundational abstraction for physical quantum channels and decoherence-sensitive interfaces. Recent work by Beukers et al.~\cite{Beukers2024RemoteEnt} and Awschalom et al.~\cite{Awschalom2021QuICS} highlights how this layer supports teleportation circuits, fidelity metrics, and quantum memory coupling. At the opposite end, Layer 8 introduces cognitive orchestration powered by LLMs, which has begun to influence adaptive routing, fidelity-aware decision-making, and semantic layer mapping for quantum-classical hybrids~\cite{Illiano2022QuantumStackSurvey, Prados2023LLMQuantum}.
We revisit each OSI layer between these boundary layers in light of quantum advancements. Layer 3, for instance, must accommodate coherence-aware routing with entanglement-based topologies~\cite{Cacciapuoti2020EntanglementRouting}, while Layer 5 must incorporate contextual session models using entangled metadata and quantum tokens~\cite{Singh2021QuantumSurvey}. Cross-layer dependencies—long treated as violations of stack purity in classical design—become essential features in quantum stack orchestration, especially given the probabilistic nature of fidelity, entropy, and decoherence timescales~\cite{Caleffi2018QuantumSDN, Vajner2022QuantumDots}.
Therefore, the proposed stack preserves layered models' clarity while introducing abstractions tailored to quantum information flow, dynamic fidelity, entropy buffering, and probabilistic orchestration. Each layer will be discussed in detail with its quantum-specific responsibilities, supporting technologies, and current implementation trends.

\begin{table}[h]
\caption{Quantum-Converged OSI Stack}
\centering
\begin{tabular}{|c|l|}
\hline
Layer & Description \\
\hline
8 & Cognitive Intent Plane – LLM-based orchestration \\
7 & Application – Quantum APIs and overlay systems \\
6 & Presentation – Quantum-native metadata formatting \\
5 & Session – Contextual handshake with quantum tokens \\
4 & Transport – Flow entropy, PQC reliability \\
3 & Network – Entanglement routing, SDN control \\
2 & Data Link – Fidelity-aware MAC, QEC \\
1 & Physical – RIS-enabled entangled channels \\
0 & Quantum Substrate – Decoherence and teleportation \\
\hline
\end{tabular}
\end{table}

\subsection{Layer 0 – Quantum Substrate}
This foundational layer provides hardware-level abstraction of quantum devices and serves as the execution base for entanglement, decoherence control, and teleportation primitives. Beukers et al.~\cite{Beukers2024RemoteEnt} introduce entanglement buffering techniques to manage memory-based photonic interfaces. Similarly, Awschalom et al.~\cite{Awschalom2021QuICS} emphasize the development of quantum interconnects (QUICS) as critical enablers for Layer 0 interoperability. Malik et al.~\cite{Malik2020ChipTeleport} demonstrate chip-to-chip teleportation in silicon circuits, offering evidence of hardware substrate readiness for stack integration.

\subsection{Layer 1 – Physical}
This layer encompasses quantum signal propagation over fiber, free-space, and non-terrestrial links enhanced by Reconfigurable Intelligent Surfaces (RIS). Picchi~\cite{Picchi2023Thesis} and Pierucci et al.~\cite{Pierucci2022Metropolitan} discuss RIS applications in LEO/GEO satellite quantum systems, demonstrating how they optimize polarization alignment and channel gain in entanglement delivery. These systems complement quantum repeaters and error-tolerant photon interfaces, reviewed in Azuma et al.~\cite{Azuma2023Repeaters}.

\subsection{Layer 2 – Data Link}
At this layer, MAC protocols are redesigned to be coherence-sensitive. Granelli et al.~\cite{Granelli2022HybridStack} propose MAC arbitration considering decoherence and fidelity drift. Wang et al.~\cite{Wang2023PPQ} introduce prioritized purification and entanglement scheduling strategies to enhance QEC resilience. QuISP and NetSquid simulations~\cite{Coopmans2021NetSquid, joubert2024quisp} provide testbed support for evaluating dynamic MAC behavior under noisy conditions.

\subsection{Layer 3 – Network}
The quantum network layer manages routing decisions based on entanglement fidelity, latency, and coherence budgets. Cacciapuoti et al.~\cite{Cacciapuoti2020EntanglementRouting} propose probabilistic routing over entangled topologies, while Zhang and Liu~\cite{Zhang2023MultipathRouting} advocate multipath segment routing to ensure robustness. Q-SDN frameworks such as those by Chiti et al.~\cite{Pierucci2024NontTerrQuantum} offer control-loop separation with programmable fidelity-aware routing.

\subsection{Layer 4 – Transport}
The transport layer ensures session coherence using entropy tracking, error budgets, and post-quantum cryptographic handshakes. Baseri et al.~\cite{Baseri2024QuantumSafeSurvey} introduce a QTLS framework integrating entropy-aware tokenization and PQC primitives. This approach departs from TCP's retransmission logic in favor of fidelity-constrained transmission with entropy-based feedback loops.

\subsection{Layer 5 – Session}
This layer initiates and maintains quantum sessions using context-encoded metadata. Grubbs et al.~\cite{Grubbs2022Middleboxes} introduce zero-knowledge transcript-based oracles to exchange quantum tokens securely. Mehic et al.~\cite{Mehic2022QKDNetworks} present handshake models with embedded entropy descriptors, enabling adaptive session negotiation under coherence constraints.

\subsection{Layer 6 – Presentation}
The quantum presentation layer enables semantic transformation of quantum streams using metadata-aware formatters. Chaoub and Elkotob~\cite{Chaoub2025MobileLLM6G} propose LLM-enhanced edge agents that encode entropy tags, fidelity expectations, and format translations for cross-domain communication. Ray~\cite{Ray2025ModelContextLLM} expands this with token-aware formatting for real-time QML/IoT streams.

\subsection{Layer 7 – Application}
This layer exposes SDKs and APIs for application-layer developers, supporting QKD, quantum conferencing, and distributed quantum learning. Romero-Álvarez et al.~\cite{RomeroAlvarez2024QaaS} offer a service-oriented stack model with modular QKD APIs. Bakar et al.~\cite{Bakar2024PostQuantumIPSec} build overlay systems supporting PQC-secured IPsec tunnels for federated quantum applications.

\subsection{Layer 8 – Cognitive Intent Plane}
\begin{figure}[htbp]
    \centering
    \includegraphics[width=0.85\linewidth]{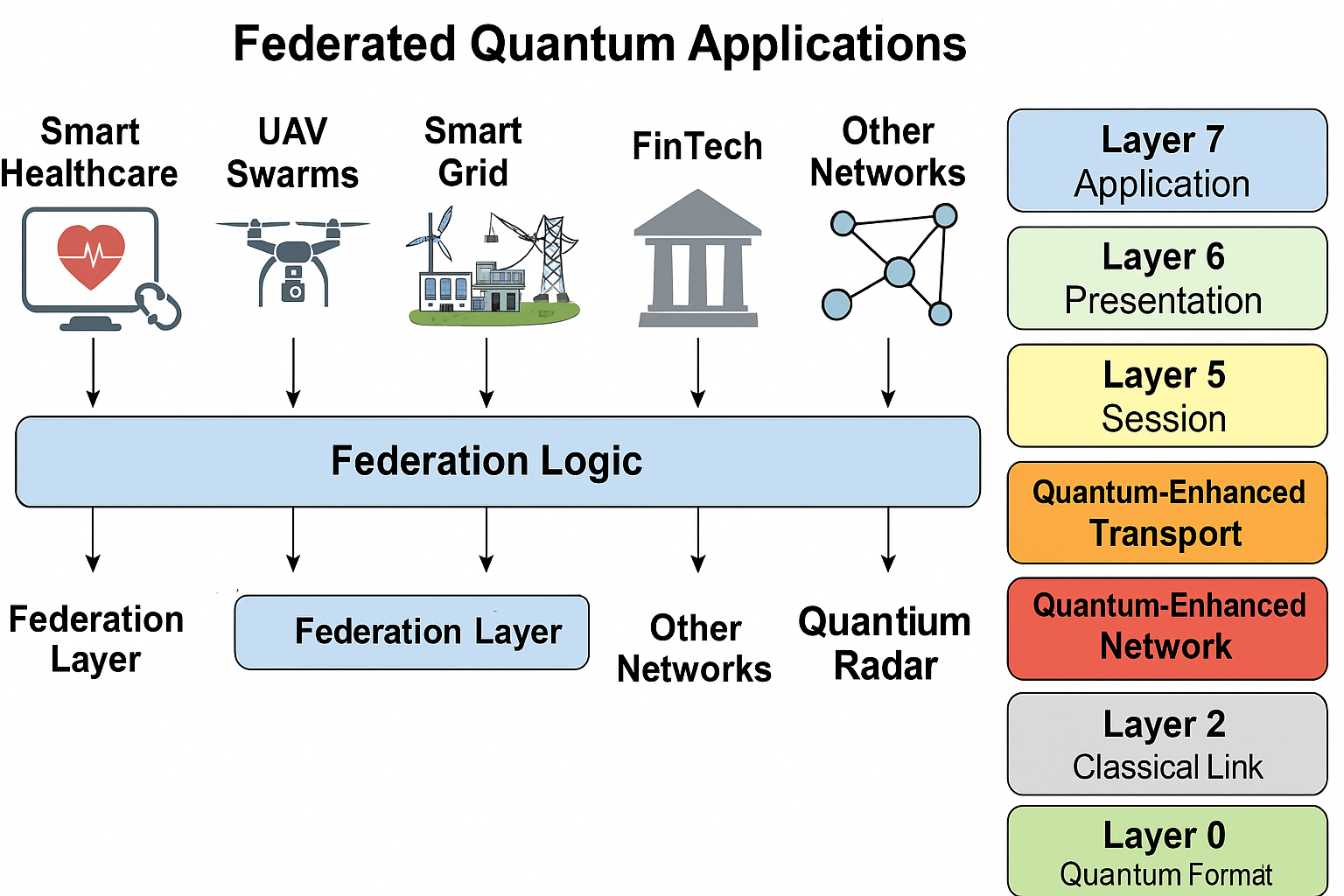}
    \caption{Federated Quantum Applications across domains such as Smart Healthcare, UAV Swarms, Smart Grids, and FinTech. These applications are orchestrated through a federation layer aligned with the proposed quantum-converged OSI model.}
    \label{fig:federated_quantum}
\end{figure}

\begin{figure}[htbp]
\centering
\includegraphics[width=0.48\textwidth]{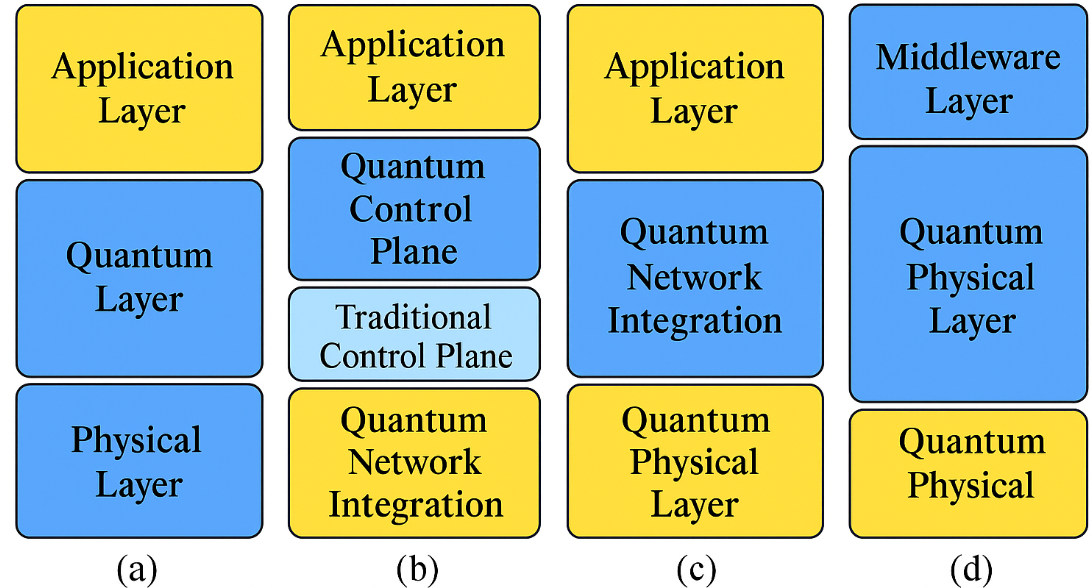} 
\caption{The proposed Quantum-Converged OSI stack including Layer 0 (Quantum Substrate) and Layer 8 (Cognitive Intent Plane).}
\label{fig:quantum-osi-stack}
\end{figure}

The cognitive layer introduces LLMs and intelligent agents to interpret user intent, manage cross-layer policies, and adapt session behavior. Prados-Garzón et al.~\cite{Prados2023LLMQuantum} demonstrate how 6G-based cognitive agents orchestrate quantum routing based on semantic awareness. Chaoub et al.~\cite{Chaoub2025MobileLLM6G} present multi-modal LLM architectures capable of real-time protocol adaptation across layers.
To illustrate the practical manifestation of our proposed quantum OSI layers, Table~\ref{tab:qnet-stack-elements} presents a categorized mapping between stack functionalities and real-world technologies. This includes entanglement-level identifiers, simulators such as QuNetSim and NetSquid \cite{diadamo2023qunetsim, deJong2020NetSquid}, quantum software frameworks like Qiskit \cite{Cross2018Qiskit} and Cirq \cite{Shi2022Cirq}, and orchestration services such as Entanglement-as-a-Service \cite{RomeroAlvarez2024QaaS}. These examples reinforce the feasibility of the stack's architectural abstraction across hardware, software, and semantic levels.
Fig.~\ref{fig:quantum-osi-stack} illustrates the proposed 9-layer Quantum-Converged OSI model, integrating quantum substrate functions and LLM-based semantic control planes.
\begin{table*}[htbp]
\centering
\caption{Quantum Network Stack Elements and Sample Technologies}
\label{tab:qnet-stack-elements}
\renewcommand{\arraystretch}{1.5}
\begin{tabular}{|p{4cm}|p{11cm}|} 
\hline
\textbf{Quantum Stack Elements} & \textbf{Sample Technologies / Concepts} \\
\hline
\textbf{Entanglement Identification} & EPR Pair IDs, Bell Pair Tracking, Quantum State Hashing \\
\hline
\textbf{Quantum Addressing} & Quantum Network Identifiers (QNID), Entangled Node ID, Quantum MAC Address \\
\hline
\textbf{Quantum Sensing} & Quantum-enhanced sensors, Single-photon detectors, Quantum radar, Spin-based magnetometers \cite{Khan2024QuantumFuture, ElMorsalani2024QuantumSensors, Lloyd2008QuantumRadar} \\
\hline
\textbf{Quantum Communication} & QKD, Teleportation Protocols, BB84, E91, MDI-QKD, QRNA, QLoop, Qubus Protocols \cite{Mehic2023, sidhu2021space, Pirker2019Modular} \\
\hline
\multicolumn{2}{|c|}{\textbf{Quantum Computation}} \\
\hline
\textbf{Hardware} & IonQ, IBM Q, Rigetti, PsiQuantum, Photonic Chips, Quantum Repeaters, Quantum Routers \cite{Taha2024Photonics, Beukers2024RemoteEnt, Azuma2023Repeaters} \\
\hline
\textbf{Software} & Qiskit, Cirq, QuNetSim, NetSquid, QuISP, Strawberry Fields, QCoDeS, ProjectQ \cite{Cross2018Qiskit, Shi2022Cirq, diadamo2023qunetsim, deJong2020NetSquid, Fujiwara2020QuISP, QCoDeS2023} \\
\hline
\textbf{Quantum Services} & Entanglement-as-a-Service (EaaS), Quantum Key-as-a-Service (KaaS), Quantum Cloud API, Quantum Orchestration Services, LLM-Aided Stack Automation \cite{RomeroAlvarez2024QaaS, Prados2023LLMQuantum} \\
\hline
\textbf{Quantum Semantics} & Quantum Context Encoding, Entanglement Metadata, Fidelity Metrics, Entropy Cost, Stack-Level Intent Modeling \cite{Illiano2022QuantumStackSurvey, Getu2025SemanticComm, Nguyen2024DriftEntropyBenchmark} \\
\hline
\end{tabular}
\end{table*}
Fig.~\ref{fig:federated_quantum} illustrates how diverse quantum domains interface via the federation logic built on top of the Quantum-Converged OSI stack’s upper layers.

\section{Layer 0: Quantum Substrate}
\begin{figure}[]
    \centering
    \includegraphics[width=0.85\linewidth]{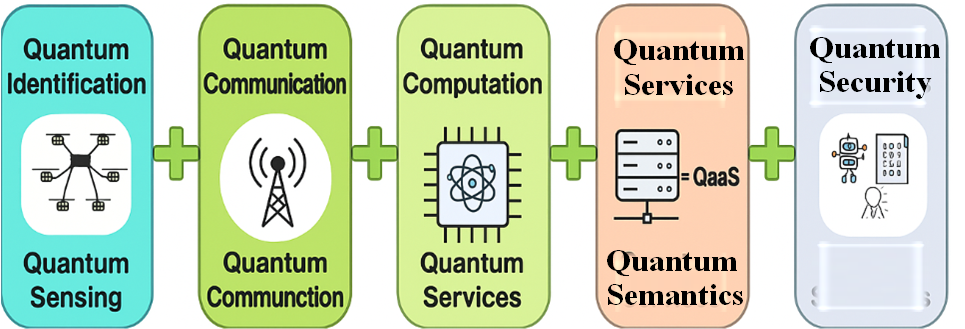}
    \caption{Representative technologies comprising the quantum substrate layer, including ion traps, NV centers, SPDC photonics, and superconducting qubits. Adapted from substrate-layer architecture surveys \cite{bradley2019ten, monroe2021programmable, shalm2015strong}.}
    \label{fig:quantum_elements}
\end{figure}

The Quantum Substrate represents the foundational stratum of the Quantum-Converged OSI architecture, directly interfacing with quantum physical hardware and entanglement distribution systems. Unlike classical OSI models, which abstract away the communication medium, Layer 0 in a quantum context must account for the inherently fragile and non-clonable nature of quantum states \cite{kimble2008quantum, wehner2018quantum}. This layer facilitates the generation, stabilization, and routing of entangled quantum states, acting as the bedrock for teleportation, QKD, and quantum memory synchronization \cite{duan2001long, simon2017towards, sangouard2011quantum}.
Entanglement fidelity, coherence time, and teleportation success rates are core performance metrics at this layer, governed by hardware constraints (e.g., qubit platform, noise) and environmental dynamics \cite{van2018multiplexed, lanyon2017efficient}. Physical implementations span trapped ions \cite{monroe2021programmable}, Nitrogen-Vacancy (NV) centers \cite{bradley2019ten}, and photonic systems leveraging Spontaneous Parametric Down-Conversion (SPDC) \cite{shalm2015strong, yin2017satellite}. These platforms enable entangled links that span local, metropolitan, and intercontinental scales. Layer 0 devices span a wide spectrum of quantum physical platforms, as illustrated in Fig.~\ref{fig:quantum_elements} and discussed in recent architectural reviews.

Several simulation and emulation platforms, including NetSquid \cite{deJong2020NetSquid}, QuISP \cite{quisp2020}, and QuNetSim \cite{schafer2020qnetsim}, have been developed to model Layer 0 behaviors under realistic noise, latency, and fidelity parameters. Protocols such as BB84 and E91 serve as practical demonstrations of quantum communication's physical-layer dynamics, though their scalability remains tightly coupled with advances in substrate-level entanglement and repeater networks \cite{bennett1984quantum, ekert1991quantum, briegel1998quantum}.
As quantum networks scale, Layer 0 is projected to evolve from monolithic links to modular, programmable substrates capable of autonomously negotiating entangled routes and self-healing decoherence-induced degradation. Integration with upper-layer orchestration, such as Layer 8's cognitive intent plane, is expected to enable AI-enhanced management of fidelity, coherence, and entropy constraints \cite{caleffi2020quantum, pirker2018modular}.
The following subsections delve deeper into the requirements, existing literature, enabling technologies, and future research directions for this crucial foundational layer.

\begin{table*}[htbp]
    \centering
    \caption{Comparison of selected quantum substrate platforms across coherence, fidelity, and scalability.}
    \label{tab:substrate_comparison}
    \begin{tabular}{|l|c|c|c|c|l|}
        \hline
        \textbf{Platform} & \textbf{Coherence Time} & \textbf{Fidelity} & \textbf{Entanglement Range} & \textbf{Scalability} & \textbf{References} \\
        \hline
        Trapped Ions & >1s & High & Local–Regional & Medium & \cite{monroe2021programmable} \\
        NV Centers   & ms  & Medium–High & Metropolitan & Medium & \cite{bradley2019ten} \\
        SPDC Photonics & ns–$\mu$s & Medium & Long-distance (fiber/satellite) & High & \cite{shalm2015strong, yin2017satellite} \\
        \hline
    \end{tabular}
\end{table*}

\subsection{Requirements}
Layer 0, the Quantum Substrate, underpins the entire quantum networking stack by handling entanglement generation, quantum state preservation, and transmission through a physical medium. Unlike the classical physical layer, the substrate must maintain fragile quantum states and mediate the probabilistic behavior of entangled communication channels. The design of this layer is constrained by physical phenomena such as decoherence, noise, and the no-cloning theorem \cite{kimble2008quantum, wehner2018quantum}.
Despite these physical constraints, a primary requirement is the ability to distribute entangled qubits across nodes, supporting entanglement swapping, purification, and teleportation \cite{duan2001long, briegel1998quantum, sangouard2011quantum}. Moreover, entanglement generation must be high-fidelity and compatible with multiple physical platforms such as trapped ions \cite{monroe2021programmable}, NV centers in diamond \cite{bradley2019ten}, superconducting qubits, and photonic systems using SPDC sources \cite{shalm2015strong, yin2017satellite}. Additionally, substrate protocols must synchronize with quantum memories to allow for storage and scheduling of quantum states \cite{simon2017towards, van2018multiplexed}.

Another key requirement is ensuring resilience against environmental perturbations. Quantum coherence must be preserved over communication channels such as optical fiber, free-space links, or satellite-ground optical paths \cite{yin2017satellite, simon2017towards}. Moreover, this layer must accommodate real-time monitoring and feedback of quantum fidelity and entropy to enable higher-layer orchestration and routing protocols \cite{caleffi2020quantum, pirker2018modular}. 
Layer 0 must also facilitate interoperation with quantum-classical hybrid devices through well-defined APIs, allowing stack-wide communication of coherence metrics, memory availability, and entanglement states \cite{deJong2020NetSquid, schafer2020qnetsim}. These properties form the basis for SDQN, which adapts fidelity-aware routing and control based on substrate-level statistics \cite{caleffi2020quantum}.
Lastly, scalability and modularity are essential to support large-scale entangled topologies. The substrate must efficiently support quantum repeater chaining, node multiplexing, and entanglement distribution protocols over heterogeneous hardware \cite{wehner2018quantum, briegel1998quantum}. These foundational functions enable higher-layer abstractions such as entanglement-aware MAC protocols, session-level handshakes, and semantic applications.

\subsection{Existing Literature}
The foundational concept of Layer 0 as a quantum substrate derives from the broader vision of a global quantum internet, where physical entanglement channels serve as the communication backbone \cite{kimble2008quantum, wehner2018quantum}. Early theoretical models such as quantum teleportation \cite{bennett1993teleporting}, quantum repeaters \cite{briegel1998quantum}, and entanglement purification \cite{dur1999quantum} established the necessity for a dedicated physical layer that manages fragile quantum states with strict coherence and fidelity requirements.
In addition, numerous experimental platforms have explored the physical instantiation of this layer. Moreover, trapped ion systems have demonstrated robust quantum coherence and modular entanglement \cite{monroe2021programmable}. In contrast, NV centers in diamond offer solid-state alternatives with long coherence times and high-fidelity memory capabilities \cite{bradley2019ten}. However, photonic systems, particularly those using SPDC, have enabled entanglement distribution over terrestrial and satellite links \cite{yin2017satellite, shalm2015strong}, highlighting the potential for long-distance, Layer-0 communication.

Architectural abstractions have emerged to formalize the role of Layer 0 within network stacks. Pirker and Dür proposed a modular quantum network framework where entanglement links are structured into stackable resource layers, implicitly corresponding to the quantum substrate \cite{pirker2018modular}. Caleffi et al. extended this view through SDQN, where Layer 0 metrics such as fidelity, error rate, and coherence time inform routing and scheduling decisions across the stack \cite{caleffi2020quantum}.

Simulation tools have become instrumental in modeling the behavior and performance of the quantum substrate. NetSquid enables discrete-event simulation of entanglement distribution, including physical-layer metrics like memory lifetimes and optical losses \cite{deJong2020NetSquid}. QuNetSim and QuISP offer modular APIs and hardware emulation layers to replicate Layer 0 behavior in hybrid classical-quantum simulations \cite{schafer2020qnetsim, quisp2020}. These platforms allow researchers to test protocols such as BB84 and E91 under real-world physical assumptions, offering critical insights into the operational characteristics of Layer 0 components.
In addition, a growing body of literature also investigates the coupling between Layer 0 and emergent quantum protocols such as QKD, quantum sensing, and distributed quantum computation. These studies emphasize the importance of accurate decoherence modeling and the need for fidelity-aware protocol stacks \cite{sangouard2011quantum, simon2017towards, van2018multiplexed}.
Collectively, this literature establishes Layer 0 not merely as a hardware interface but as an intelligent substrate capable of supporting adaptive, entanglement-centric services that enable scalable, secure, and resilient quantum communication infrastructures.

\subsection{Technologies and Challenges}
The technological foundation of Layer 0 consists of a diverse array of quantum hardware platforms and communication primitives, each tailored to different coherence, fidelity, and distance trade-offs. Prominent physical qubit technologies include trapped ions \cite{monroe2021programmable}, NV centers in diamond \cite{bradley2019ten}, superconducting transmon qubits \cite{kjaergaard2020superconducting}, and photonic systems using single-photon sources and entangled photon pairs via SPDC \cite{shalm2015strong, yin2017satellite}.
Entanglement generation is typically achieved through SPDC, quantum dot emission, or ion-photon entanglement techniques, while entanglement swapping and purification require high-fidelity Bell-state measurement (BSM) circuits and quantum memories \cite{briegel1998quantum, sangouard2011quantum}. These memories—often realized via atomic ensembles, cryogenic solid-state systems, or spin-photon interfaces—must maintain coherence over durations compatible with round-trip communication delays in multi-hop networks \cite{van2018multiplexed, simon2017towards}.

Quantum repeaters represent a cornerstone technology for extending entanglement across long distances. However, their implementation remains constrained by high-loss channels, gate infidelity, and memory decoherence \cite{dur1999quantum, duan2001long}. Most repeater architectures also suffer from low throughput and significant latency due to the probabilistic nature of entanglement generation and the need for heralding and feedback.

Another critical challenge is managing quantum decoherence—losing information in a quantum system due to environmental interaction. While NV centers and trapped ions provide relatively long coherence times under cryogenic or ultra-high vacuum conditions, they are challenging to integrate into large-scale, mobile, or power-efficient networks \cite{bradley2019ten, monroe2021programmable}. On the other hand, photonic systems offer room-temperature operation and higher scalability but face issues like photon loss, indistinguishability, and mode mismatch \cite{shalm2015strong, pan2012multiphoton}.
Additionally, the Layer 0 substrate must also incorporate low-loss optical fibers, free-space optical systems, and satellite-ground links to distribute entanglement across diverse geographic and topological domains \cite{yin2017satellite, Liao2017SatelliteMicius}. These links are subject to attenuation, atmospheric turbulence, and alignment errors, particularly in satellite-based QKD and intercontinental quantum networks.

From a systems perspective, integrating substrate devices into larger programmable stacks introduces abstraction, interface definition, and control synchronization challenges. Tools like NetSquid and QuNetSim support emulation of such hardware constraints and provide fidelity-aware API interfaces to upper layers \cite{deJong2020NetSquid, schafer2020qnetsim, quisp2020}. However, aligning real-time substrate conditions (e.g., coherence degradation or memory availability) with stack-wide orchestration mechanisms (e.g., SDN control planes) remains an open problem.
Finally, security concerns arise even at Layer 0, particularly in adversarial environments. Although quantum communication is inherently resistant to eavesdropping due to the no-cloning theorem and quantum measurement collapse \cite{bennett1984quantum, ekert1991quantum}, denial-of-service attacks targeting entanglement resources, memory bottlenecks, or link synchronization must be considered in future implementations.

\subsection{Future Directions}
The evolution of Layer 0 hinges on disruptive hardware innovation and intelligent orchestration architectures that can adapt to quantum system fragility. One significant research trajectory involves the development of next-generation quantum repeaters with real-time fidelity tracking, error mitigation capabilities, and entanglement-aware path selection \cite{briegel1998quantum, van2018multiplexed}. Such systems could leverage entanglement swapping with AI-optimized scheduling algorithms, improving throughput and reducing decoherence exposure across long-distance quantum links \cite{caleffi2020quantum}.
Another critical research frontier is the expansion of space-based quantum communication. Missions such as the Micius satellite \cite{yin2017satellite, Liao2017SatelliteMicius} have demonstrated feasibility for Layer 0 entanglement distribution over thousands of kilometers. Future constellations of LEO satellites may enable persistent and globally scalable quantum substrate coverage. These networks will demand lightweight, radiation-resistant photonic payloads and inter-satellite quantum links, all governed by cross-layer feedback mechanisms \cite{wehner2018quantum, simon2017towards}.

Advancements in quantum memory fidelity and lifetime are equally vital. Emerging platforms such as rare-earth-doped crystals and hybrid atom-photon systems offer coherence times that extend from milliseconds to seconds \cite{bradley2019ten, simon2010quantum}. Coupling these memories with heralded entanglement generation schemes and cryogenic photonic circuits could significantly enhance the reliability of Layer 0 channels \cite{pan2012multiphoton, sangouard2011quantum}.
Thus, there is growing interest in integrating Layer 0 devices with higher-layer orchestration frameworks via middleware APIs. Moreover, quantum substrate control can be augmented by machine learning agents residing in Layer 8 (Cognitive Intent Plane), which use substrate telemetry—fidelity, QBER, entropy rate—as input for predictive link adaptation and quantum resource allocation \cite{caleffi2020quantum, pirker2018modular}. These control systems must support real-time, decentralized decision-making to maintain end-to-end entanglement viability.

Additionally, Layer 0 must evolve to support quantum-digital twin synchronization, distributed quantum sensing, and eventually, quantum internet nodes capable of multiplexed service hosting \cite{wehner2018quantum, kimble2008quantum}. Research is increasingly focusing on protocols that allow co-packaged transmission of quantum and classical metadata, reducing handoff latency and improving coordination between quantum-classical interface layers \cite{deJong2020NetSquid, schafer2020qnetsim}.
Finally, the substrate must eventually support programmability, composability, and modular certification to facilitate ecosystem interoperability. Emerging standards such as QIR and SDQN paradigms offer a glimpse into programmable Layer 0 interfaces that can be reconfigured dynamically for different qubit technologies and topological conditions \cite{caleffi2020quantum, quisp2020}.

\section{Layer 1: Physical Layer}
Layer 1 in the Quantum-Converged OSI architecture represents the core transmission layer that interfaces directly with the entanglement substrate (Layer 0) and is the conduit for raw quantum signal propagation across quantum links. While Layer 0 manages qubit states and entanglement generation, Layer 1 addresses the physical mechanisms for transmitting those quantum states over fiber, free-space, chip-scale, or satellite-mediated links. This includes encoding, modulation, transmission protocols, and hardware transduction across optical and matter qubit systems.
In contrast to classical physical layers, Layer 1 in a quantum context must address fundamentally new challenges: quantum state fragility, probabilistic signal propagation, quantum measurement collapse, and non-deterministic transmission success. This demands protocols and hardware that maintain quantum coherence while facilitating low-loss, noise-resilient propagation over diverse environments \cite{wehner2018quantum, simon2017towards}.

Multiple transmission media are employed in modern quantum physical layers, including optical fiber, satellite-ground free-space optics, Photonic Integrated Circuits (PICs), chip-to-chip quantum buses, and waveguides \cite{yin2017satellite, Liao2017SatelliteMicius, pan2012multiphoton}. Quantum repeaters, quantum frequency converters, and electro-optical modulators are essential physical-layer elements, supporting wavelength-multiplexing, signal routing, and noise filtering across entangled quantum networks.
A complementary innovation gaining traction is the integration of RIS — programmable metasurfaces capable of dynamically shaping quantum wavefronts and adapting to environmental conditions. While RIS has been extensively studied in classical wireless networks \cite{ahmed2023sky, ahmed2022blockage}, its adaptation for quantum channels, particularly in photonic entanglement routing and decoherence suppression, presents a promising frontier for Layer 1 enhancements \cite{basar2021quantumris, ren2023quantumris}.
This section explores the layered design of the quantum physical layer, surveying its foundational requirements, supporting literature, implementation technologies, and emerging research directions. In the following subsections, we present a structured analysis encompassing encoding techniques, photon sources, waveguide materials, RIS-enhanced transmission, and the future trajectory of programmable quantum communication media.
Table~\ref{tab:quantum-phy-protocols} summarizes major physical-layer quantum protocols including BB84, E91, MDI-QKD, and satellite-based implementations like Micius \cite{Pirker2019Modular, Mehic2022QKDNetworks, Liao2017SatelliteMicius, Khan2024QuantumFuture}.
\begin{table*}[htbp]
\centering
\caption{Characteristics of Quantum PHY Protocols}
\label{tab:quantum-phy-protocols}
\renewcommand{\arraystretch}{1.4}
\begin{tabular}{|p{3.2cm}|p{2.2cm}|p{2.6cm}|p{2.8cm}|p{4.5cm}|}
\hline
\textbf{Quantum PHY Protocol} & \textbf{Qubit Medium} & \textbf{Encoding Technique} & \textbf{Operating Band / Spectrum} & \textbf{Error Control Method} \\
\hline
BB84 (Fiber) \cite{Mehic2023, Pirker2019Modular} & Photons & Polarization Encoding & 1550 nm (C-band) & Decoy State + Error Estimation \\
\hline
E91 \cite{Pirker2019Modular} & Photons & Entangled Pairs & 1310/1550 nm & Bell Test Verification \\
\hline
MDI-QKD \cite{Mehic2022QKDNetworks} & Photons & Time-bin or Phase & 1550 nm & BSM \\
\hline
Qubits over NV Centers \cite{Khan2024QuantumFuture, ElMorsalani2024QuantumSensors} & Electron Spins & Spin-Based Optical Readout & 637 nm (ZPL) & QEC \\
\hline
Continuous-Variable QKD \cite{Pirker2019Modular} & Coherent States & Gaussian Modulation & 1550 nm & Reconciliation + Homodyne Detection \\
\hline
Satellite QKD (Micius) \cite{Liao2017SatelliteMicius, sidhu2021space} & Photons (Space) & Polarization & Free-Space Optical & Adaptive Optics + Post-selection \\
\hline
Quantum Teleportation (Fiber) \cite{Pirker2019Modular, Khan2024QuantumFuture} & Entangled Photons & Dual-Rail & 1310/1550 nm & Bell Measurement + Feedforward \\
\hline
\end{tabular}
\end{table*}

\begin{figure}[]
    \centering
    \includegraphics[width=0.95\linewidth]{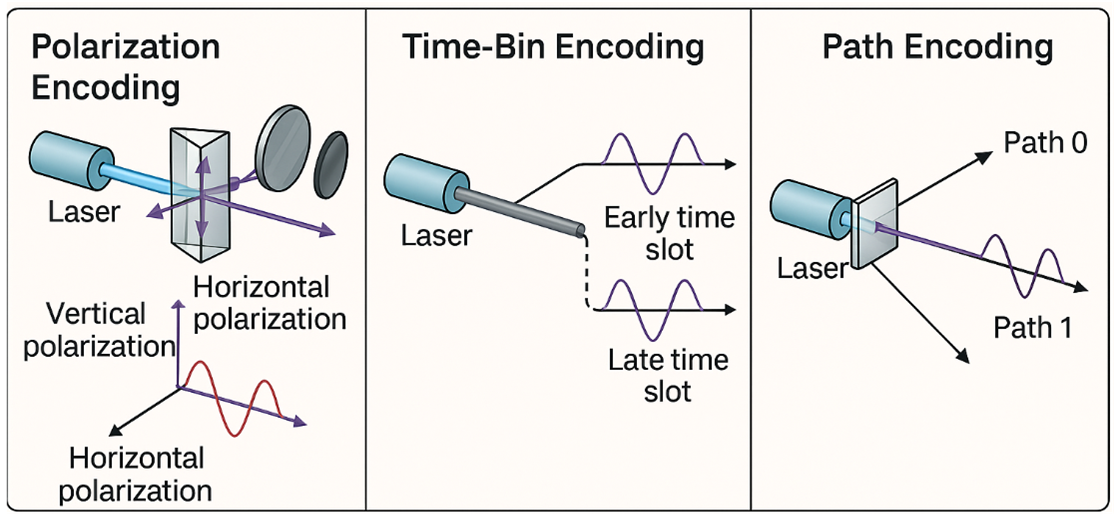}
    \caption{Encoding modalities used in physical-layer quantum transmission, including polarization, time-bin, and path encoding. Source: summarized from \cite{gisin2002quantum, ursin2007entanglement}.}
    \label{fig:layer1_encoding}
\end{figure}

\subsection{Requirements}
Layer 1 of the Quantum-Converged OSI model is tasked with the reliable transmission of quantum information between network nodes through physical channels. Its requirements are fundamentally distinct from classical physical layers due to the quantum-specific constraints of state fragility, measurement collapse, and non-clonability \cite{wehner2018quantum, kimble2008quantum}.
First and foremost, the layer must support the transmission of qubits across various quantum media while preserving their coherence and fidelity. These channels include single-mode and multi-mode optical fibers, free-space optical links, on-chip photonic waveguides, and satellite-ground communication paths \cite{yin2017satellite, pan2012multiphoton, Liao2017SatelliteMicius}. The medium must be low-loss and engineered to mitigate decoherence, polarization drift, dispersion, and mode mismatch—factors that can collapse quantum states in transit \cite{simon2017towards}. Various encoding schemes—shown in Fig.~\ref{fig:layer1_encoding}—are tailored for specific quantum media and system constraints \cite{gisin2002quantum, ursin2007entanglement}.

Second, the layer must incorporate encoding mechanisms suited to the physical qubit modality in use—polarization encoding for photonic qubits, time-bin encoding for fiber-based networks, or path encoding in on-chip systems \cite{gisin2002quantum, ursin2007entanglement}. These methods must be compatible with entanglement-based protocols (e.g., E91), prepare-and-measure QKD (e.g., BB84), and teleportation schemes, requiring precise timing synchronization and interferometric stability \cite{bennett1984quantum, ekert1991quantum}.
Additionally, Layer 1 must also facilitate wavelength multiplexing and filtering to enable scalable transmission over shared channels and across heterogeneous quantum devices. Quantum frequency converters (QFCs) are crucial for bridging different photonic frequencies between memory nodes and transmission channels \cite{ikuta2018frequency, zaske2012visible}.

This layer must interface with dynamic environmental awareness mechanisms to improve adaptability and physical-layer programmability. One promising technology in this direction is the RIS, which can be used to reshape quantum wavefronts, enhance transmission probabilities, and suppress scattering in urban or turbulent atmospheric settings \cite{basar2021quantumris, ren2023quantumris}. Integration with RIS requires a control interface capable of low-latency feedback between measurement outcomes and surface actuation.
Finally, Layer 1 must expose fidelity and loss metrics to higher layers, especially Layer 3 (Network) and Layer 8 (Intent Plane), to enable quantum-aware routing, dynamic link reconfiguration, and performance prediction \cite{caleffi2020quantum}. This demands standardized telemetry APIs, real-time coherence diagnostics, and stack-wide entanglement resource management integration.

\subsection{Existing Literature}
The physical transmission of quantum states across a network has been a central focus of quantum communication research since the inception of QKD. Early experimental demonstrations such as BB84 \cite{bennett1984quantum} and E91 \cite{ekert1991quantum} laid the groundwork for secure quantum information transfer using polarization, phase, and time-bin encoded photons. These protocols inherently rely on physical channels' robustness and their ability to maintain quantum coherence over distance.
A key milestone in practical quantum communication was the demonstration of entanglement-based transmission over 144 km of free space between La Palma and Tenerife \cite{ursin2007entanglement}. This experiment validated the feasibility of long-range free-space quantum communication and provided early empirical insight into atmospheric decoherence and photon loss. Similar challenges have been explored in optical fiber-based implementations, where losses become significant beyond 100–200 km due to absorption and scattering, even with ultra-low-loss fibers \cite{gisin2002quantum}.

Satellite-based quantum communication represents a significant leap in the physical-layer capabilities of quantum networks. The Chinese Micius satellite has successfully distributed entangled photon pairs over 1200 km and achieved intercontinental QKD between ground stations in China and Austria \cite{yin2017satellite, Liao2017SatelliteMicius}. These demonstrations underscore the importance of free-space optics and orbital geometries as part of the physical-layer transmission infrastructure.
Moreover, photon generation and encoding technologies have also advanced significantly. Systems based on SPDC remain a primary source for entangled photon pairs \cite{shalm2015strong}. In contrast, quantum dot and NV center systems offer deterministic single-photon emission under cryogenic conditions \cite{bradley2019ten}. These emitters are typically coupled to waveguides or optical fibers and require sub-nanosecond synchronization with detection systems.

PICs and chip-to-chip quantum buses have emerged as scalable alternatives to bulk optics, enabling low-loss, polarization-maintaining, and temperature-stable transmission on planar substrates \cite{wang2020integrated}. Integrated devices allow quantum gates, beam splitters, and phase shifters to coexist on a single chip, facilitating miniaturized quantum networks and localized Layer 1 operations.
In addition, QFC has become increasingly critical for enabling inter-device compatibility across disparate operating wavelengths. Techniques have been demonstrated for converting visible photon emissions (e.g., 637 nm from NV centers) into telecom-band wavelengths (e.g., 1550 nm) suitable for long-distance fiber transmission \cite{zaske2012visible, ikuta2018frequency}.

In parallel, the rise of RIS introduces new possibilities for physical-layer optimization in quantum systems. These metasurfaces, typically composed of programmable sub-wavelength elements, can steer and shape quantum wavefronts to suppress multipath effects, atmospheric scattering, and photon loss \cite{basar2021quantumris}. Recent studies have proposed metasurfaces that operate in low-photon and entangled-state regimes, adapting their reflectivity based on photon detection feedback \cite{ren2023quantumris}.
Simulation environments such as NetSquid and QuNetSim have modeled Layer 1 constraints like link loss, detector efficiency, and timing jitter, thereby providing insight into physical-layer throughput under varying quantum device conditions \cite{deJong2020NetSquid, schafer2020qnetsim}. These tools help bridge theoretical models with practical engineering constraints.
\begin{figure}[]
    \centering
    \includegraphics[width=0.85\linewidth]{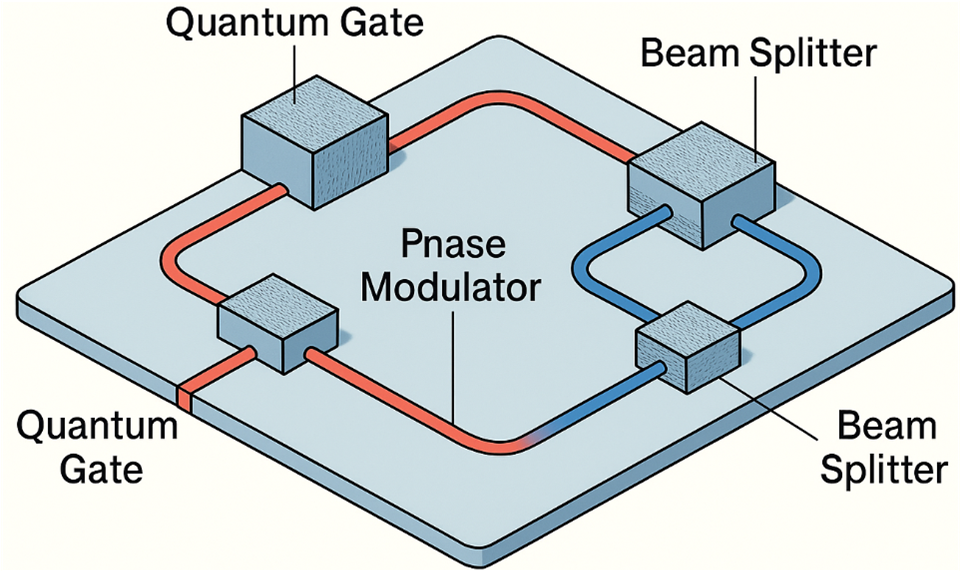}
    \caption{Integrated photonic circuits enable on-chip quantum gates, beam splitters, and modulators for scalable Layer 1 transmission. Based on photonic hardware layouts discussed in \cite{wang2020integrated}.}
    \label{fig:pic_layer1}
\end{figure}

\subsection{Technologies and Challenges}
The quantum physical layer has undergone significant technological evolution, driven by the need for scalable, high-fidelity, and adaptable transmission media. One of the key advancements is the development of integrated photonic platforms that support on-chip quantum state generation, manipulation, and routing. These systems leverage silicon photonics and lithium niobate modulators to provide phase stability, compact form factors, and polarization maintenance \cite{wang2020integrated}. PICs reduce system complexity and allow co-location of beam splitters, interferometers, and detectors—effectively miniaturizing Layer 1 components.

In long-distance terrestrial and satellite quantum communication, fiber and free-space optics remain dominant transmission channels. However, issues such as fiber attenuation, chromatic dispersion, and atmospheric turbulence still pose major challenges. Quantum signals degrade rapidly over long distances due to exponential photon loss and temporal decoherence, especially in fiber channels beyond 100 km. In free-space and satellite-ground paths, beam divergence and weather-dependent signal degradation introduce high variability in qubit arrival rates \cite{li2021weather}.

QFC technologies are increasingly being adopted to bridge quantum sources emitting at non-telecom wavelengths with low-loss fiber-optic bands near 1550 nm. Recent implementations have shown high-efficiency, low-noise conversion compatible with entanglement-preserving operations \cite{maring2020quantum}. These interfaces are essential for hybrid networks, enabling interoperability between memory nodes, photon sources, and detectors across disparate spectral domains.
RIS are emerging as a transformative tool for quantum signal propagation, especially in free-space and indoor quantum systems. RISs are artificial surfaces composed of sub-wavelength meta-atoms that can be programmed to modulate the phase, amplitude, and polarization of incoming quantum light \cite{ren2023quantumris}. They promise to mitigate multipath effects, enable dynamic wavefront control, and enhance entanglement distribution in noisy environments \cite{zhou2023risquantumnetworks}. The challenge lies in making RIS devices compatible with single-photon regimes and integrating them into stack-aware control loops. PICs, as depicted in Fig.~\ref{fig:pic_layer1}, offer scalable on-chip quantum communication primitives \cite{wang2020integrated}. As shown in Fig.~\ref{fig:pic_layer1}, integrated photonic circuits implement on-chip quantum gates, beam splitters, and phase modulators to enable scalable Layer 1 quantum transmission. Such architectures are foundational for compact and programmable quantum hardware platforms.

Another critical area of concern is the development of low-jitter, high-efficiency Single-Photon Detectors (SPDs) suitable for use in Layer 1. Modern superconducting nanowire single-photon detectors (SNSPDs) have demonstrated detection efficiencies above 90\% and timing jitter below 50 ps. Yet, their operation requires cryogenic cooling and complex readout circuits, limiting scalability and deployment in mobile environments \cite{zhao2020snsdreview}.
Hardware heterogeneity across quantum networks also introduces interoperability issues. Variations in photon emission wavelengths, gate operation fidelities, and clock synchronization can affect the integrity of Layer 1 operations unless mitigated by protocol-level compensation and adaptive interface layers \cite{schoute2021chipstackintegration}.
Lastly, the tight coupling between physical hardware and stack-wide orchestration introduces standardization and real-time control challenges. Exposing Layer 1 metrics—such as transmission loss, QBER, and coherence decay—to higher layers requires the development of telemetry APIs and lightweight middleware, notably to support Layer 8’s cognitive orchestration plane \cite{du2022quantumstackinsights}.

\subsection{Future Directions}
The future trajectory of the quantum physical layer is marked by the convergence of programmable optics, integrated quantum photonics, and environmental adaptivity. A key direction is the advancement of large-scale PICs that integrate hundreds of quantum components—such as interferometers, phase shifters, and detectors—on a single chip. These systems will form the foundation of chip-based Layer 1 modules, enabling compact, scalable, and reconfigurable quantum routers and repeaters \cite{wang2020integrated}.
Another emerging paradigm involves embedding physical-layer intelligence directly into the transmission medium using programmable metasurfaces and RIS technologies. Unlike traditional passive optics, RIS platforms are expected to adapt dynamically to atmospheric turbulence, multipath reflections, and environmental variability in real time. Future RIS arrays optimized for single-photon regimes may incorporate embedded control logic to modulate entangled wavefronts, phase-conjugate loss paths, or support adaptive polarization correction \cite{ren2023quantumris, zhou2023risquantumnetworks}.
In addition, QFC is poised to evolve into a standardized interface layer between heterogeneous quantum hardware. Future QFC modules will likely support entanglement-preserving conversion with sub-dB loss, enabling memory-photon interoperability between disparate quantum nodes \cite{maring2020quantum}. These modules will become critical in multi-domain network topologies integrating trapped ions, NV centers, and quantum dots, each operating at distinct frequencies.
\begin{figure}[]
    \centering
    \includegraphics[width=0.9\linewidth]{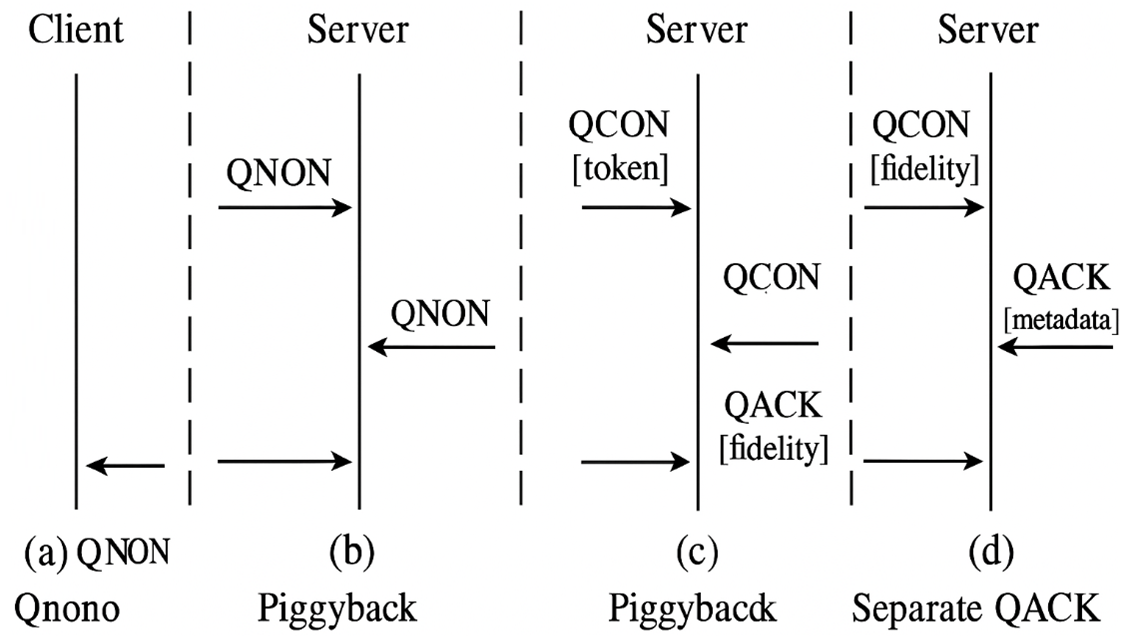}
    \caption{Quantum link-layer operations including MAC arbitration, entanglement feedback, and classical coordination. Entanglement success is probabilistic and governed by decoherence and QBER constraints \cite{yu2021qmac, bhaskar2022classicalintegration}.}
    \label{fig:quantum_mac_flow}
\end{figure}

Co-packaged photon sources and superconducting detectors may also be integrated with CMOS control electronics to achieve Layer 1 nodes with native support for real-time telemetry. This will automatically expose fidelity, loss, and timing data to higher stack layers (e.g., Layer 8) for feedback-driven orchestration \cite{zhao2020snsdreview, du2022quantumstackinsights}. As quantum networks scale, such cross-layer signaling will be essential for proactive route adaptation and entanglement-aware scheduling.
In addition to terrestrial advances, future Layer 1 systems will extend to space-based entanglement infrastructure—featuring constellations of RIS-equipped satellites capable of aligning, correcting, and propagating entangled photons across orbital tiers. These systems will require Layer 1 protocols that operate in relativistic and Doppler-shifted regimes, creating demand for novel calibration, timing synchronization, and frequency stabilization techniques \cite{li2021weather}.
Finally, a growing frontier lies in modular physical-layer APIs that abstract hardware details and present standardized interfaces to orchestration agents and Layer 2 protocols. These APIs will unify metrics such as QBER, polarization drift, and path length variance, enabling multi-vendor compatibility and dynamic configuration of Layer 1 behaviors based on network context \cite{du2022quantumstackinsights}.

\begin{table*}[htbp]
    \centering
    \caption{Comparison of quantum transmission media for Layer 1 physical-layer implementations.}
    \label{tab:layer1_media}
    \begin{tabular}{|l|c|c|c|c|l|}
        \hline
        \textbf{Medium} & \textbf{Distance} & \textbf{Decoherence} & \textbf{Env. Sensitivity} & \textbf{Scalability} & \textbf{Examples} \\
        \hline
        Optical Fiber & ~100 km & Medium & Low & High & Fiber-based QKD \\
        Free-Space Optics & 100–1000 km & High & Very High & Medium & Satellite-Ground Links \\
        On-Chip PIC & ~cm–m & Low & Low & Very High & Quantum Routers \\
        RIS-Enhanced Channels & 100 m–km & Adaptive & Moderate & Experimental & Smart Indoor Quantum Links \\
        \hline
    \end{tabular}
\end{table*}

\section{Layer 2: Data Link}
Layer 2 in the Quantum-Converged OSI model performs the essential role of managing quantum link access, channel control, and error resilience between directly connected quantum nodes. Unlike its classical counterpart, which handles framing, MAC, and error checking, the quantum data link layer must address quantum-specific constraints such as decoherence, measurement collapse, and non-deterministic entanglement success \cite{yu2021qmac, muralidharan2022qmemorymgmt}.
Fundamentally, Layer 2 enables fidelity-aware coordination of quantum link usage, employing quantum MAC protocols that accommodate probabilistic entanglement generation and time-varying quantum channel characteristics \cite{liu2023adaptiveqmac, bauml2023qmacsurvey}. It must determine when and how quantum frames (entangled states, Bell pairs, or single qubits) can be transmitted, stored, or reused based on fidelity thresholds and network-wide coherence demands \cite{du2022quantumstackinsights}. Quantum MAC and link-layer operations, shown in Fig.~\ref{fig:quantum_mac_flow}, depend on hybrid signaling between qubit memory state and classical feedback acknowledgments \cite{yu2021qmac, bhaskar2022classicalintegration}.

A core function at this layer is QEC, which attempts to preserve quantum state integrity across noisy channels and imperfect quantum memories. Unlike classical parity checks or CRCs, QEC must operate without directly measuring quantum data, instead using ancilla qubits and syndrome extraction methods to detect and correct errors. Recent advances in surface codes, low-overhead QEC schemes, and fault-tolerant protocols have reshaped the practical viability of Layer 2 error resilience \cite{fowler2021highthreshold, hao2023modularqec}.
Additionally, Layer 2 is responsible for quantum memory scheduling, {entanglement buffering}, and {link-layer handshaking} that informs higher-layer session management (Layer 5) and quantum routing (Layer 3). This includes time-synchronized qubit release, decoherence-aware memory replacement, and hybrid MAC signaling for entanglement confirmation and qubit reordering \cite{muralidharan2022qmemorymgmt}.

Hybrid classical–quantum control signaling is often required to coordinate qubit readiness, memory state, and MAC-level arbitration in time-sensitive protocols. Classical feedback is essential for entanglement heralding, error correction acknowledgment, and MAC arbitration \cite{bhaskar2022classicalintegration}.
Emerging approaches also incorporate AI-enhanced MAC protocols and {fidelity prediction models}, which adjust transmission behavior based on entropy metrics, QBER trends, and coherence decay forecasts. These feedback mechanisms integrate with Layer 8 orchestration planes to optimize link usage and maintain end-to-end fidelity under constrained quantum conditions \cite{du2022quantumstackinsights, zhao2023aiqnet}.

\subsection{Requirements}
Layer 2 of the Quantum-Converged OSI model serves as the intermediate coordination layer between raw quantum transmission (Layer 1) and quantum routing (Layer 3). It is responsible for organizing access to quantum communication channels, maintaining fidelity through error correction mechanisms, and scheduling entanglement resources in real time. Unlike classical MAC and link control, the quantum data link layer must operate under severe physical constraints, including measurement-induced collapse, decoherence, and non-clonability of quantum data.
One fundamental requirement is the deployment of fidelity-aware MAC protocols capable of scheduling entangled pairs or qubit transmissions based on the instantaneous quality of quantum links. These protocols must operate probabilistically, since entanglement generation is inherently non-deterministic and success rates vary with channel loss and noise. MAC decisions must therefore consider qubit readiness, quantum memory availability, and fidelity thresholds for downstream application layers \cite{yu2021qmac, liu2023adaptiveqmac}.

A second requirement is the integration of QEC mechanisms to mitigate errors induced by thermal noise, gate imperfections, and qubit decoherence. Unlike classical error detection, QEC must operate without collapsing the quantum state. This demands the implementation of syndrome-based methods using ancilla qubits, such as surface codes, repetition codes, and topological codes with minimal overhead \cite{fowler2021highthreshold, hao2023modularqec}.
Layer 2 must also manage quantum memory buffers at each node. These buffers hold entangled states for future use or until routing decisions are made by Layer 3. Therefore, the layer requires synchronization protocols for memory refresh, storage lifetime management, and garbage collection of stale or degraded qubits \cite{muralidharan2022qmemorymgmt}.

Another requirement is hybrid classical-quantum link coordination. As qubit transmissions and entanglement distribution depend on quantum and classical signaling (e.g., heralding, feedback, acknowledgments), Layer 2 must maintain tight coupling with classical control channels. Latency in these channels can affect synchronization and frame reordering, particularly in session-based protocols or teleportation workflows \cite{bhaskar2022classicalintegration}.
Moreover, Layer 2 must expose link-layer performance metrics—such as QBER, fidelity scores, and entanglement aging rates—to upper layers. These metrics are essential for network-level path selection (Layer 3), session initiation (Layer 5), and orchestration planes (Layer 8). The interface should support telemetry APIs and standard fidelity thresholds that allow for automatic protocol adaptation based on environmental conditions \cite{du2022quantumstackinsights}.
Finally, to support scaling, Layer 2 must be modular and hardware-agnostic, functioning across diverse quantum technologies (e.g., photonic, ion trap, NV center) and supporting interoperation with QKD systems, quantum repeater nodes, and co-located quantum processors.

\subsection{Existing Literature}
Recent literature has seen rapid development in Layer 2 components for quantum networks, particularly in the domains of quantum MAC protocols and QEC. As quantum networking transitions from experimental platforms to early deployments, the reliability of link-layer operations under probabilistic, noisy, and decoherence-prone conditions is a core concern.
A major focus has been on the design of {quantum MAC protocols} that coordinate entanglement distribution and qubit transmission over shared links. Yu and Devitt proposed Q-MAC, a probabilistic MAC protocol that accounts for channel decoherence, memory availability, and entanglement generation success rate \cite{yu2021qmac}. Liu et al. further extended this model by developing an adaptive MAC protocol that dynamically reconfigures link arbitration based on coherence aging and real-time fidelity estimation \cite{liu2023adaptiveqmac}. A recent survey by Bäuml and Dahlberg categorizes quantum MAC strategies into contention-based, reservation-based, and hybrid approaches, noting the lack of standardization and the need for machine learning-driven arbitration \cite{bauml2023qmacsurvey}.

In parallel, {QEC} research has moved from theoretical constructs to hardware-aware codes optimized for real-world use. Fowler et al. demonstrated that surface codes can now achieve fault-tolerant thresholds compatible with near-term quantum hardware \cite{fowler2021highthreshold}. Hao et al. introduced modular QEC schemes that combine syndrome-based error detection with localized, low-overhead decoding strategies suitable for networked architectures \cite{hao2023modularqec}.
{Quantum memory scheduling and buffering}, critical for Layer 2 resource management, has also received attention. Muralidharan et al. proposed fidelity-aware quantum memory management protocols that prioritize qubit refresh and discard policies based on estimated coherence lifetime and teleportation latency \cite{muralidharan2022qmemorymgmt}. Their results highlight the necessity of coupling memory operations with QEC and routing strategies to prevent fidelity loss due to aging.

A further line of research addresses {classical–quantum coordination}, which is essential for synchronizing heralded entanglement, link-layer acknowledgment, and syndrome communication. Bhaskar and Roy showed that efficient classical signaling protocols can significantly reduce frame reordering and latency in distributed QKD and teleportation networks \cite{bhaskar2022classicalintegration}.
More recently, {telemetry-driven MAC and QEC orchestration} has been proposed as a means to unify Layer 2 behaviors with upper-layer decision-making. Du et al. introduced a telemetry API framework that exposes quantum link-layer metrics—such as QBER, qubit freshness, and memory state—to Layer 8 orchestration agents \cite{du2022quantumstackinsights}. This has enabled early implementations of AI-driven fidelity prediction models that adjust entanglement attempts based on historical link behavior \cite{zhao2023aiqnet}.
\begin{table*}[htbp]
    \centering
    \caption{Comparison of selected MAC and QEC strategies for Layer 2 implementation.}
    \label{tab:mac_qec}
    \begin{tabular}{|l|c|l|l|c|}
        \hline
        \textbf{Protocol} & \textbf{Type} & \textbf{Strengths} & \textbf{Limitations} & \textbf{Ref} \\
        \hline
        Q-MAC & Probabilistic & Entanglement-aware, Fidelity-driven & Local scope only & \cite{yu2021qmac} \\
        Adaptive MAC & Dynamic & Real-time reallocation, Memory-sensitive & High control complexity & \cite{liu2023adaptiveqmac} \\
        Surface Code & Topological & High fault tolerance & Requires many ancilla & \cite{fowler2021highthreshold} \\
        Modular QEC & Localized & Hardware-adaptive, Minimal overhead & Platform-specific tuning & \cite{hao2023modularqec} \\
        \hline
    \end{tabular}
\end{table*}

\begin{figure}[]
    \centering
    \includegraphics[width=0.85\linewidth]{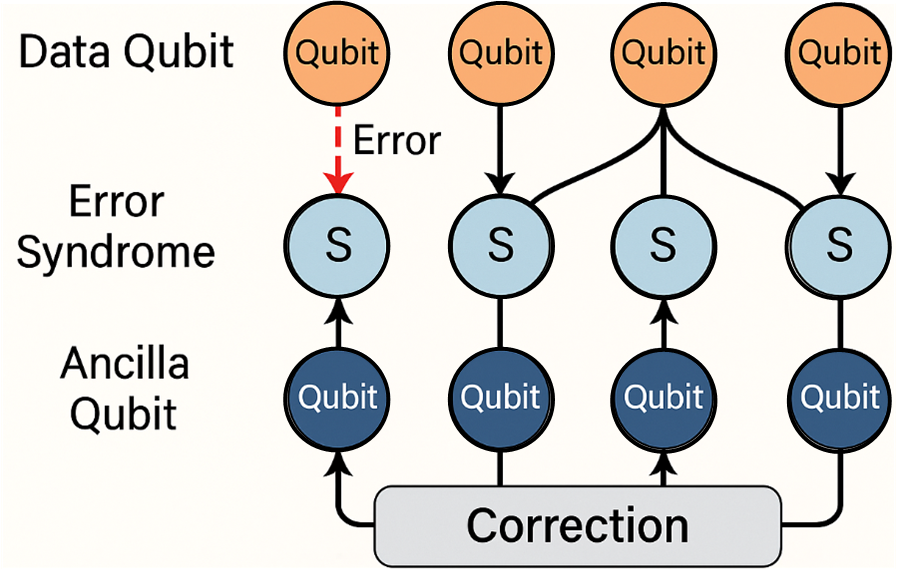}
    \caption{Illustration of link-layer QEC using syndrome-based detection and ancilla qubits for non-invasive error recovery. Based on surface and modular QEC code principles \cite{fowler2021highthreshold, hao2023modularqec}.}
    \label{fig:qec_link}
\end{figure}

\subsection{Technologies and Challenges}
Implementing a scalable, high-fidelity Layer 2 in quantum networks entails significant engineering and architectural challenges. At the forefront is the need for reliable quantum MAC protocols that can coordinate access to entanglement links while accounting for stochastic generation success, memory latency, and environmental decoherence. Contemporary approaches, such as adaptive MAC frameworks, use real-time feedback and entropy-aware channel assessment to avoid resource contention and minimize QBER \cite{liu2023adaptiveqmac, bauml2023qmacsurvey}. However, these systems remain largely simulation-based and lack standardized implementation interfaces.

Another critical technological pillar is {QEC}. Modern implementations focus on low-overhead, hardware-aware codes such as surface codes and repetition codes with minimal qubit overhead. However, fault-tolerant quantum memories remain an obstacle, as the operational fidelity of syndrome measurements and ancilla qubit entanglement must be maintained across distributed hardware \cite{hao2023modularqec}. Even recent high-threshold surface code architectures demand extremely low gate error rates and precise timing synchronization between nodes \cite{fowler2021highthreshold}.
Quantum memory management poses its own set of challenges. Physical quantum memories (e.g., rare-earth crystals, trapped ions, spin-photon interfaces) exhibit finite coherence windows, requiring Layer 2 to implement intelligent memory aging and discard policies. Fidelity must be continuously tracked at the qubit level, enabling the system to retire, refresh, or re-route memory resources based on probabilistic decoherence forecasts \cite{muralidharan2022qmemorymgmt}. As illustrated in Fig.~\ref{fig:qec_link}, link-layer QEC employs ancilla qubits and syndrome extraction to detect and correct errors without collapsing the quantum state. This layer safeguards coherence during transmission over noisy links by integrating modular and surface code-based strategies.

A central limitation across current implementations is the tight coupling between quantum and classical control layers. For example, classical signaling must coordinate entanglement heralding, acknowledge syndrome data, and inform higher layers of successful teleportation. Delays in classical channels can lead to link desynchronization, frame reordering, and quantum memory expiration \cite{bhaskar2022classicalintegration}. Bridging this hybrid protocol layer remains a core challenge in terrestrial and satellite quantum links.
Emerging trends in stack-aware telemetry and AI-assisted fidelity prediction are promising but underdeveloped. Du et al. introduced middleware APIs that expose real-time fidelity metrics and memory states to orchestration agents, enabling stack-wide policy enforcement and adaptive MAC control \cite{du2022quantumstackinsights}. These systems must balance telemetry overhead with the need for sub-millisecond response times to maintain coherence during routing and error correction. More advanced systems are exploring AI-driven agents that predict decoherence trajectories and proactively reserve transmission windows to minimize entanglement loss \cite{zhao2023aiqnet}.
\begin{figure}[]
    \centering
    \includegraphics[width=0.75\linewidth]{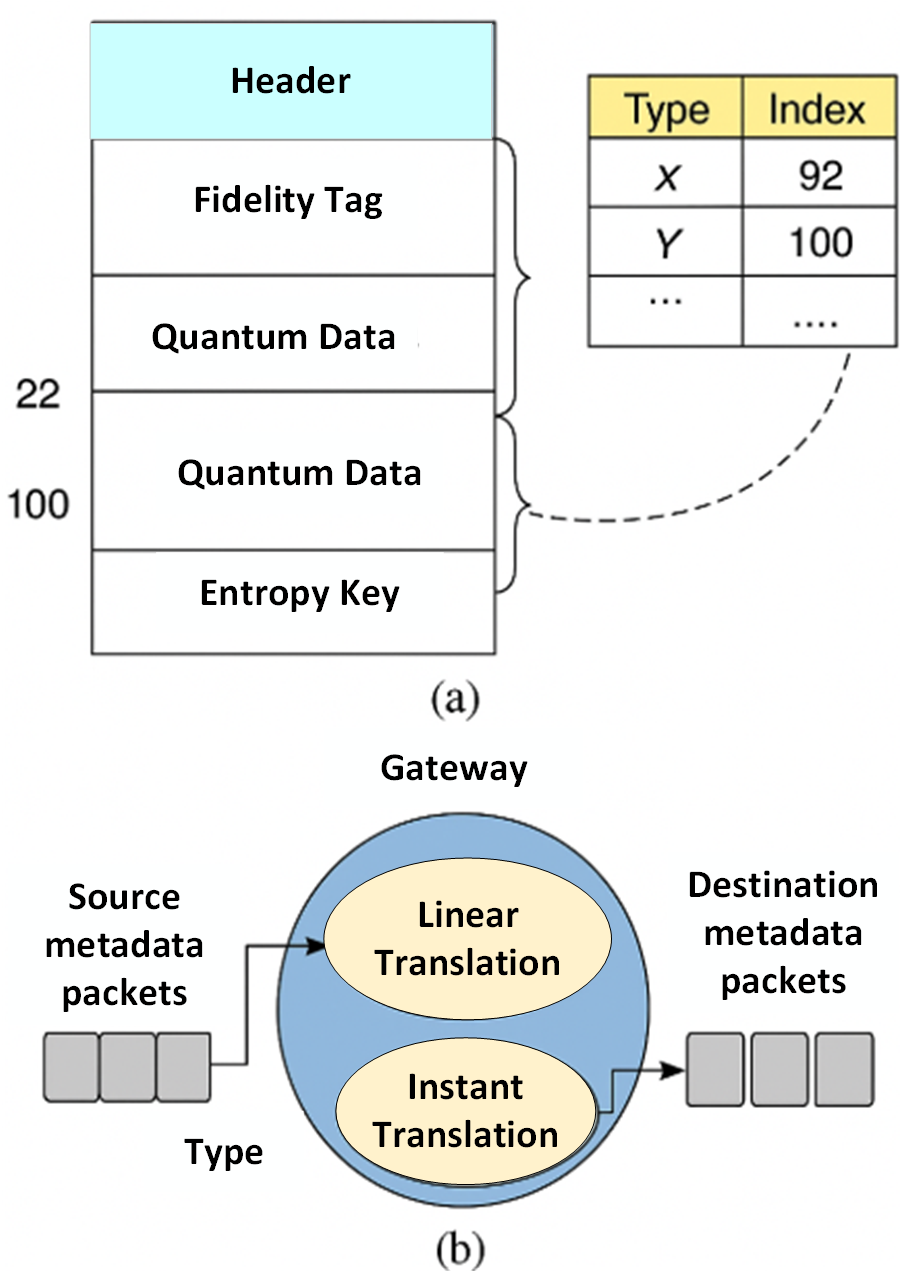}
    \caption{Cross-layer telemetry and control flow integrating MAC arbitration, fidelity monitoring, and orchestration feedback from Layer 8. Adapted from telemetry-aware quantum network control frameworks \cite{du2022quantumstackinsights, zhao2023aiqnet}.}
    \label{fig:telemetry_loop}
\end{figure}

Hardware heterogeneity is an overarching barrier. Variability in photon sources, memory lifetimes, and gate fidelities requires Layer 2 to include hardware abstraction interfaces and modular MAC-QEC pipelines. Without this modularity, system-wide fidelity degrades as qubits traverse incompatible network domains.
Another notable dimension is the integration of security mechanisms within Layer 2 operations. Protocols must address denial-of-service risks, rogue entanglement requests, and coherence flooding attacks. These threats are particularly acute in multi-tenant or federated quantum networks where trust boundaries span across operators. As depicted in Fig.~\ref{fig:telemetry_loop}, modern Layer 2 systems expose fidelity metrics to orchestrators for telemetry-informed link scheduling and MAC optimization \cite{du2022quantumstackinsights, zhao2023aiqnet}.

\subsection{Future Directions}
The future of Layer 2 in quantum networks will be defined by its ability to adapt dynamically to fidelity constraints, environmental variability, and hardware heterogeneity. A major research direction is the development of {AI-assisted MAC protocols} that integrate machine learning models to predict decoherence, entanglement failure probabilities, and QBER trends across time. These intelligent schedulers can enable proactive link arbitration, buffer-aware entanglement queuing, and early detection of channel degradation \cite{zhao2023aiqnet}.
Equally important is the evolution of {low-overhead QEC} schemes that are optimized for distributed, heterogeneous networks. Traditional surface and topological codes, while powerful, often require significant ancillary resources. Future QEC protocols will need to balance error correction strength with circuit depth and memory utilization, particularly in resource-constrained or mobile quantum platforms \cite{hao2023modularqec}. Hybrid QEC schemes combining localized syndrome decoding with real-time hardware feedback may reduce latency and extend quantum memory lifetimes in multi-hop scenarios.

As quantum networks grow in scale and complexity, {modular stack-aware memory management} will become essential. Memory control algorithms must consider not only fidelity and coherence decay but also upper-layer context—such as pending session initiation (Layer 5) or predicted teleportation demand (Layer 6). Emerging approaches explore telemetry-exposed quantum buffers that automatically signal qubit state freshness, buffer saturation, and scheduling preferences to orchestration agents \cite{du2022quantumstackinsights, muralidharan2022qmemorymgmt}.
Future Layer 2 systems will also integrate {middleware-defined abstraction layers} that normalize link-layer operations across diverse quantum hardware types (e.g., photonic, ion trap, NV center). This will enable MAC-QEC pipelines to dynamically load device-specific parameters while exposing unified APIs to Layers 3 and 8 \cite{bhaskar2022classicalintegration}. The result is a virtualized data link layer capable of cross-technology routing and entanglement forwarding.

Another promising direction lies in {telemetry-informed security at Layer 2}. Entanglement abuse, flooding attacks, or memory exhaustion in federated networks could be mitigated by monitoring entanglement request patterns, buffer misuse, and fidelity spoofing attempts. Secure MAC scheduling combined with quantum-safe authentication at link handshakes will support trust frameworks in cross-operator scenarios.
Lastly, Layer 2 research will likely intersect with {quantum digital twins}—virtual replicas of network nodes that simulate qubit state, entropy accumulation, and fidelity drift. These twins can act as predictive controllers for MAC and QEC processes, using real-time feedback from telemetry APIs and Layer 8 agents to preconfigure entanglement schedules and memory usage strategies \cite{zhao2023aiqnet, du2022quantumstackinsights}.
\begin{figure}[]
    \centering
    \includegraphics[width=0.60\linewidth]{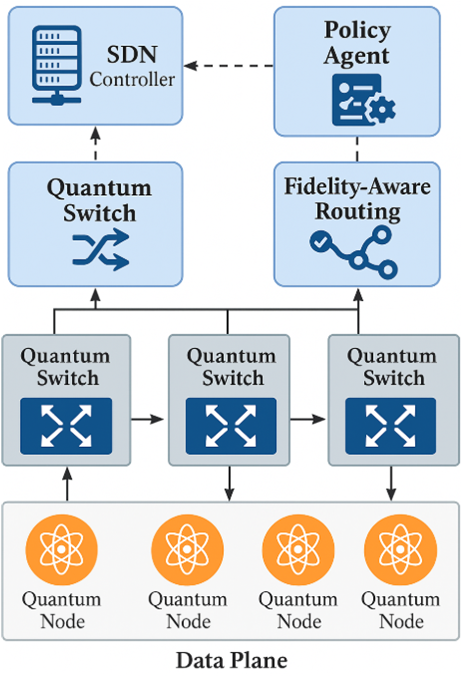}
    \caption{Software-Defined Quantum Network architecture for Layer 3, showing centralized (or distributed) orchestration with fidelity-aware path control and multi-hop entanglement routing. Adapted from \cite{caleffi2020sdqn, du2022quantumstackinsights}.}
    \label{fig:sdqn_routing}
\end{figure}

\section{Layer 3: Network}
Layer 3 in the Quantum-Converged OSI model is the network control and entanglement routing layer, responsible for path selection, fidelity monitoring, and qubit forwarding across multi-hop quantum networks. Unlike classical routing, where packets can be duplicated or rerouted with minimal impact, quantum routing must manage fragile, non-clonable qubits whose viability depends on coherence time, memory scheduling, and probabilistic entanglement success \cite{caleffi2020sdqn, chakraborty2021entanglementrouting}.
The primary function of this layer is to coordinate the establishment and maintenance of entangled links between non-adjacent nodes, enabling quantum teleportation, distributed quantum computation, and end-to-end QKD. This involves orchestrating entanglement swapping, purification, and routing updates across dynamically evolving topologies, often in tandem with classical signaling and hybrid control protocols.

A key evolution in this space is the advent of SDQN. Inspired by SDN paradigms, SDQN introduces a logically centralized controller (or cognitive plane at Layer 8) that dynamically adjusts routing decisions based on real-time fidelity, buffer state, and network entropy conditions \cite{du2022quantumstackinsights}. This enables adaptive and programmable quantum path selection, link repair, and policy enforcement across heterogeneous quantum devices.
Additionally, Layer 3 must also accommodate quantum-aware routing metrics reflecting decoherence rates, link reliability, memory aging, and entanglement fidelity—not just hop count or latency. Multi-path entanglement, probabilistic forwarding, and coherence-time-constrained routing require algorithmic innovations far beyond classical shortest-path protocols \cite{fu2022qaware, jiang2023quantumrouting}. Fig.~\ref{fig:sdqn_routing} illustrates an SDQN architecture where Layer 3 controllers adaptively route entangled links under fidelity and coherence constraints \cite{caleffi2020sdqn, du2022quantumstackinsights}.
In upcoming quantum Internet deployments, Layer 3 will interface directly with physical-layer telemetry (Layer 1), memory coordination (Layer 2), and orchestration agents (Layer 8), forming the backbone of scalable, programmable, and survivable quantum infrastructures.
The following subsections provide a structured analysis of Layer 3’s requirements, supporting literature, enabling technologies, and forward-looking challenges.

\subsection{Requirements}
Layer 3 in a quantum network stack fundamentally differs from its classical counterpart due to quantum information's probabilistic and fragile nature. Instead of routing deterministic data packets, it must manage the creation, swapping, and consumption of entangled links across a distributed set of quantum nodes. As such, the Layer 3 quantum network must satisfy unique functional and non-functional requirements.
The most critical requirement is support for {fidelity-aware entanglement routing}. Unlike classical paths optimized for hop count or latency, quantum routes must be chosen based on dynamic coherence time, QBER, and available entanglement fidelity at each hop. Routing metrics must reflect time-sensitive degradation of quantum states, probabilistic entanglement success, and memory buffer limitations at intermediary nodes \cite{fu2022qaware, chakraborty2021entanglementrouting}.
The layer must also coordinate {entanglement swapping} operations across nodes that do not share direct quantum links. This requires precise orchestration of intermediate Bell measurements and synchronization with classical channels and Layer 2 buffers. Failures or delays in these operations can lead to decoherence or invalid routing paths, necessitating real-time rerouting mechanisms \cite{caleffi2020sdqn}.
Additionally, Layer 3 must implement a {probabilistic path computation engine}. This engine must adapt to frequent entanglement availability and fidelity changes, using statistical models or RL to discover and update viable routing paths. Traditional shortest-path algorithms like Dijkstra or Bellman-Ford are inadequate due to their lack of coherence sensitivity and assumption of stable link states \cite{jiang2023quantumrouting}.

Another key requirement is integration with SDQN controllers. These centralized or logically distributed agents receive telemetry from lower layers (e.g., qubit coherence, buffer status, path aging) and compute network-wide routing and policy decisions. This mandates Layer 3 APIs for real-time telemetry ingestion, intent translation, and reconfiguration under programmable logic \cite{du2022quantumstackinsights, caleffi2020sdqn}.
Layer 3 must also be compatible with {multi-domain quantum network architectures}, where entanglement flows across different administrative regions or technologies (e.g., fiber vs. satellite). This implies support for inter-domain routing policies, fidelity-based trust constraints, and entanglement reservation systems similar to classical Quality-of-Service (QoS) reservation protocols.
Lastly, Layer 3 must support {entanglement-based session continuity}. As quantum states cannot be cloned or buffered indefinitely, routing decisions must be tightly coupled with session initiation (Layer 5) and entanglement handshakes. The network layer must expose qubit readiness, path fidelity scores, and routing success probabilities to upper layers, enabling predictive session binding and application-level QoS enforcement.

\subsection{Existing Literature}
The study of Layer 3 in quantum networks has rapidly advanced with the rise of entanglement-based communications and the need for coherence-aware routing. The classical concepts of shortest-path and deterministic forwarding are insufficient for quantum systems, where entanglement is probabilistic, fidelity decays over time, and memory coherence constraints play a central role.
Caleffi et al. introduced the concept of {SDQN}, proposing a layered abstraction model where control logic is decoupled from the data plane, allowing global fidelity- and intent-aware routing decisions \cite{caleffi2020sdqn}. Their architecture facilitates entanglement orchestration via a centralized controller informed by live metrics such as QBER and memory aging.
Chakraborty et al. modeled {entanglement routing as a stochastic process} involving BSM operations and fidelity-aware swap path assembly. Their framework provided formal performance guarantees under probabilistic link success, defining entanglement throughput and decoherence tradeoffs \cite{chakraborty2021entanglementrouting}.
Fu et al. proposed a fidelity-aware dynamic routing protocol that predicts end-to-end path viability based on real-time fidelity degradation and temporal coherence limits \cite{fu2022qaware}. Their results show that classical routing metrics (e.g., hop count, latency) lead to substantial fidelity loss in large-scale networks.

Zhang et al. proposed {entanglement flow optimization models} that integrate traffic shaping, buffer availability, and path reliability scores into a RL framework \cite{zhang2022reinforcement}. Similarly, Jiang et al. implemented a deep Q-learning model to learn probabilistic routing behaviors that maximize teleportation success in fluctuating quantum topologies \cite{jiang2023quantumrouting}.
Van Meter et al. developed {quantum repeater-based architectures} to facilitate routing in hybrid fiber-satellite infrastructures, identifying policy conflicts between fidelity maximization and entanglement distribution latency \cite{vanmeter2021hybridrouting}. Their architectural models emphasize path provisioning mechanisms using telemetry-based fidelity scoring.
Yuan et al. proposed a {multi-tenant quantum routing policy engine} that supports concurrent entanglement sessions using isolated path-state buffers and link arbitration \cite{yuan2023multitenant}. Their work introduced a trust-scoped path allocation scheme to prevent cross-session coherence interference.
Du et al. provided an API-layer abstraction for telemetry-integrated routing stacks, where Layer 3 continuously ingests qubit fidelity metrics, swap attempt success rates, and classical handshaking feedback from Layer 2 \cite{du2022quantumstackinsights}. Their model allows intent-driven orchestration agents (Layer 8) to influence real-time route selection.
Zhao et al. presented an AI-assisted stack controller that integrates RL and entropy forecasting to preemptively reconfigure routing paths in anticipation of fidelity collapse, forming the basis for {cognitive entanglement routing} \cite{zhao2023aiqnet}.
\begin{table*}[htbp]
    \centering
    \caption{Comparison of Layer 3 routing protocols in quantum networks.}
    \label{tab:layer3_routing}
    \begin{tabular}{|l|c|l|l|l|}
        \hline
        \textbf{Protocol} & \textbf{Type} & \textbf{Strengths} & \textbf{Limitations} & \textbf{Reference} \\
        \hline
        SDQN & Centralized & Policy-based, telemetry-driven & Control latency & \cite{caleffi2020sdqn} \\
        Bell-Chain Probabilistic & Stochastic & Coherence-aware path modeling & Fragile swap dependencies & \cite{chakraborty2021entanglementrouting} \\
        Deep RL Routing & AI-Based & Adaptive to real-time fidelity drift & Convergence time & \cite{jiang2023quantumrouting} \\
        Hybrid RL-Heuristic & Combined & Balance speed and precision & Requires policy tuning & \cite{zhang2022reinforcement} \\
        \hline
    \end{tabular}
\end{table*}

\subsection{Technologies and Challenges}
Deploying Layer 3 capabilities in quantum networks introduces a new class of architectural and technological challenges that differ fundamentally from classical routing. At its core, Layer 3 must coordinate entanglement routing over time-variant, probabilistic links while maintaining coherence and fidelity across multi-hop paths.
One major challenge is the design of {fidelity-aware routing algorithms} that operate under entanglement uncertainty and hardware variability. Traditional link-state or distance-vector protocols assume deterministic link metrics and persistent connectivity. In contrast, quantum links are ephemeral, and their success depends on stochastic entanglement generation, qubit coherence times, and memory freshness. Thus, routing engines must dynamically recompute paths based on temporal fidelity degradation and qubit expiration forecasts \cite{fu2022qaware, chakraborty2021entanglementrouting}.
{Entanglement swapping orchestration} remains a fragile and error-prone operation that Layer 3 must manage. Swapping operations require precise timing, classical feedforward coordination, and error detection between intermediate nodes. Errors in swap order or classical feedback can lead to fidelity collapse or loss of end-to-end connectivity \cite{vanmeter2021hybridrouting}. Ensuring synchronization between Layer 3 and Layer 2 is essential but challenging, particularly in high-traffic or inter-domain environments.
Another technological bottleneck lies in {SDQN control planes}. While SDQN architectures promise flexibility, real-time path orchestration, and intent-driven routing, practical implementations are still early. The need for quantum-aware telemetry APIs that expose fidelity, coherence decay, and memory statistics remains unmet in most experimental testbeds \cite{du2022quantumstackinsights, caleffi2020sdqn}. Furthermore, control latency in SDQN systems could exceed coherence windows, making centralized decisions impractical without edge-level delegation.

{Scalability} is another open challenge. As networks grow beyond tens of nodes, entanglement routing becomes exponentially complex due to the combinatorics of possible entangled paths and fidelity permutations. RL-based routing agents have been proposed to address this, but their convergence time and real-time applicability remain limited \cite{jiang2023quantumrouting, zhang2022reinforcement}. Hybrid AI-heuristic models may help balance route quality and decision latency.
Quantum networks also suffer from {heterogeneous hardware integration}. Differences in qubit types (e.g., NV centers vs. ion traps), transmission wavelengths, and quantum memory lifetimes require routing decisions considering hardware-specific fidelity drop-off functions. Layer 3 must include a hardware abstraction layer to normalize routing metrics across diverse platforms \cite{yuan2023multitenant}.
{Security concerns} at the network layer are particularly novel in the quantum domain. Quantum denial-of-service (QDoS) threats—such as link saturation with false entanglement requests—can waste limited quantum memory and destroy active sessions. Secure entanglement reservation and rate-limiting policies are not yet standardized, especially for multi-tenant federated quantum networks \cite{yuan2023multitenant}.
Lastly, terrestrial and satellite quantum network interoperability presents physical and protocol-level obstacles. Timing synchronization, Doppler shift correction, and dual-mode routing stacks must be incorporated into Layer 3 for hybrid path establishment and failover \cite{vanmeter2021hybridrouting}.

\subsection{Future Directions}
Future advancements in Layer 3 of quantum networks will focus on increasing scalability, adaptability, and resilience by integrating AI-based control, distributed orchestration, and cross-domain interoperability. Entanglement's fragility and probabilistic nature require a fundamental rethinking of how network control functions are designed, provisioned, and adapted over time.
One promising direction is the evolution of {decentralized SDQN}. While current SDQN models rely on a centralized control plane, future implementations will likely adopt hierarchical or federated models to reduce latency and mitigate control bottlenecks. Distributed SDQN controllers will locally manage entanglement resources and routing policies based on regional fidelity and memory metrics while remaining synchronized with global orchestration agents \cite{caleffi2020sdqn, du2022quantumstackinsights}.

\begin{figure}[]
    \centering
    \includegraphics[width=0.9\linewidth]{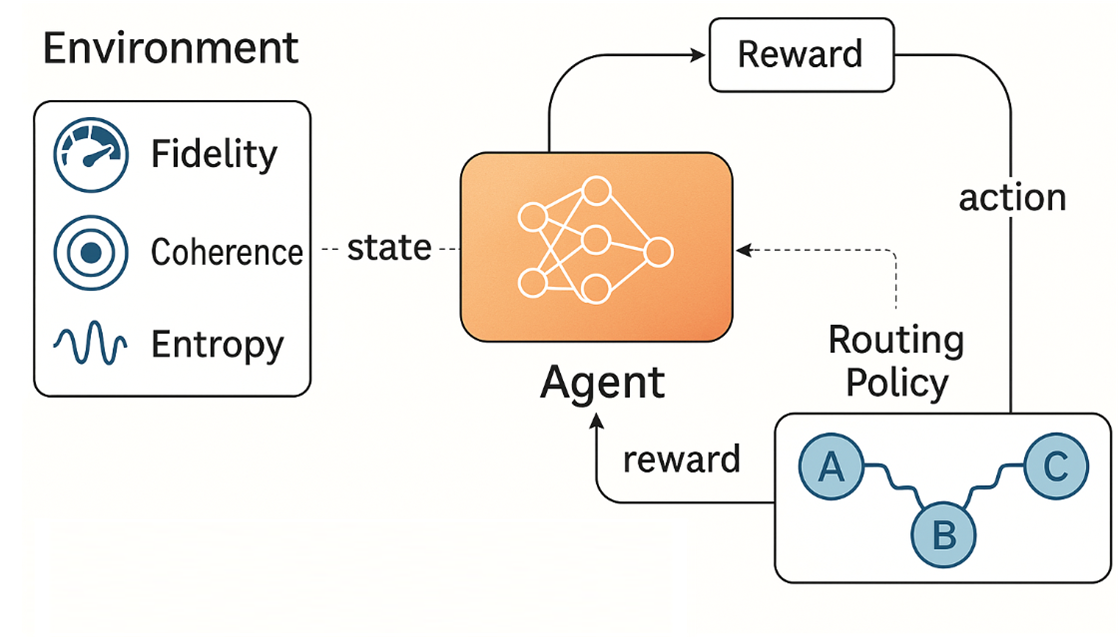}
    \caption{RL-assisted quantum routing model that uses fidelity, coherence, and entropy telemetry to learn optimal entanglement paths. Based on \cite{jiang2023quantumrouting, zhao2023aiqnet}.}
    \label{fig:rl_routing}
\end{figure}

{RL-assisted routing} is expected to become foundational in Layer 3 decision-making. Agents can be trained on real-time feedback from quantum telemetry APIs to predict optimal routes, perform load balancing, and anticipate fidelity degradation. Hybrid RL-heuristic models may provide a balance between convergence speed and routing precision, enabling proactive entanglement path provisioning and route failover strategies \cite{jiang2023quantumrouting, zhang2022reinforcement, zhao2023aiqnet}.
Another important development is the emergence of {intent-driven quantum networking}, expected to impact Layer 3 operations.. In this paradigm, applications express quantum intent—such as “prepare entanglement between Alice and Bob with >90\% fidelity in 20 ms”—and the network layer computes the optimal path based on live resources and environmental conditions. This will require deep integration of Layer 3 with orchestration APIs and predictive fidelity models \cite{du2022quantumstackinsights}.
As shown in Fig.~\ref{fig:rl_routing}, RL agents can train on stack telemetry to discover dynamic, fidelity-aware quantum routes \cite{jiang2023quantumrouting, zhao2023aiqnet}.
As depicted in Fig.~\ref{fig:rl_routing}, the RL-assisted quantum routing model leverages real-time telemetry—such as fidelity, coherence time, and entropy metrics—to learn optimal paths for entanglement distribution. This approach enables adaptive routing under dynamic quantum network conditions.

Another frontier lies in {multi-domain and multi-technology routing}. Future quantum networks will span heterogeneous regions—combining fiber, satellite, and chip-scale entanglement domains. Layer 3 must interoperate across differing quantum platforms, trust boundaries, and routing policy frameworks. Federated route negotiation protocols and inter-domain telemetry exchange will become essential \cite{yuan2023multitenant, vanmeter2021hybridrouting}.
{Programmable path redundancy and fault tolerance} are also critical. Quantum links are inherently unreliable; therefore, Layer 3 must be capable of provisioning disjoint entangled paths and switching between them when QBER thresholds are exceeded. Routing stacks may embed fidelity-driven link prediction modules that flag weakening paths before they fail, similar to preemptive failover in classical MPLS networks \cite{fu2022qaware}.
Lastly, {security-aware routing} in federated networks will require quantum-aware access control, rate limiting, and entanglement reservation systems. These measures are necessary to prevent denial-of-service conditions in shared memory pools and to isolate entanglement flows by session, tenant, or security clearance \cite{yuan2023multitenant}.

\section{Layer 4: Transport}
Layer 4 in the Quantum-Converged OSI model introduces the mechanisms to establish, maintain, and gracefully terminate quantum communication sessions between distributed nodes. Unlike its classical counterpart—which relies on sequence numbers, acknowledgments, and retransmission to ensure end-to-end reliability—the quantum transport layer must preserve entanglement coherence and fidelity over dynamic network paths without violating no-cloning constraints or introducing irreversible quantum measurement.
This layer coordinates {quantum session control}, including handshake protocols, entanglement usage negotiation, and session termination. Additionally, it must manage temporal resource allocation, enforce fidelity agreements, and adapt to link variability during live sessions. Moreover, in QKD, it ensures the secure completion of key generation rounds. Furthermore, in quantum teleportation, it guarantees the delivery of Bell pairs with integrity throughout their lifecycle.

A key function of Layer 4 is to handle {flow control and session binding} across probabilistic, fidelity-sensitive connections. Since quantum information cannot be buffered, retransmitted, or cloned, the transport layer must implement fidelity-aware scheduling, timeout negotiation, and application-driven qubit reservation. These capabilities are tightly integrated with lower layers—particularly Layer 3 (Routing) and Layer 2 (Buffers and MAC)—through bidirectional signaling and telemetry.
Additionally, Layer 4 must provide abstraction mechanisms that hide physical-layer instability from upper-layer applications. Through session continuity management, Layer 4 can remap entanglement paths transparently in response to fidelity degradation, memory expiration, or route changes triggered by the SDQN controller or Layer 8 cognitive agents.
Emerging proposals for this layer include teleportation protocol wrappers, entanglement time budgeting, and error-conditioned qubit flushing policies that allow higher layers to remain agnostic of underlying quantum dynamics. This section explores the core requirements, research literature, technological enablers, and long-term challenges for enabling a secure, scalable, and programmable quantum transport layer.

\subsection{Requirements}
Layer 4 of the Quantum-Converged OSI model enables reliable, session-aware, and fidelity-constrained quantum communication across distributed nodes. Unlike classical transport layers, which use mechanisms like packet sequencing, acknowledgments, and retransmissions, the quantum transport layer must operate without violating the no-cloning theorem or collapsing fragile qubit states. As such, it must support quantum-native reliability and flow control primitives tailored to the coherence constraints of entangled information.
A foundational requirement is {quantum session control}, including handshake protocols that negotiate fidelity thresholds, timeout durations, and entanglement consumption policies before information exchange. These sessions may involve multi-hop entanglement chains, and thus require contextual integration with Layer 3 routing updates and Layer 2 buffer availability \cite{vincent2023sessionq, schoute2022qtransport}.
In addition, the layer must also implement {flow control mechanisms} adapted to quantum links' non-deterministic and lossy nature. These include qubit transmission windows, coherence-constrained pacing policies, and application-level entanglement budgeting strategies. Because qubits cannot be retransmitted, flow control must balance delivery timing against qubit degradation to maintain usable fidelity at the receiver \cite{fu2022qaware}.

Another critical requirement is support for {fidelity-aware transport adaptation}. Layer 4 must be able to react to telemetry inputs such as entanglement lifetime, QBER, or link failure probability. It must adjust session parameters accordingly, possibly renegotiating session terms, switching entangled paths (via Layer 3), or releasing reserved qubits preemptively to conserve memory \cite{du2022quantumstackinsights}.
Additionally, {Session continuity and abstraction} are also essential. Layer 4 must decouple logical quantum sessions from the underlying entangled paths. If fidelity along a given route degrades or if a quantum memory expires, the transport layer should transparently rebind the session to a new entangled path while maintaining logical continuity. This capability supports higher-layer consistency in teleportation-based workflows and QKD key streams \cite{vincent2023sessionq}.
The transport layer must also enforce {application-layer QoS constraints}, such as minimum fidelity, latency bounds for teleportation, or key refresh rates in QKD. These constraints must be negotiated during session setup and enforced throughout the session’s lifetime, triggering feedback to Layer 8 orchestration agents when violations occur \cite{schoute2022qtransport}.
Finally, Layer 4 must support {multiplexed session coordination}, allowing multiple logical quantum tasks to share entanglement resources across the same physical infrastructure. This necessitates tagging, isolation, and ordering mechanisms to prevent interference between concurrent quantum sessions.

\subsection{Existing Literature}
The development of transport-layer protocols for quantum networks is a recent but critical area of research. Traditional TCP/IP-based models are inadequate in quantum contexts, where retransmission, buffering, and packet duplication are physically impossible due to the no-cloning theorem and coherence sensitivity. As a result, researchers have proposed transport-layer architectures centered on quantum session control, entanglement scheduling, and fidelity-aware flow policies.

Vincent et al. introduced one of the first formal models of {quantum session abstraction}, proposing a session-aware protocol that coordinates entanglement acquisition and maintenance under fidelity and coherence constraints \cite{vincent2023sessionq}. Their work emphasizes decoupling logical session states from physically entangled links, enabling applications to operate transparently even as fidelity degrades or paths are re-allocated.
Schoute and Wehner proposed a layered framework for {quantum transport services}, including teleportation session setup, memory-aware qubit flushing, and fidelity-constrained delivery guarantees \cite{schoute2022qtransport}. They define transport-layer contracts that allow applications to request specific coherence durations, minimum entanglement fidelity, and path continuity, all while leveraging telemetry feedback from lower layers.
Fu et al. contributed to the conversation from a routing-adjacent angle, introducing fidelity-aware timing and pacing constraints for teleportation flows over unreliable links \cite{fu2022qaware}. Their results showed that link-layer metrics—like entanglement age and qubit freshness—must be explicitly considered at the transport level to prevent session violation or fidelity loss.
Du et al. proposed a stack-integrated architecture that incorporates telemetry APIs exposing memory state, QBER, and routing behavior to Layer 4 session managers \cite{du2022quantumstackinsights}. This design enables the transport layer to trigger session renegotiation, initiate handoffs, or terminate sessions when underlying qubit states degrade beyond threshold values.

Another line of research has addressed {teleportation-lifecycle-aware transport design}. Nasir et al. modeled quantum transport as a lifecycle system, where each qubit pair undergoes entanglement generation, fidelity verification, teleportation usage, and measurement collapse \cite{nasir2021teleportflow}. Their framework suggests buffering pre-verified Bell pairs and integrating application-level fidelity policies directly into session controllers.
Zhou et al. extended these ideas by simulating {transport-layer error flushing policies}, where qubits that fall below fidelity thresholds are automatically released or redirected to lower-priority tasks \cite{zhou2023quantumflush}. Their simulation study highlights the risk of fidelity “deadlocks” where low-quality qubits block session progress unless actively flushed or replaced.
Emerging work by Huang et al. explores {multiplexed quantum session scheduling} across shared memory pools and routing interfaces, proposing tagging-based qubit isolation for concurrent transport flows \cite{huang2023multiplexed}. Their scheme supports prioritization, inter-session handoffs, and teleportation result coordination.

\subsection{Technologies and Challenges}
Realizing a robust and scalable quantum transport layer demands novel technologies that are deeply aware of quantum mechanical constraints while overcoming key architectural and coordination challenges. The fragility of entanglement, non-deterministic link behavior, and qubit irreversibility make Layer 4 design fundamentally different from classical transport mechanisms.
One key challenge is enabling {session-level abstraction over volatile entanglement paths}. Physical links may suffer from fidelity collapse or memory expiration mid-session, yet applications expect uninterrupted logical service. Transport protocols must dynamically remap or renegotiate session routes using Layer 3 telemetry while hiding these transitions from the application layer \cite{vincent2023sessionq, schoute2022qtransport}.
\begin{table*}[htbp]
    \centering
    \caption{Comparison of quantum transport-layer strategies and their capabilities.}
    \label{tab:layer4_transport}
    \begin{tabular}{|l|l|l|l|}
        \hline
        \textbf{Protocol / Feature} & \textbf{Objective} & \textbf{Limitation} & \textbf{Reference} \\
        \hline
        SessionQ & Session abstraction & Path rebind complexity & \cite{vincent2023sessionq} \\
        QTransport Contract & SLA enforcement & Requires deep telemetry & \cite{schoute2022qtransport} \\
        Entanglement Budgeting & Qubit scheduling & SLA enforcement overhead & \cite{du2022quantumstackinsights} \\
        Fidelity-Aware Flow Control & Session integrity & Irreversible losses on error & \cite{fu2022qaware} \\
        Multiplexing & Concurrent sessions & Flow interference, tagging & \cite{huang2023multiplexed} \\
        \hline
    \end{tabular}
\end{table*}

A second major issue is {flow control over non-bufferable quantum data}. Since qubits cannot be cloned or retransmitted, transport-layer flow control must operate on timing windows and probabilistic delivery guarantees rather than sequence buffers. Memory over-commitment or misaligned scheduling can result in fidelity drops, requiring live qubit rejection or swapout strategies \cite{zhou2023quantumflush, du2022quantumstackinsights}.
Additionally, transport-layer protocols must also manage {session timing and lifetime constraints} that account for qubit coherence windows, teleportation handshakes, and entanglement scheduling. Moreover, teleportation success is highly sensitive to the freshness of Bell pairs; therefore, session managers must interact closely with memory aging metrics from Layer 2 and fidelity prediction models from Layer 3 \cite{nasir2021teleportflow, fu2022qaware}.

A persistent technical hurdle is {integrating real-time telemetry and feedback loops}. While recent telemetry APIs provide access to metrics like QBER, link uptime, and qubit freshness, incorporating this into transport decisions with millisecond responsiveness remains challenging. Session-layer logic must consume, interpret, and act on these metrics without exceeding coherence delays \cite{du2022quantumstackinsights}.
{Multiplexed session scheduling} is another emerging challenge. Layer 4 must support concurrent quantum applications sharing memory and routing infrastructure as networks scale. Isolating these flows while preserving fairness, latency guarantees, and fidelity targets introduces complex session tagging, priority queues, and quantum namespace management \cite{huang2023multiplexed}.
Finally, hardware-standardized support for session primitives—such as teleportation completion callbacks, memory state signaling, or fidelity-aware qubit reservation—makes deployment across vendor ecosystems difficult. Transport layer abstractions must eventually decouple logical flows from platform-specific hardware behaviors. As shown in Fig.~\ref{fig:rl_routing}, qubit expiration and link volatility drive flow-control adjustments to maintain teleportation fidelity and session viability \cite{du2022quantumstackinsights, zhou2023quantumflush}.

\subsection{Future Directions}
The evolution of the quantum transport layer is central to enabling stable, scalable quantum applications over complex, probabilistic infrastructure. Future developments will prioritize session resilience, cross-layer adaptability, and the ability to abstract volatile quantum resources into predictable service contracts.
A significant research trajectory is the design of {predictive session binding}, in which the transport layer proactively reserves entanglement paths and qubit memory slots based on application intent and real-time fidelity forecasts. Using telemetry from lower layers, Layer 4 can pre-bind sessions to resources expected to satisfy fidelity, coherence, and latency constraints over the session duration \cite{du2022quantumstackinsights, vincent2023sessionq}.
Another key direction is {teleportation lifecycle orchestration}, where the transport protocol manages the complete qubit journey—from entanglement generation to Bell measurement and classical transmission. Session-aware transport stacks will monitor this lifecycle for timing violations, failed swaps, or fidelity drops and trigger rerouting or reinitialization if session integrity is at risk \cite{schoute2022qtransport, nasir2021teleportflow}.
Moreover, {session virtualization} is also emerging to scale quantum flows. This involves abstracting transport sessions from their physical entanglement routes, enabling dynamic session migration across fidelity-compatible paths without upper-layer disruption. Such virtualization would support load balancing, preemption, and failure recovery—especially in multi-hop or hybrid (e.g., fiber-satellite) networks \cite{zhou2023quantumflush, vincent2023sessionq}.
\begin{figure}[]
    \centering
    \includegraphics[width=0.7\linewidth]{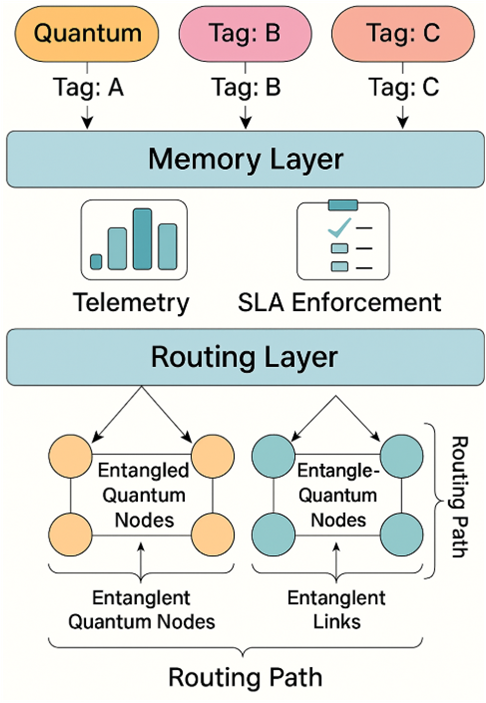}
    \caption{Concurrent quantum sessions multiplexed over shared memory and routing layers, with tagging for isolation and telemetry for SLA enforcement. Based on designs in \cite{huang2023multiplexed}.}
    \label{fig:quantum_multiplex}
\end{figure}

To meet the demands of increasingly multi-tenant quantum networks, future transport stacks must implement {flow isolation and multiplexed qubit tagging}. These mechanisms would enable concurrent session management, ensuring that QKD flows, quantum computing tasks, and teleportation services do not interfere with each other despite sharing memory pools or entangled links \cite{huang2023multiplexed}.
Another frontier involves {intent-aware transport orchestration}, where applications specify fidelity-SLA (Service Level Agreement) objectives like: “deliver 10 Bell pairs with $\geq$95\% fidelity within 50 ms.” In coordination with Layer 8, the transport layer would translate these goals into real-time session setup, memory provisioning, and retry strategies, reacting dynamically to decoherence or QBER drift \cite{du2022quantumstackinsights}.
Finally, {hardware-agnostic transport interfaces} will become essential. Future transport APIs must normalize behavior across a broad spectrum of quantum devices—photonic chips, NV centers, trapped ions—allowing universal session control over heterogeneous quantum infrastructure.
Fig.~\ref{fig:quantum_multiplex} visualizes multi-session multiplexing in the transport layer, where tagging and resource fencing enable concurrent entangled flows \cite{huang2023multiplexed}.

\section{Layer 5: Session }
\begin{figure}[]
    \centering
    \includegraphics[width=0.55\linewidth]{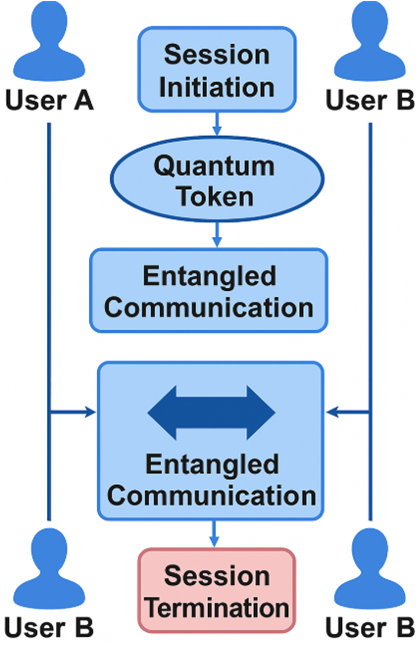}
    \caption{Session-layer role negotiation across trust domains. The sender, verifier, and receiver exchange authenticated metadata to bind entangled states with session logic. Based on identity resolution models in \cite{rao2023identityrouting, chen2023teleportationfabric}.}
    \label{fig:session_roles}
\end{figure}

Layer 5 in the Quantum-Converged OSI model serves as the orchestration hub for distributed quantum application sessions. It is responsible for establishing, coordinating, and maintaining logical quantum sessions between applications or user agents over physical and virtual quantum links. Unlike the transport layer, which deals with qubit delivery and entanglement continuity, the session layer operates at a higher abstraction level—aligning session logic with task roles, identities, and end-to-end service agreements.
A primary function of this layer is to support {quantum service binding}, where application-layer requests (e.g., QKD, blind quantum computation, distributed quantum sensing) are mapped to underlying entanglement sessions and resource topologies. This includes synchronizing pre-entangled pairs, managing logical identities, and orchestrating handshake protocols that enforce session type, task role, and fidelity constraints.
Quantum sessions may involve multi-party coordination (e.g., GHZ-based voting, quantum conference keying) or multi-domain entanglement spanning heterogeneous infrastructures. Thus, Layer 5 must also manage {cross-domain role coordination}, identity resolution, and protocol dispatch—ensuring that all participants operate under a common semantic contract despite network variability. Fig.~\ref{fig:session_roles} shows cross-domain role binding with session-layer trust propagation \cite{rao2023identityrouting, chen2023teleportationfabric}.

An additional responsibility of Layer 5 is maintaining session state under fidelity constraints. It must monitor and adjust for decoherence risks, timeouts, or fidelity violations that may require session renegotiation, pausing, or termination. This includes coupling to Layer 8 orchestration services and Layer 4 transport policies to dynamically respond to changes in qubit availability or routing conditions.
Emerging research in session-layer abstractions introduces ideas such as fidelity-scoped session classes, identity-aware qubit tagging, Quantum Session Managers (QSMs), and cross-platform session brokers. These innovations aim to make the session layer robust to environmental drift while enabling high-level automation and trust-driven collaboration across quantum nodes.
The subsections that follow outline the design requirements, literature foundations, implementation technologies, and forward-looking challenges of Layer 5 in quantum network architectures.

\subsection{Requirements}
The session layer in a quantum network architecture plays a pivotal role in abstracting low-level entanglement flows into application-aligned service contexts. Layer 5 must coordinate the creation, maintenance, and teardown of logical quantum sessions between endpoints, while accommodating quantum-specific constraints such as coherence time, fidelity drift, and identity-sensitive task allocation.
A core requirement is the ability to perform {quantum service binding}, where high-level tasks—such as QKD, quantum secure multiparty computation (QSMC), or delegated quantum computing—are mapped to active entanglement sessions with the appropriate fidelity guarantees, timing constraints, and participant roles \cite{vincent2023sessionq, goyal2023quantumsession}. This requires awareness of transport-layer capabilities and dynamic negotiation with the orchestration layer (Layer 8).
Additionally, Layer 5 must support {identity-aware session coordination}. Many quantum applications, especially those involving asymmetric trust (e.g., blind quantum computation or federated sensing), depend on role-specific entanglement handling. The session layer must resolve identities across quantum nodes, assign task-specific roles (e.g., sender, verifier, receiver), and enforce identity-scoped quantum contracts throughout the session lifecycle \cite{rao2023identityrouting}.

It must also provide mechanisms for {cross-domain session control}, enabling multi-party and multi-administrative entanglement sessions to operate coherently. In such scenarios, Layer 5 must manage protocol synchronization, role verification, and trust-boundary enforcement to ensure that participants adhere to a unified session policy despite underlying network heterogeneity \cite{goyal2023quantumsession, yin2022sessionroles}.
A further requirement is the implementation of {fidelity-scoped session contracts}. The session layer must continuously monitor fidelity metrics (e.g., entanglement quality, decoherence rate, qubit expiration) and ensure that session-specific Service Level Objectives (SLOs) are maintained. This demands integration with telemetry APIs from Layer 2 and Layer 3 and the ability to trigger re-binding or graceful termination if fidelity thresholds are violated \cite{du2022quantumstackinsights}.
The session layer must also expose {session lifecycle management interfaces} to higher layers. These include APIs for session creation, pause, resumption, and teardown, as well as support for soft state transitions—such as session migration or role reassignment—in response to environmental changes or application behavior.
Lastly, Layer 5 must support {concurrent quantum session isolation and tagging}. As quantum networks scale to support multiple simultaneous services, the session layer must ensure that qubits, entanglement buffers, and identity roles are scoped correctly to their originating session. This prevents information leakage, role confusion, and session interference in shared memory environments.
\begin{table*}[htbp]
    \centering
    \caption{Session-layer role-binding models and their operational characteristics.}
    \label{tab:session_roles}
    \begin{tabular}{|l|l|c|c|l|}
        \hline
        \textbf{Model} & \textbf{Use Case} & \textbf{Identity Binding} & \textbf{Scope} & \textbf{Reference} \\
        \hline
        Teleportation Triad & TaaS, Circuit Injection & Yes & Node-local & \cite{chen2023teleportationfabric} \\
        QKD Bidirectional & Secure Key Setup & Yes & Pairwise Link & \cite{vanmeter2021qkdservices} \\
        Stateless Session & Blind Quantum Compute & Optional & Single-flow & \cite{tavakoli2023distributedqml} \\
        Federated Mapping & DQML, Entangled Voting & Yes & Distributed Role-Aware & \cite{rao2023identityrouting} \\
        \hline
    \end{tabular}
\end{table*}

\subsection{Existing Literature}
Recent advances in quantum networking have catalyzed interest in session-layer architectures that abstract entanglement flows into secure, role-aware, and service-consistent quantum sessions. The session layer has been reconceptualized not simply as a communication convenience, but as a crucial coordination and binding layer for application-level tasks such as QKD, delegated computing, and distributed consensus.
Vincent et al. introduced a formal framework for {session-aware entanglement management}, where quantum sessions are constructed from a pool of transport-level Bell pairs and maintained with coherence-aware consistency rules. Their approach defines sessions as dynamic, role-scoped containers that include fidelity thresholds and validity windows to support advanced applications like teleportation-as-a-service \cite{vincent2023sessionq}.
Goyal et al. proposed one of the first {cross-domain session orchestration models}, highlighting the problem of heterogeneous session binding across administrative boundaries. They introduced Quantum Session Descriptors (QSDs) to capture role assignments, fidelity contracts, and policy constraints, which can be exchanged and verified across domains for secure session instantiation \cite{goyal2023quantumsession}.

Yin et al. focused on {role-aware coordination}, showing that many multi-party quantum protocols—e.g., GHZ-based voting or conference keying—require persistent and verifiable role assignments (e.g., leader, sender, verifier). Their proposed session layer implements a quantum role registry and session authorization logic to ensure identity-role consistency \cite{yin2022sessionroles}.
Rao and Kim addressed the challenge of {identity resolution and secure session admission}. They proposed an identity-based session negotiation protocol where node identifiers are cryptographically bound to session requests, ensuring that entanglement-based services are accessible only to authenticated entities with pre-established role profiles \cite{rao2023identityrouting}.
Du et al. extended the SDQN stack with {session-aware telemetry interfaces}. Their architecture enables Layer 5 to query fidelity scores, memory state, and link usage history to validate session readiness and make live session migration decisions in response to fidelity degradation \cite{du2022quantumstackinsights}. This model connects Layer 5 directly to Layer 8 cognitive agents that manage orchestration policies.
Nasir et al. explored {fidelity-scoped session contracts} for quantum computing services. They introduced contract primitives that bound the entanglement lifetime, QBER thresholds, and qubit freshness expectations for session-bound tasks like delegated gate execution and error-verified quantum state sharing \cite{nasir2021teleportflow}.
Lastly, Huang et al. demonstrated scalable {session tagging and multiplexing} in quantum transport systems. Their system assigns session identifiers to qubits and memory buffers, enabling concurrent quantum applications to safely share infrastructure while maintaining session integrity and isolation \cite{huang2023multiplexed}.

\subsection{Technologies and Challenges}
Implementing a functional quantum session layer presents a new class of technological and architectural challenges, owing to the fragility of quantum information, dynamic fidelity conditions, and the need for secure, role-aware coordination across distributed domains.
A fundamental challenge is the development of {session orchestration engines} capable of aligning multiple entanglement paths, fidelity constraints, and participant roles into coherent session containers. These engines must negotiate session setup requests, verify identity-role bindings, and provision transport-level entanglement under fidelity-aware deadlines \cite{vincent2023sessionq, goyal2023quantumsession}.
Maintaining {identity and role consistency} across a session lifecycle is particularly difficult in federated quantum environments. Many quantum services require strong guarantees about sender, receiver, and intermediary identities—especially in scenarios involving asymmetric trust or multiparty computation. However, most quantum systems today lack standardized APIs for cryptographically verifiable identity-role assertions or revocable session claims \cite{rao2023identityrouting}.

{Fidelity-scoped contract enforcement} remains a key limitation. Session-layer contracts often require minimum QBER thresholds, entanglement freshness, or temporal guarantees. Enforcing these contracts requires tight integration with real-time telemetry systems from Layer 2 and 3, and the ability to trigger automatic session reconfiguration or teardown if fidelity metrics fall below defined SLOs \cite{du2022quantumstackinsights, nasir2021teleportflow}.
Another major hurdle is {session multiplexing across shared quantum resources}. As the number of concurrent quantum services grows, the session layer must support tagging and isolation of logical flows across shared buffers, memory, and routing interfaces. Ensuring that quantum sessions do not interfere, leak information, or overwrite each other is non-trivial in today’s hardware-constrained environments \cite{huang2023multiplexed}.

{Session migration and continuity} are also unsolved challenges. Entangled links may degrade mid-session due to decoherence, route failure, or memory expiration. The session layer must decide whether to pause, reroute, or re-establish a session while preserving its application-level semantics. This requires live negotiation with Layer 4 and Layer 8, as well as awareness of role integrity and task state \cite{vincent2023sessionq}.
Furthermore, {cross-domain interoperability} adds complexity. Session descriptions, identity schemas, and service policies vary across quantum domains, leading to session negotiation failures or partial agreement. Without standardized session descriptors or role translation layers, federated sessions may require manual configuration or a trusted overlay to mediate \cite{goyal2023quantumsession}.
Lastly, the lack of {hardware abstraction and programmable session APIs} impedes rapid development. Current session-layer logic is often hardcoded to specific hardware capabilities (e.g., NV centers, photonic chips), making deployment brittle. A robust API framework is needed to expose a common set of session operations (create, pause, modify, terminate) across a heterogeneous quantum substrate.

\subsection{Future Directions}
As quantum networks evolve toward supporting long-lived, multi-party, and policy-bound services, the session layer must advance from basic role mapping to a full-featured service binding framework. Future developments will focus on dynamic coordination, application intent integration, and secure abstraction over volatile entanglement substrates.
A key direction is the emergence of {programmable quantum session contracts}. These contracts will function as machine-readable descriptors specifying service-level constraints—including fidelity thresholds, session lifetime bounds, entanglement rate floors, and authorized role mappings. Session controllers will validate these contracts in real time using telemetry feeds and orchestration feedback, adapting contracts as environmental conditions change \cite{vincent2023sessionq, du2022quantumstackinsights}.
Another promising trajectory involves {federated session brokers}, which mediate session setup across multi-operator or multi-domain environments. These brokers will interpret cross-domain session descriptors (for instance, QSDs), perform role-mapping translations, and negotiate resource commitments that respect each domain’s trust boundaries and platform constraints \cite{goyal2023quantumsession}.

{Cognitive session agents} will likely play a pivotal role at the Layer 5–Layer 8 boundary. These agents will use predictive analytics to determine session survivability, recommend session pausing or migration, and optimize fidelity-aware qubit allocation based on traffic prediction and usage context \cite{du2022quantumstackinsights, yin2022sessionroles}.
Another critical area is the design of {role-agnostic session flows}. Many existing systems hardcode roles such as sender or receiver. Future session architectures will support dynamic role reassignment and abstract session templates that adapt to runtime decisions, such as changing initiator or renegotiating session priorities \cite{rao2023identityrouting}.
In the context of resource sharing, scalable {session tagging and traffic classification systems} will emerge. These systems will tag qubits with session identifiers and priority labels, allowing routers, buffers, and orchestration agents to differentiate and prioritize flows under load or degradation scenarios \cite{huang2023multiplexed}.
Finally, {session-layer APIs} will be standardized to enable composability and cross-stack automation. These APIs will include lifecycle primitives (e.g., sessionCreate, sessionBind, sessionModify, sessionMigrate) and be exposed via quantum middleware to developers and orchestrators alike, enabling full stack-aware automation of quantum services.

\section{Layer 6 – Presentation}
Layer 6 in the Quantum-Converged OSI model is a semantic intermediary between quantum sessions and their informational representations. It transforms raw quantum states, entangled qubits, or classical-quantum hybrid data streams into application-appropriate formats while preserving quantum coherence and minimizing overhead.
The core functions of this layer include {quantum encoding and decoding}, {state compression}, and {fidelity-preserving format translation}. To support these functions, quantum encoding schemes—such as basis rotation, hybrid state construction, and error-tolerant modulation—must be dynamically selected based on channel noise, task requirements, and available qubit resources.

For example, applications such as QKD, quantum fingerprinting, and distributed quantum sensing may impose different constraints on encoding complexity, fidelity retention, and circuit depth. Layer 6 ensures that qubit representations are application-aware and hardware-compliant, aligning semantic requirements with physical-layer constraints.
Another essential function of this layer is also the {interpretation of measurement results and classical metadata} from quantum protocols. This includes reconstructing logical messages from qubit collapse events, applying phase/frame corrections, and preparing qubit payloads for use in distributed computations.
In more advanced networks, Layer 6 may also negotiate {cross-platform encoding translation}, allowing information between quantum systems with differing encoding standards (e.g., photonic vs. ion-trap vs. NV-center). Thus, this layer acts as a quantum codec and normalization interface in such scenarios.
The following subsections examine the specific requirements, key literature, implementation challenges, and forward-looking developments surrounding Layer 6 design in quantum internet stacks.

\subsection{Requirements}
Layer 6 in quantum network architecture must translate quantum state representations between physical encoding and application-level semantics while minimizing entropy growth and coherence loss. As such, it introduces several non-trivial requirements that differ sharply from classical presentation models.
First, the layer must support {adaptive quantum encoding} based on application context, channel noise, and device capabilities. Encoding schemes—such as dual-rail, time-bin, orbital angular momentum, or hybridized encodings—must be selected to preserve fidelity while remaining compatible with the receiver’s decoding protocol \cite{khaneja2021hybridmodulation}. These encodings must be dynamically negotiated between peers or orchestrated through metadata provided by Layer 5.
A second requirement is the implementation of {quantum data compression and redundancy minimization}. Unlike classical compression, quantum compression must operate without measurement, often leveraging reversible transformations or entanglement-assisted protocols. Therefore, presentation-layer compressors must optimize qubit usage without compromising the integrity of superposition or entanglement states \cite{pirandola2020quantumcompression}.
Another key function of this layer is to provide {semantic translation across encoding domains}. As quantum networks increasingly span heterogeneous devices, Layer 6 must mediate differences in basis, qubit interpretation, and encoding protocols. This includes translation between photonic and matter-based encodings, time-bin vs. polarization, or logical vs. physical qubit representations \cite{calderon2022semantictranslation, tan2023presentationlayer}.

A further requirement is supporting {format abstraction for quantum-classical hybrid protocols}. Many quantum applications (e.g., QKD, teleportation-based messaging) rely on classical-quantum interaction where metadata and qubit payloads must be jointly interpreted. Thus, Layer 6 must implement parsing, tagging, and encoding rules that ensure that classical annotations are coupled correctly with quantum data streams.
Another vital requirement is {application-driven presentation tuning}. Quantum applications may have varying latency, fidelity, or encoding overhead sensitivity. For example, QKD might favor robustness, whereas quantum fingerprinting may prioritize efficiency. Hence, Layer 6 must expose knobs or profiles to optimize encoding paths accordingly, with fallback mechanisms to select alternate compression or modulation styles when channel conditions degrade.
Lastly, Layer 6 must preserve {fidelity-awareness during all transformations}. Transformations at the presentation layer—encoding, compressing, or normalizing—must track how operations affect entanglement fidelity, decoherence rates, and overall quantum utility. Layers 2 and 3 telemetry should inform these transformations to ensure minimum-loss operation.
\begin{table*}[htbp]
    \centering
    \caption{Quantum encoding schemes used at the presentation layer across hardware platforms.}
    \label{tab:layer6_encodings}
    \begin{tabular}{|l|l|c|c|l|l|}
        \hline
        \textbf{Encoding} & \textbf{Platform} & \textbf{Fidelity} & \textbf{Transformability} & \textbf{Use Cases} & \textbf{Ref} \\
        \hline
        Polarization & Photonic & Medium & High & QKD, DQS, QVote & \cite{khaneja2021hybridmodulation} \\
        Spin-based & NV/Ion Trap & High & Low & Memory, Logic Circuits & \cite{tan2023presentationlayer} \\
        Orbital Angular Momentum & FSO/Photonic & High & Low & Satellite & \cite{calderon2022semantictranslation} \\
        Dual-Rail & Photonic & Medium & High & Short-range QKD, Sensors & \cite{pirandola2020quantumcompression} \\
        \hline
    \end{tabular}
\end{table*}

\subsection{Existing Literature}
Research on the presentation layer in quantum networks is still emerging, but recent studies have begun formalizing the translation, encoding, and compression mechanisms required to support semantically consistent quantum communications. These efforts ensure that quantum state representations can be interpreted accurately across heterogeneous devices and use cases while preserving fidelity.
Gyongyosi introduced an {adaptive quantum encoding framework} that dynamically selects encoding bases depending on noise levels and channel conditions. This work demonstrated how presentation-layer flexibility in encoding (e.g., polarization, dual-rail, or time-bin) could significantly improve fidelity in long-distance quantum communication networks.
Khaneja et al. further expanded this line of research by proposing {hybrid modulation schemes} that combine photonic and matter qubit states using device-specific optimization strategies. Their results support the idea of a presentation layer that acts as a codec between different quantum hardware platforms while preserving superposition and coherence integrity \cite{khaneja2021hybridmodulation}.
Pirandola and colleagues contributed to the theoretical foundations of {quantum compression}, exploring entanglement-assisted lossless protocols. Their study highlights that quantum compression at the presentation layer must balance circuit depth and logical qubit count while considering entropy bounds imposed by quantum mutual information \cite{pirandola2020quantumcompression}.

Calderon and Lee explored the concept of {semantic-aware data translation} between application contexts. Their presentation-layer system classifies quantum flows (e.g., QKD, sensing, computation) and maps them to encoding schemes and fidelity tolerances suited for each task. Their architecture introduces translation descriptors as metadata that enable interoperability across multi-service platforms \cite{calderon2022semantictranslation}.
Tan and Zhou proposed a modular presentation layer model for quantum networks that separates {logical representation from physical encoding}. Their architecture supports cross-stack interoperability by defining translation APIs that adapt qubit data for different quantum processors and memory models. This separation of concerns allows upper layers to remain agnostic to encoding hardware \cite{tan2023presentationlayer}.

\subsection{Technologies and Challenges}
Designing a quantum presentation layer involves addressing unique constraints imposed by coherence, non-clonability, and hardware diversity. It must enable quantum state interoperability, semantic abstraction, and resource optimization while preserving the integrity of entanglement and measurement outcomes.
A primary challenge is the {diversity of encoding schemes} used across quantum platforms. Photonic systems may use polarization or time-bin encoding, while solid-state systems rely on spin states or charge configurations. The presentation layer must bridge these formats via encoding translation modules or quantum codecs, yet standardized interfaces for this functionality remain immature \cite{khaneja2021hybridmodulation, tan2023presentationlayer}.

{Dynamic encoding adaptation} is another unsolved problem. Channel noise, gate fidelity, and link decoherence rates change over time, requiring presentation-layer logic to optimize encoding bases and modulation schemes continuously. While research has proposed adaptive encoding controllers, integrating them with stack-wide telemetry and making them hardware-agnostic remains an open challenge.
Quantum {data compression} introduces several significant challenges related to circuit overhead and fidelity preservation. Entanglement-assisted compression schemes require ancilla qubits and precise timing, while the no-cloning theorem and measurement avoidance constrain irreversible compression. Thus, Presentation-layer compressors must balance resource utilization with fidelity constraints without introducing entropy growth or decoherence \cite{pirandola2020quantumcompression}.

{Semantic translation} between application-layer protocols and hardware-level encodings remains underdeveloped. For instance, a distributed quantum sensing task and a QKD flow may have different fidelity needs, encoding preferences, and measurement post-processing logic. Therefore, Presentation-layer modules must classify flows and apply encoding transformations based on flow semantics—a capability not supported by current stacks in real-time \cite{calderon2022semantictranslation}.
Another challenge is {format abstraction and interoperability}. A single end-to-end session may traverse multiple qubit types as quantum networks scale, requiring different encoding-decoding pipelines. Without a unified presentation-layer abstraction, fidelity may be lost at encoding boundaries, or session setup may fail due to incompatible encoding assumptions \cite{tan2023presentationlayer}.
Furthermore, the absence of {standard APIs for presentation-layer tasks} limits development and automation. No common language for quantum encoding selection, compression toggles, or encoding metadata exchange severely restricts multi-vendor or hybrid system integration.
Finally, all presentation-layer operations must be {fidelity-aware and telemetry-driven}. Without constant feedback from Layer 2 (error rates, qubit state) and Layer 3 (path viability), encoding and compression decisions risk degrading rather than enhancing end-to-end utility.

\subsection{Future Directions}
The presentation layer of quantum networks is poised to become a key locus of innovation, transforming from a passive encoding interface into an intelligent, programmable mediator of quantum semantics and fidelity. Future directions point to the need for contextual encoding, real-time optimization, and cross-platform abstraction in increasingly heterogeneous and service-diverse quantum infrastructures.
One of the most critical advances will be the development of {context-aware encoding frameworks}. These systems will analyze service profiles (e.g., QKD vs. distributed computing), environmental parameters (noise, QBER, coherence time), and hardware constraints to select optimal encoding schemes dynamically. Such encoding logic must be informed by telemetry from Layers 2–3 and orchestrated by policy rules from Layer 8 \cite{tan2023presentationlayer}.

Another key trajectory involves {fidelity-preserving quantum transcoding}. Future networks will span photonic, superconducting, NV-center, and trapped-ion nodes, each requiring unique qubit representations. The presentation layer must introduce codec-style translation modules that convert between logical formats while minimizing fidelity loss—functioning similarly to video transcoding in classical networks \cite{khaneja2021hybridmodulation}.
Additionally, {semantic translation APIs} will become increasingly important as quantum applications demand differentiated fidelity, latency, and security guarantees. Future presentation-layer modules will interface directly with Layer 5 and Layer 8 to determine the service type, then select encoding schemes or compression techniques that align with that service’s priorities—e.g., robustness for QKD, efficiency for sensing \cite{calderon2022semantictranslation}.
Moreover, compression strategies will also evolve into {application-driven and entropy-adaptive schemes}. Compression modules will use insights from quantum information theory and entropy-aware scheduling to adjust granularity, ancilla usage, and entanglement-assistance parameters based on current memory state and session fidelity goals \cite{pirandola2020quantumcompression}.

Another emerging direction is {encoding-agnostic session abstraction}, where the presentation layer provides virtualization features that allow quantum sessions to persist across encoding domain boundaries. This would enable seamless interoperation between vendors, regions, and technologies, possibly backed by encoding negotiation protocols embedded in session metadata \cite{tan2023presentationlayer}.
Finally, we anticipate {standardization of presentation-layer APIs} across stacks. These APIs will include primitives such as \texttt{encode(fidelity_profile)}, \texttt{transcode(target_device)}, \texttt{compress(mode)}, and \texttt{getSemanticDescriptor()}. Such interfaces will empower middleware systems and Layer 8 agents to manage fidelity-aware data transformation, tagging, and delivery programmatically.

\section{Layer 7: Application}
Layer 7 of the Quantum-Converged OSI model constitutes the interface layer between quantum network infrastructure and end-user quantum services. It defines the operational models, abstract service types, and quantum-aware application interfaces that enable a broad spectrum of quantum-native and quantum-enhanced use cases to function reliably over entanglement-based infrastructure. Unlike classical networking environments, where application layers interact with deterministic packet streams and error-resilient channels, quantum applications must operate on non-clonable, coherence-limited qubit resources that degrade over time and are subject to probabilistic success rates.
This layer supports various application domains, including QKD, quantum teleportation, Distributed Quantum Computing (DQC), quantum-enhanced sensing and imaging, and Quantum Federated Learning (QFL). Each application class requires unique handling of fidelity thresholds, timing constraints, and circuit depth limitations and must integrate closely with the underlying session, transport, and presentation layers to maintain quantum correctness. For instance, QKD applications must support session persistence, entanglement recycling, and error reconciliation, while teleportation-based services depend on real-time Bell-pair coordination and classical control feedback. 

The application layer must also provide well-defined interfaces that allow developers and orchestrators to declare application-level intents, such as minimum fidelity, routing preferences, entanglement rates, or security policies. These intents are passed to orchestration agents in Layer 8, translating them into policy bindings and telemetry-driven adjustments across the quantum stack. Example interfaces may include primitives such as \texttt{initQKDSession()}, \texttt{submitQuantumJob()}, or \texttt{bindRole()}, which enable developers to express use-case-specific resource needs and fidelity constraints declaratively.
Furthermore, Layer 7 is a semantic boundary between classical control logic and quantum state behavior. It must mediate between quantum service definitions and the underlying mechanisms of entanglement distribution, qubit synchronization, and quantum memory allocation. In this role, Layer 7 not only facilitates use-case execution but also enforces policy constraints and trust models, particularly in multi-tenant or cross-domain quantum environments.
As the range and sophistication of quantum applications expand, the application layer must evolve to support modular and programmable service architectures capable of dynamically binding qubit resources, adapting to fidelity degradation, and interfacing seamlessly with centralized and edge-deployed orchestration planes. The following subsections present a detailed analysis of the application layer's design requirements, review the state of the art in quantum service modeling, examine enabling technologies and protocol frameworks, and explore emerging developments aimed at making Layer 7 programmable, scalable, and resilient under practical quantum constraints.

\subsection{Requirements}
Layer 7 must support quantum applications that rely on fragile, probabilistic entanglement as the substrate for computation, communication, and control. Unlike its classical counterpart, which can assume deterministic delivery and abstract network behavior, the quantum application layer must explicitly manage fidelity dependencies, resource scarcity, session timing, and multi-role logic in coordination with the rest of the stack.
The first requirement is expressing and enforcing {application-level quantum service intents}. These intents include minimum acceptable fidelity, maximum tolerated QBER, coherence window durations, and identity-role bindings. For example, a QKD session must specify key refresh intervals and entanglement freshness thresholds, while a distributed quantum computation job may require synchronized Bell-pair delivery and predefined circuit depth limits. These intents must be communicated to orchestration agents in Layer 8 and reflected in stack-wide policy enforcement \cite{kalb2021quantumservices, chen2023teleportationfabric}.
Another core requirement is the provision of {service-specific abstractions and control APIs}. The application layer must provide callable primitives such as \texttt{initQKDSession()}, \texttt{requestEntanglement()}, or \texttt{submitQuantumCircuit()}, which are mapped to session, presentation, and transport operations below. These APIs must be dynamically configurable and support metadata extension to accommodate evolving fidelity models, trust constraints, and latency sensitivities \cite{vanmeter2021qkdservices}.

Application layer must also support {application-role negotiation and task binding}. Many quantum services, particularly those involving teleportation or collaborative computing, require defined sender, receiver, and verifier roles. Layer 7 must bind these roles to authenticated identities, enforce exclusivity or delegation policies, and maintain consistent task-role mapping throughout the session lifecycle \cite{rao2023identityrouting}.
In multi-service environments, Layer 7 must enable {service isolation and multi-tenant execution}. Applications must be logically partitioned to ensure that session fidelity, memory allocation, and entanglement buffers are scoped exclusively to their corresponding tasks. This requires that the application layer interact with Layer 5 session tagging and Layer 4 flow control to prevent resource interference between concurrent jobs \cite{kalb2021quantumservices}.

A further requirement is the support for {fidelity-aware execution logic}. Applications must be capable of adapting their behavior based on real-time telemetry: suspending tasks during fidelity collapse, rerouting entanglement under memory failure, or downgrading operation modes under high QBER. The application layer must consume such telemetry and expose control knobs that adjust fidelity, throughput, or reliability tradeoffs accordingly \cite{du2022quantumstackinsights, tavakoli2023distributedqml}.
Lastly, Layer 7 must offer {interoperability and cross-platform abstraction}. Applications should function independently of the underlying quantum hardware (e.g., ion traps, NV centers, superconducting circuits), relying on the stack to virtualize qubit control, encoding schemes, and entanglement protocols. This abstraction is critical for building portable, service-oriented quantum applications that scale across heterogeneous infrastructures.
\begin{table*}[htbp]
    \centering
    \caption{Quantum application classes and their operational parameters at the application layer.}
    \label{tab:layer7_apps}
    \begin{tabular}{|l|c|c|c|l|l|}
        \hline
        \textbf{Application} & \textbf{Fidelity} & \textbf{Timing} & \textbf{Role Binding} & \textbf{API Primitive} & \textbf{Ref} \\
        \hline
        QKD & High & Medium & Yes & \texttt{initQKDSession()} & \cite{vanmeter2021qkdservices} \\
        Teleportation Service & Very High & High & Yes & \texttt{requestTeleportation()} & \cite{chen2023teleportationfabric} \\
        DQML & Medium–High & Tight & Yes & \texttt{submitQuantumCircuit()} & \cite{tavakoli2023distributedqml} \\
        Quantum Sensing & Medium & Loose & Optional & \texttt{startSensorStream()} & \cite{wehner2018quantum} \\
        Quantum Consensus & High & Real-time & Yes & \texttt{initConsensusTask()} & \cite{rao2023identityrouting} \\
        \hline
    \end{tabular}
\end{table*}

\subsection{Existing Literature}
The quantum application layer is a rapidly evolving domain, with ongoing research efforts seeking to define standardized service models, orchestrated application logic, and high-level interfaces for quantum communication and computation. While Layer 7 in classical networks traditionally handles formatting and session logic for stable packet flows, the quantum application layer must coordinate with the entire quantum stack to ensure coherent, fidelity-bound delivery of fragile quantum states.
One of the foundational visions was presented by Wehner et al., who outlined the conceptual framework for a global quantum internet and emphasized the need for modular, role-aware applications that could adapt to probabilistic and noisy entanglement-based links \cite{wehner2018quantum}. This work has inspired efforts to define use-case-specific application frameworks for quantum networks.
Kalb et al. advanced the field by proposing a {service-oriented architecture for quantum applications}. Their model separates application logic from entanglement distribution mechanisms, enabling scalable, pluggable service execution across nodes and qubit technologies. This layered design supports the abstraction of fidelity targets, resource reservations, and task-binding logic within the application layer \cite{kalb2021quantumservices}.

Van Meter et al. developed orchestration models for {QKD as a service}, formalizing how Layer 7 applications could express fidelity requirements, entanglement throughput demands, and session parameters via service descriptors. Their orchestration stack integrates tightly with transport, session, and orchestration layers, and defines contract-driven QKD deployment in multi-tenant networks \cite{vanmeter2021qkdservices}.
Chen et al. introduced a framework for {Teleportation-as-a-Service (TaaS)} in the context of teleportation-based applications. Their architecture defines application-level APIs for teleportation sessions, classical control exchange, and fidelity threshold management. They show how TaaS primitives can be reused across quantum messaging, distributed computation, and secure state injection workflows \cite{chen2023teleportationfabric}.
Tavakoli et al. explored the domain of {Distributed Quantum Machine Learning (DQML)}. Their analysis outlines how Layer 7 applications must support role-bound models for data encoding, entangled parameter updates, and decentralized inference while maintaining synchronization with stack-level telemetry to handle fidelity degradation and memory drift \cite{tavakoli2023distributedqml}.
Additional work by Du et al. integrated {telemetry-driven APIs} into the application layer, allowing services to adjust behavior in real-time based on memory coherence, link statistics, or session path updates. Their work enables fidelity-aware adaptation directly at the application interface and forms the basis for dynamic quantum workflows and trust-aware application policy management \cite{du2022quantumstackinsights}.

\subsection{Technologies and Challenges}
Implementing Layer 7 in a quantum network stack presents unique technological and architectural challenges that differ fundamentally from those in classical distributed systems. Quantum applications must contend with probabilistic delivery, decoherence-prone entanglement resources, and dynamic fidelity constraints while integrating into a coherent stack-wide orchestration model.
A core challenge is the lack of {standardized quantum application interfaces and programming models}. While classical applications benefit from well-defined APIs and protocol stacks, quantum service definitions remain ad hoc and tied to specific hardware platforms or middleware abstractions. Developers currently lack common primitives for initiating, monitoring, and terminating quantum service sessions—such as \texttt{requestTeleportation()}, \texttt{submitDistributedCircuit()}, or \texttt{setQKDPolicy()}—which hinders modular design and portability across quantum infrastructures \cite{kalb2021quantumservices, vanmeter2021qkdservices}.
Another significant barrier lies in {fidelity-aware service orchestration}. Quantum applications must make real-time decisions based on coherence windows, entanglement age, and QBER—all time-varying and resource-dependent. Application logic must be coupled with telemetry streams and orchestration control loops, yet current systems lack the fine-grained hooks and abstractions necessary for dynamic fidelity adaptation \cite{du2022quantumstackinsights}.

{Multi-role and identity binding} in distributed quantum applications is also challenging. Applications involving teleportation, blind computation, or federated learning must coordinate authenticated identities across roles like sender, verifier, and receiver. However, quantum platforms lack identity-binding protocols at the application level, making role negotiation and inter-domain trust enforcement difficult to manage securely \cite{rao2023identityrouting, chen2023teleportationfabric}.
Additionally, the layer also lacks robust support for {cross-platform service abstraction}. As applications increasingly span heterogeneous infrastructures (e.g., photonic circuits, NV centers, superconducting qubits), Layer 7 must operate independently of encoding schemes, session policies, and device-specific constraints. Without universal abstractions for service semantics and control, developers must hardcode infrastructure assumptions, undermining scalability and vendor neutrality \cite{tan2023presentationlayer}.
Another hurdle is the absence of {service composability and reusability}. In classical systems, microservices can be composed using containers and APIs. Quantum applications lack composable primitives for reusing entanglement workflows, layering quantum jobs, or chaining teleportation operations. The application layer must evolve to provide container-like wrappers that allow quantum services to be assembled, monitored, and migrated \cite{kalb2021quantumservices}.
Finally, the lack of {stack-aware simulators and development environments} limits testing and deployment. Application developers have minimal access to simulation tools that reflect stack-wide behavior (e.g., how Layer 3 routing failure impacts Layer 7 fidelity guarantees). Without such platforms, debugging distributed quantum applications and verifying fidelity resilience remains infeasible \cite{tavakoli2023distributedqml}.

\subsection{Future Directions}
As quantum networks evolve from experimental links to programmable, service-centric infrastructures, the application layer must transition from static task execution to a dynamic, intent-driven orchestration environment. Future Layer 7 architectures will integrate fidelity-awareness, identity governance, and resource abstraction into programmable APIs supporting modular, reusable, and distributed quantum services.
One major trajectory is the development of {intent-driven quantum service APIs}. Future quantum applications will express high-level objectives—such as “deliver 10 Bell pairs with >90\% fidelity in 100 ms” or “execute a 4-party QKD session with distributed basis sifting”—through declarative interfaces. These APIs will be processed by orchestration agents in Layer 8, which translate intent into routing, session binding, and fidelity management operations across the lower stack \cite{kalb2021quantumservices, chen2023teleportationfabric}.

Another promising direction is the emergence of {role-aware multi-party application protocols}. These protocols will extend existing identity bindings by enabling real-time negotiation of application roles (e.g., sender, controller, verifier) across trust boundaries and administrative domains. Future service architectures will use authenticated role delegation, threshold agreements, and zero-knowledge identity proofs to manage participation in teleportation, federated learning, and consensus tasks \cite{rao2023identityrouting}.
We will also see the rise of {composable quantum service containers}, allowing applications to be modularized into reusable functions—e.g., a QKD handshake module, a teleportation buffering stage, or an entanglement health check agent. These containers will expose standardized metadata interfaces, enabling developers to assemble workflows and orchestrate them over programmable, multi-tenant stacks \cite{tavakoli2023distributedqml, vanmeter2021qkdservices}.

Another frontier is the development of {application-side fidelity forecasting and failover logic}. By incorporating telemetry feedback loops from lower layers (memory age, QBER, path viability), Layer 7 applications autonomously pause, reroute, or downgrade fidelity requirements based on predictive models. Such self-adaptive behavior will be especially important in distributed computation, where gate fidelity or qubit coherence determines success \cite{du2022quantumstackinsights}.
Interoperability will also drive innovation in {cross-platform service abstraction}. Future Layer 7 systems will rely on encoding-independent session descriptors and qubit-neutral APIs, ensuring that applications can be deployed across photonic, superconducting, or NV-center infrastructures without modification. Quantum middleware will mediate these abstractions and expose compatibility layers for heterogeneous entanglement schemes \cite{tan2023presentationlayer}.
Finally, as quantum service ecosystems mature, we anticipate the creation of {stack-aware Quantum Development Environments (QDEs)}. These platforms will simulate not only qubit behavior but full stack-wide interactions—routing volatility, entanglement decoherence, session negotiation, and transport feedback—giving developers tools to test, verify, and debug complex Layer 7 workflows before deployment \cite{tavakoli2023distributedqml}.

\section{Layer 8: Orchestration and Cognitive Plane}
Layer 8 in the Quantum-Converged OSI model represents the orchestration and cognitive control plane that governs stack-wide decision-making, service management, and dynamic adaptation. It functions as the policy interpreter, telemetry aggregator, and intent-based controller, enabling programmable, resilient, and scalable quantum networking.
Unlike the traditional OSI model, which stops at Layer 7, this additional layer addresses the complex interplay between application logic, network state, hardware variability, and service guarantees in quantum systems. Quantum applications often have stringent fidelity, latency, and coherence constraints that require real-time response to environmental and systemic fluctuations—capabilities beyond the scope of individual protocol layers.
The orchestration plane receives high-level service intents from Layer 7—such as fidelity thresholds, role trust boundaries, or circuit execution goals—and translates them into coordinated actions across the lower layers. These include selecting entanglement paths, adapting encoding formats, initiating session migrations, and modifying flow priorities. To do so, Layer 8 consumes telemetry from memory buffers, link-layer QBER monitors, transport-layer qubit state managers, and routing engines.

A key function of this layer is {intent-aware orchestration}, which uses AI/ML-based decision models to match service requests with optimal resource allocations under probabilistic link dynamics. Orchestration agents may run centralized, distributed, or federated, depending on network scale, and must support interoperability across domains.
Layer 8 is also responsible for {policy enforcement and trust coordination}. It manages security policies, identity resolution, and SLAs governing quantum job execution. For example, it may enforce entropy budgeting for multi-tenant quantum memory, rate-limit QKD flows, or isolate untrusted domains via qubit firewalling and routing exclusion.
Moreover, this layer supports {closed-loop telemetry feedback}, allowing the system to adaptively reconfigure based on changing network and hardware states. It enables cognitive control functions such as dynamic path re-selection, encoding downgrading, teleportation retry scheduling, and predictive entanglement path aging mitigation.
In hybrid quantum-classical environments, the orchestration layer may interact with cloud-native management systems (e.g., Kubernetes-like orchestrators for quantum jobs), allowing seamless integration into future edge-cloud quantum computing ecosystems.

\subsection{Requirements}
Layer 8 is the orchestrating and cognitive control plane that unifies quantum service intent, real-time telemetry, and multi-layer protocol behavior into a coherent decision-making framework. Unlike lower layers, which execute discrete functional roles, the orchestration plane coordinates resource allocation, policy enforcement, and adaptive response across the entire quantum stack.
The first core requirement is interpreting and acting on {intent-driven service descriptions} from Layer 7. Quantum applications express goals—such as latency sensitivity, minimum fidelity, entanglement bandwidth, or identity trust requirements—that must be decomposed into operational configurations across Layers 1–7. Orchestration agents must map these intents to session setup parameters, encoding selection, routing preferences, and qubit handling policies \cite{zanin2021quantumorchestration, chen2023teleportationfabric}.

A second major requirement is deploying a {closed-loop telemetry architecture}. To respond to fidelity degradation, memory expiration, or QBER fluctuation, Layer 8 must ingest real-time metrics from quantum memories, link monitors, qubit state managers, and routing stacks. These inputs inform dynamic reconfiguration—such as triggering teleportation retries, rerouting entanglement paths, or pausing sessions at Layer 5 \cite{du2022quantumstackinsights, fu2022qaware}.
Orchestration must also provide {fidelity-aware optimization and control logic}. Using AI/ML models or rule-based systems, Layer 8 must decide how to prioritize tasks under entanglement scarcity, schedule qubit transfers, allocate memory buffers, or adjust transport-layer pacing. For example, routing agents may need to minimize cumulative entanglement age across a multi-hop path while meeting a target QBER threshold \cite{tavakoli2023distributedqml}.

The orchestration plane must support {multi-domain coordination and policy enforcement}. Quantum networks may span trust boundaries, jurisdictions, and hardware vendors. Orchestration controllers must enforce session isolation, quantum firewall rules, and SLAs that restrict how entanglement can be distributed or consumed across domains. This includes identity validation, entropy usage tracking, and cross-domain role binding protocols \cite{caleffi2020sdqn}.
Another requirement is the provisioning of a {programmable orchestration interface}. Stack developers and administrators need access to declarative APIs that define orchestration rules, register applications, set policy triggers, and monitor network state. These interfaces may resemble Kubernetes-style descriptors, enabling quantum services to be managed like containerized microservices on classical cloud platforms \cite{du2022quantumstackinsights}.
Lastly, Layer 8 must incorporate {resilience and failover logic}. In volatile quantum networks, where fidelity can collapse abruptly, or sessions can expire unexpectedly, the orchestration plane must preemptively prepare alternate paths; redundant entanglement flows, or downgrade modes for quantum services. This capability is essential for maintaining service continuity in mission-critical or time-sensitive applications \cite{fu2022qaware, chen2023teleportationfabric}.

\subsection{Existing Literature}
The orchestration and cognitive control plane for quantum networks is a relatively nascent yet rapidly developing area of research. With the increasing complexity of quantum services and infrastructure, recent efforts have focused on building programmable, telemetry-aware orchestration frameworks that mirror classical cloud-native platforms' flexibility and automation capabilities.
Caleffi et al. proposed one of the earliest formalizations of SDQN. Their architecture separates the control plane from the data plane in quantum networks and introduces programmable orchestration interfaces for entanglement path selection, memory provisioning, and fidelity management. SDQN enables controllers to dynamically configure quantum flows based on network policies and telemetry updates, making it one of the foundational approaches to cognitive quantum networking \cite{caleffi2020sdqn}.
\begin{table*}[htbp]
    \centering
    \caption{Orchestration and cognitive control functions at Layer 8.}
    \label{tab:layer8_capabilities}
    \begin{tabular}{|l|p{5.5cm}|c|l|}
        \hline
        \textbf{Function} & \textbf{Description} & \textbf{Affected Layers} & \textbf{Reference} \\
        \hline
        Intent Compilation & Translate application-level SLAs into stack configurations & L1–L7 & \cite{zanin2021quantumorchestration} \\
        Closed-loop Reconfiguration & Dynamic policy updates from telemetry triggers & L1–L5 $\biconditional$ L8 & \cite{du2022quantumstackinsights} \\
        Cognitive Control (AI) & Learn fidelity patterns and predict service degradation & L2–L5 & \cite{tavakoli2023distributedqml} \\
        Trust Domain Enforcement & Map identity to entanglement policy domains & L5, L7 & \cite{chen2023teleportationfabric} \\
        SLA Enforcement & Guarantee application-level entropy/fidelity contracts & L4–L7 & \cite{fu2022qaware} \\
        Declarative API Interface & Expose orchestrator control as programmable service descriptors & L8 Admin & \cite{caleffi2020sdqn} \\
        \hline
    \end{tabular}
\end{table*}

Zanin and Calafate expanded on this vision by introducing an {intent-based orchestration framework} tailored to quantum networks. Their work allows applications to express high-level service objectives—such as latency or fidelity goals—compiled into network-level actions via orchestration policies. This model supports programmable service chaining and policy-driven rerouting in response to fidelity collapse or qubit expiration \cite{zanin2021quantumorchestration}.
Du et al. presented a practical implementation of a {real-time telemetry infrastructure} for orchestrating quantum services. Their stack provides APIs for accessing link fidelity, qubit memory state, entanglement usage statistics, and routing metrics. These telemetry feeds serve as inputs for orchestration agents at Layer 8, enabling stack-wide reconfiguration in response to qubit quality drift or resource bottlenecks \cite{du2022quantumstackinsights}.
Fu et al. explored {fidelity-aware orchestration logic} that adjusts transport-layer behavior and routing decisions based on QBER and link lifetime predictions. Their fidelity estimation model supports proactive entanglement path rerouting and qubit drop policies that maintain application-level service guarantees under dynamic network conditions \cite{fu2022qaware}.
Chen et al. introduced a specialized orchestration framework for {TaaS}, showing how cognitive control loops can govern Bell-pair provisioning, qubit synchronization, and teleportation retries. Their architecture exposes a declarative API for orchestrators to initiate teleportation sessions, negotiate fidelity requirements, and trigger entanglement reconnections during service degradation \cite{chen2023teleportationfabric}.
In distributed quantum machine learning, Tavakoli et al. demonstrated the necessity of {cognition-enabled orchestration} to schedule qubit sharing, gradient merging, and training session synchronization across distributed nodes. Their model highlights how orchestrators must manage fidelity decay, memory expiration, and entanglement contention while maintaining global model convergence \cite{tavakoli2023distributedqml}.

\subsection{Technologies and Challenges}
Deploying a robust orchestration and cognitive control plane in quantum networks presents a new class of challenges—rooted in the probabilistic nature of entanglement, rapid decoherence, non-clonability of qubits, and lack of stack-wide abstractions for fidelity-aware coordination. Unlike classical SDN systems, orchestration in quantum contexts must integrate real-time telemetry, quantum resource constraints, and application-level fidelity demands to make viable decisions under uncertainty.
A central technological challenge is the development of {programmable orchestration agents}. These agents must interpret Layer 7 intents and enforce configuration decisions across multiple quantum subsystems—ranging from routing to transport pacing, session resets, and encoding adaptation. While frameworks such as SDQN have proposed controller abstractions \cite{caleffi2020sdqn}, real-world implementations remain limited due to lack of unified orchestration APIs and programmable telemetry interfaces \cite{du2022quantumstackinsights}.
Another critical issue is building {low-latency telemetry integration}. Orchestration layers must ingest fidelity metrics, entanglement age, QBER, memory state, and route uptime with sub-second precision. However, quantum hardware platforms often lack reliable telemetry export mechanisms, and measurement intervals can conflict with qubit lifetimes. Designing a telemetry pipeline that is fast, reliable, and non-invasive remains an open engineering problem \cite{fu2022qaware}.

The orchestration layer must also resolve {cross-domain policy conflicts and trust coordination}. In multi-operator environments, orchestration agents need mechanisms to verify identity roles, enforce resource boundaries, and negotiate entanglement paths without exposing session metadata or compromising SLAs. Today’s orchestration models lack cryptographic identity frameworks and fine-grained trust boundary management required for inter-domain quantum workflows \cite{zanin2021quantumorchestration}.

A further challenge is orchestrating {coherence-constrained qubit lifecycles}. Unlike classical packets, qubits cannot be stored indefinitely or re-sent once corrupted. Thus, orchestration systems must act on predictive fidelity models and entanglement aging forecasts to trigger teleportation retries, schedule memory reuse, or downgrade sessions before coherence loss becomes terminal \cite{chen2023teleportationfabric}.
The orchestration layer must also support {AI/ML-based cognitive control}. With quantum state behavior being partially observable and highly dynamic, rule-based orchestration quickly becomes insufficient. Future orchestration platforms must incorporate RL or Bayesian models that optimize entanglement throughput and service resilience over time—especially for use cases such as distributed quantum computing or federated learning \cite{tavakoli2023distributedqml}.
Finally, there is a lack of {stack-aware orchestration development environments}. Currently, no simulators or testbeds allow developers to test orchestration policies against realistic fidelity degradation, memory exhaustion, and session churn across quantum layers. This hinders reproducibility, benchmarking, and rapid innovation in cognitive control strategies.

\subsection{Future Directions}
As quantum networks mature and support multi-tenant, cross-domain, and application-diverse workloads, the orchestration layer will evolve from reactive flow management to an autonomous, predictive, and programmable control architecture. Future orchestration systems will blend real-time telemetry, AI cognition, policy enforcement, and semantic understanding into a unified decision-making framework.
A major direction is the development of {closed-loop cognitive orchestration}. Future orchestrators will use AI/ML models to continuously ingest fidelity metrics, qubit aging data, and session performance logs to optimize entanglement routing, teleportation retries, and circuit dispatch in real-time. RL agents could train on fidelity loss patterns to proactively reassign qubit buffers or reconfigure sessions before coherence collapse \cite{tavakoli2023distributedqml, fu2022qaware}.
Another key trend will be the deployment of {intent-compiling orchestration planes}. These platforms will translate high-level service intents from Layer 7 into stack-wide configuration commands. For instance, a request for “multi-party teleportation with >95\% fidelity” could be decomposed into session-layer role assignments, Layer 3 path binding, and Layer 2 memory reservation, all governed by orchestration policies \cite{zanin2021quantumorchestration}.
Future platforms will also support {federated orchestration architectures}. As quantum networks span administrative and geographic boundaries, orchestrators must interoperate securely. Federated cognitive controllers will negotiate trust boundaries, exchange encrypted session policies, and support role mapping across domains—similar to multi-cloud control planes in classical SDN \cite{caleffi2020sdqn}.
A further innovation will be {entropy budget enforcement and fidelity-centric SLAs}. Quantum orchestrators will monitor entanglement consumption, memory freshness, and fidelity drift to enforce per-application or per-session SLAs. These contracts may specify tolerable QBER, allowable teleportation retries, or maximum decoherence age, with orchestration logic dynamically adjusting resources to maintain compliance \cite{chen2023teleportationfabric}.
The orchestration plane will also become {programmable via declarative policy APIs}. Developers and administrators will use quantum-intent languages to define orchestration behavior, bind services to specific resources, or enable automatic fallback scenarios under fidelity violation. These APIs will serve as the interface between quantum cloud platforms and orchestration engines \cite{du2022quantumstackinsights}.
Finally, a critical enabler for these directions will be the emergence of {stack-wide orchestration simulators}. These platforms will model fidelity-aware entanglement flows, session churn, path degradation, and telemetry loops—allowing validation of orchestration strategies before live deployment. This will accelerate the development of reliable, intelligent control planes for quantum service networks. 
\begin{figure}[]
    \centering
    \includegraphics[width=0.95\linewidth]{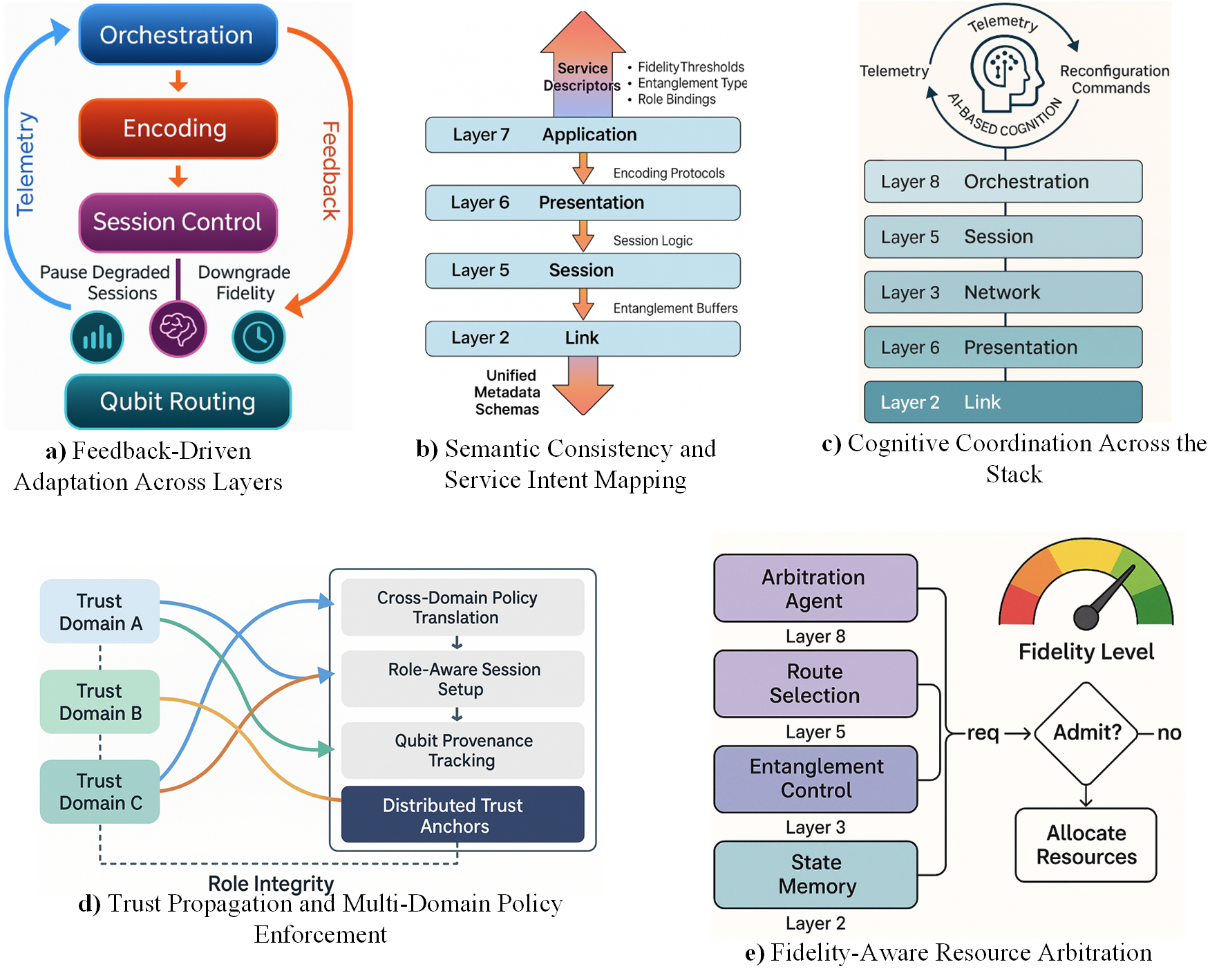}
    \caption{
    Cross-layer architectural roles in quantum networks: 
    \textbf{(a)} Feedback loops drive adaptation by integrating telemetry and orchestrator commands \cite{du2022quantumstackinsights}; 
    \textbf{(b)} Semantic intent descriptors propagate through layers to ensure encoding/session coherence \cite{zanin2021quantumorchestration}; 
    \textbf{(c)} Cognitive control loops enable AI-driven reconfiguration via telemetry feedback \cite{tavakoli2023distributedqml}; 
    \textbf{(d)} Trust domains coordinate role enforcement and provenance tracking \cite{rao2023identityrouting}; 
    \textbf{(e)} Fidelity-aware arbitration determines session viability and qubit resource allocation \cite{fu2022qaware}.
    }
    \label{fig:cross_layer_composite}
\end{figure}

\section{Cross-Layer Design Considerations}
While the Quantum-Converged OSI model emphasizes clear modularity across functional layers, real-world implementations demand coordinated interactions to adapt to quantum-specific constraints—especially fidelity degradation, memory expiration, and role-sensitive entanglement policies. Effective quantum networking thus requires intelligent, real-time, and trust-aware cross-layer control. Fig.~\ref{fig:cross_layer_composite} illustrates the key mechanisms of cross-layer coordination in quantum networks. From feedback-driven adaptation (Fig.~\ref{fig:cross_layer_composite}a) to semantic consistency enforcement (Fig.~\ref{fig:cross_layer_composite}b) and AI-powered orchestration (Fig.~\ref{fig:cross_layer_composite}c), the stack enables dynamic and intent-aware quantum communication. Trust propagation (Fig.~\ref{fig:cross_layer_composite}d) and fidelity-based arbitration (Fig.~\ref{fig:cross_layer_composite}e) further ensure that session integrity and quantum resource viability are preserved across multiple domains and layers.

\subsection{Feedback-Driven Adaptation Across Layers}
Quantum systems exhibit volatile fidelity and coherence lifetimes that fluctuate in response to environmental and system-level factors. To cope, lower-layer telemetry (e.g., QBER, memory age, link uptime) must be continuously fed into higher-layer orchestration logic at Layer 8. These feedback loops allow orchestrators to pause degraded sessions (Layer 5), downgrade encoding fidelity (Layer 6), or initiate qubit rerouting (Layer 3) before services fail \cite{du2022quantumstackinsights, fu2022qaware}.
This feedback-driven architecture is critical for enabling just-in-time recovery actions—especially in time-sensitive applications like distributed quantum computing and federated sensing, where preemptive adaptation is key to maintaining logical consistency \cite{tavakoli2023distributedqml}.

\subsection{Semantic Consistency and Service Intent Mapping}
Service-layer contracts declared at Layer 7—such as fidelity thresholds, entanglement types (Bell vs GHZ), and role-to-identity bindings—must propagate consistently across all protocol layers. Misalignments (e.g., mismatched session roles or qubit tagging inconsistencies) can break entangled workflows or introduce silent fidelity collapse \cite{zanin2021quantumorchestration, chen2023teleportationfabric}.
To ensure semantic consistency, unified metadata schemas and service descriptors must inform every layer, from encoding protocols at Layer 6 to entanglement buffer handling at Layer 2. Future orchestration systems may enforce such mappings via policy validation engines embedded in cognitive control agents \cite{caleffi2020sdqn}.

\subsection{Trust Propagation and Multi-Domain Policy Enforcement}
In quantum networks spanning federated or adversarial domains, identity-aware entanglement constraints must be enforced from the application through the physical layers. Blind computation, verifiable teleportation, and QKD all require session roles, trust domains, and identity tokens to propagate securely across layers \cite{rao2023identityrouting, chen2023teleportationfabric}.
This mandates stack-wide trust enforcement, which includes qubit provenance tracking at Layer 2, role-aware session setup at Layer 5, and cross-domain policy translation at Layer 8. Emerging models propose distributed trust anchors and entanglement ACLs to maintain role integrity in hybrid service scenarios \cite{zanin2021quantumorchestration}.

\subsection{Cognitive Coordination Across the Stack}
Future orchestration layers will incorporate AI-based cognition that integrates telemetry from all layers to forecast entropy depletion, predict session failure, and recommend policy reallocation \cite{tavakoli2023distributedqml}. These systems will dynamically reconfigure routing paths, encoding strategies, and session priorities based on real-time network conditions and long-term performance learning \cite{fu2022qaware}.
Such cognitive loops must operate in stack-aware simulators to safely test adaptation logic under dynamic quantum constraints. This will unlock closed-loop orchestration where decisions propagate downward as reconfiguration commands, and telemetry flows upward for model refinement \cite{du2022quantumstackinsights}.

\subsection{Fidelity-Aware Resource Arbitration}
Cross-layer arbitration is essential since fidelity degrades differently across layers—due to memory aging, noisy links, and encoding overhead. For example, session-layer entanglement requests must be coordinated with available memory states at Layer 2 and route viability at Layer 3. Without this coordination, sessions may be initiated with doomed fidelity profiles \cite{fu2022qaware, du2022quantumstackinsights}.
Orchestration agents at Layer 8 must enforce arbitration policies that adjust admission, routing, and session pacing to preserve coherence, ensuring resource allocation reflects global fidelity budgets and application-specific entropy limits \cite{caleffi2020sdqn}.

\begin{table*}[H]
    \centering
    \caption{Cross-layer functional dependencies in quantum networks.}
    \label{tab:cross_layer_summary}
    \begin{tabular}{|l|l|l|}
        \hline
        \textbf{Cross-Layer Theme} & \textbf{Layers Involved} & \textbf{Description} \\
        \hline
        Telemetry Propagation & L2–L8 & Vertical QBER, coherence, routing state sharing \cite{du2022quantumstackinsights} \\
        Intent Compilation & L7 → L1–L6 & Service goal decomposition and reconfiguration \cite{zanin2021quantumorchestration} \\
        Trust Mapping & L5 → L2–L3 & Identity propagation and domain exclusion \cite{rao2023identityrouting} \\
        Fidelity-Aware Pacing & L2–L4–L8 & Session, memory, transport tuning based on fidelity \cite{fu2022qaware} \\
        Role-Aware Routing & L5–L3 & Entanglement routing based on session logic \cite{chen2023teleportationfabric} \\
        Semantic Encoding Tuning & L6–L8 & Format tuning by app context and fidelity goals \cite{tan2023presentationlayer} \\
        \hline
    \end{tabular}
\end{table*}

\section{Enabling Technologies}
\begin{figure}[htbp]
    \centering
    \includegraphics[width=0.7\linewidth]{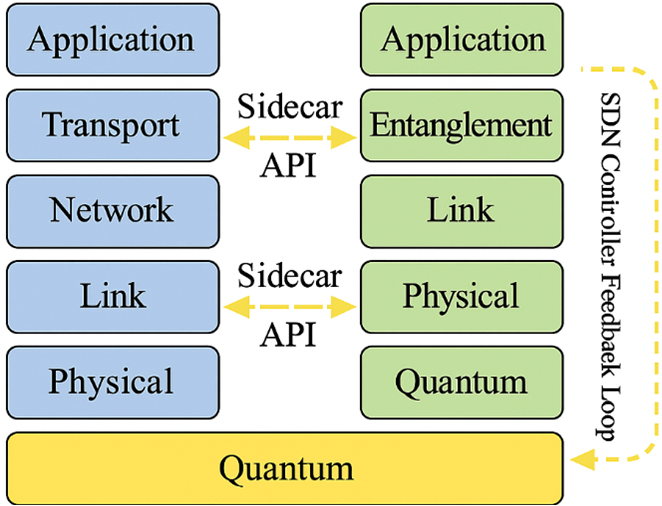}
    \caption{Hybrid quantum-classical architecture showing parallel protocol stacks and sidecar APIs for interoperability. The SDN feedback loop dynamically aligns stacks using quantum-physical insights.}
    \label{fig:hybrid_stack_api}
\end{figure}

The realization of the proposed quantum-converged OSI stack depends critically on enabling technologies that span physical interfaces, cryptographic resilience, simulation environments, and AI-driven orchestration. These technologies do not exist in isolation; they function as interconnected components that empower fidelity preservation, coherence control, semantic translation, and intent-driven policy orchestration across layers.
\textit{QML} and \textit{LLMs} have emerged as transformative tools in adaptive quantum networking. These models enable the interpretation of user-level intent, real-time adaptation of stack configurations, and predictive orchestration based on qubit coherence patterns and fidelity metrics. Recent efforts in LLM-aided intent translation within 6G architectures show promise for quantum stack alignment as well~\cite{Chaoub2025MobileLLM6G}.
Additionally, at the physical layer, \textit{RIS} are leveraged to control quantum signal propagation dynamically. These surfaces can mitigate polarization drift, optimize entanglement channel alignment, and extend the reach of quantum links in free-space or urban environments~\cite{Pierucci2022Metropolitan}.
Furthermore, \textit{QKD} forms the foundation of secure quantum communication. However, its limitations under noise and distance have prompted the integration of \textit{PQC} into hybrid stacks. PQC algorithms serve as a fallback and augmentation to QKD systems, ensuring end-to-end encryption resilience~\cite{Baseri2024QuantumSafeSurvey}.
Moreover, to ensure trust and tamper-evident integrity in quantum control layers, \textit{Blockchain} technology is increasingly employed. Distributed ledger infrastructures support decentralized identity management, quantum token auditing, and secure metadata coordination—particularly within federated QKD services~\cite{Barros2025ProofHumanity}.
In addition, simulation environments are essential for validating cross-layer interactions. Tools like \textit{NetSquid}~\cite{NetSquid2021}, \textit{QuNetSim}~\cite{9465750}, and \textit{QuISP}~\cite{joubert2024quisp} allow researchers to model decoherence, teleportation, entanglement routing, and QEC mechanisms with stack-level granularity. They enable scenario testing for network reliability, entropy propagation, and protocol efficiency in the presence of realistic noise models.
These technologies play a pivotal role in enabling reliable and scalable quantum communication systems. The following subsections provide an in-depth analysis of their operational roles and implications for protocol design at each OSI layer.
Fig.~\ref{fig:hybrid_stack_api} illustrates dual-stack coexistence through sidecar APIs that synchronize classical and quantum protocol behaviors aligned with the SDN feedback control loop.

\subsection{Quantum Hardware and Physical Interfaces}
The foundation of any quantum communication system begins at the hardware layer, which is responsible for the physical realization of qubits, entanglement channels, and signal fidelity. A dominant trend is using \textit{photonic qubits} for long-distance transmission due to their resilience to decoherence and compatibility with fiber-optic infrastructure. State-of-the-art implementations use single-photon sources derived from quantum dots~\cite{Neuwirth2023QDRepeater} and integrate high-efficiency Superconducting Nanowire Single-Photon Detectors (SNSPDs), which offer ultra-low dark counts and high timing resolution~\cite{Taha2024Photonics}. These technologies enable entanglement distribution over metropolitan and satellite-based quantum networks.
\textit{Quantum repeaters} play a pivotal role in extending the operational range of quantum networks. These devices are responsible for entanglement swapping and fidelity preservation over long distances. Recent proposals emphasize memory-based architectures to buffer quantum states during long-haul transfer and satellite-assisted repeaters for global coverage~\cite{Beukers2024RemoteEnt, Bernien2022Interconnects}. Integrating photonic–spin interfaces further enhances the hybridization between matter qubits and photonic channels~\cite{McMahon2015SpinPhoton}.
Moreover, \textit{RIS} are being investigated as a transformative mechanism to control the propagation of quantum signals in dynamic environments. By leveraging adaptive tuning of the scattering profile of RIS panels, entanglement fidelity and channel coherence can be improved even under non-line-of-sight and atmospheric distortion scenarios~\cite{Chehimi2025RIS, Zeydan2025QuantumRIS}. RIS-enhanced free-space optical links are now proposed for entanglement distribution in urban and inter-satellite quantum links.
These advancements in quantum hardware components—from superconducting detectors and entangled photon sources to programmable RIS and quantum repeaters—form the essential physical substrate upon which higher-layer quantum networking protocols can operate. As quantum OSI stacks evolve, the precision and tunability offered by these hardware technologies will directly influence stack design choices, layer convergence, and end-to-end system coherence.
Table~\ref{tab:quantum-phy-protocols} summarizes major physical-layer quantum protocols including BB84, E91, MDI-QKD, and satellite-based implementations like Micius \cite{Pirker2019Modular, Mehic2022QKDNetworks, Liao2017SatelliteMicius, Khan2024QuantumFuture}.

\subsection{Quantum-Classical Interface Protocols}
The design of robust and scalable quantum communication systems depends heavily on well-engineered quantum-classical interface protocols, which ensure seamless cooperation between quantum subsystems and classical control or transport planes. One critical element of this interface is the deployment of \textit{PQC} in parallel with QKD. PQC algorithms, such as lattice- and code-based schemes, provide an additional layer of cryptographic assurance—particularly in scenarios where QKD link integrity is compromised or unavailable. Zeydan et al.~\cite{Zeydan2025QuantumRIS} explore this hybridization in beyond-6G quantum systems, while Minoli and Occhiogrosso~\cite{Minoli2023QuantumSecurity} detail how PQC augments QKD in practical deployments across heterogeneous networks.

Another cornerstone of quantum-classical integration is the emergence of \textit{hybrid Software-Defined Networking (SDN/Q-SDN)}. In this architecture, the classical SDN controller operates alongside a quantum-aware management plane capable of dynamic entanglement-aware routing. Granelli et al.~\cite{Granelli2022HybridStack} propose an SDN-inspired framework that optimizes fidelity-aware routing and quantum-classical traffic orchestration. Such architectures allow the network to adapt in real-time to entanglement quality, coherence times, and decoherence drift changes.
Additionally, \textit{blockchain technologies} are also an emerging technology to play a pivotal role in bridging control and security mechanisms between classical and quantum layers. Rahman and Wang~\cite{Rahman2022QuantumBlockchain} highlight how distributed ledgers can track QKD session integrity, key lifecycle status, and authentication history across decentralized quantum infrastructures. In addition, Lazirko~\cite{Lazirko2023QBlockchain} discusses blockchain-based access control frameworks for quantum APIs and simulation environments, ensuring tamper-resilient identity management and trust propagation in distributed networks.
These interface protocols and architectural elements establish a reliable, auditable, and reconfigurable bridge between quantum protocols and the classical ecosystems they must interoperate with. Their presence is indispensable in building the multi-layered resilience and modularity envisioned in a quantum-converged OSI model.

\subsection{Quantum Software Frameworks}
As quantum networking advances toward layered architectures, software frameworks and simulation platforms have become indispensable for protocol development, validation, and stack optimization. Among these, \textit{NetSquid}, \textit{QuNetSim}, and \textit{QuISP} have emerged as leading discrete-event simulation tools. NetSquid supports noise modeling, entanglement queueing, and teleportation delay analysis at various stack layers~\cite{Coopmans2021NetSquid}. Similarly, QuNetSim, written in Python, provides a modular and pedagogically-oriented platform for quantum MAC protocols, session initiation models, and routing logic~\cite{9465750}. Meanwhile, QuISP uses a C++-based back-end tailored to full-stack simulation, including physical-layer noise, link-level fidelity tracking, and hybrid classical-quantum channel dependencies~\cite{joubert2024quisp}.
In parallel, development kits like \textit{Qiskit}, \textit{Cirq}, and \textit{Amazon Braket} enable programmers to design and deploy quantum algorithms in hardware-agnostic environments. These SDKs support application-layer integration with OSI layers 6–8, including quantum data compression, API abstraction, and teleportation-based conferencing. Cross et al.~\cite{Cross2018Qiskit} introduced Qiskit as a modular suite with transpiler access and quantum circuit composers, while Shi et al.~\cite{Shi2022Cirq} describe Cirq’s strengths in NISQ simulation and calibration workflows.

Moreover, the \textit{QIR} developed by Microsoft standardizes quantum compiler output for portability across backends, enabling interoperability between upper-layer software and hardware substrates. Similarly, \textit{QCoDeS}, an open-source data acquisition and device control framework, provides seamless lab instrumentation integration and experiment reproducibility. These tools make it feasible to link real hardware to network-layer abstractions for quantum stack testing in live and hybrid environments~\cite{QIRSpec, QCoDeS2023}.
These software frameworks form the backbone of modern quantum protocol design, serving as validation grounds, orchestration platforms, and application interfaces for layered quantum networks.

\subsection{AI and LLM-Assisted Orchestration}
As quantum networks grow in complexity and scale, adaptive orchestration becomes critical to maintaining entanglement quality, routing coherence, and session continuity across dynamically evolving topologies. AI—and in particular, \textit{QML} and \textit{LLMs}—has begun to redefine the orchestration paradigm for layered quantum systems.
LLMs such as GPT-4 and their 6G-specialized derivatives are being used as cognitive agents that interpret high-level user goals, perform semantic translation of network policies, and generate configuration updates for protocol stacks~\cite{Chaoub2025MobileLLM6G}. These models can synthesize network telemetry (fidelity scores, decoherence rates) and user intent to recommend stack reconfiguration or service migration, effectively forming the quantum OSI model's cognitive plane (Layer 8).

At lower layers, \textit{RL} and supervised QML models support fidelity-aware routing, congestion avoidance, and hybrid classical-quantum flow control. For instance, Baum et al.~\cite{Baum2021DeepRLPRX} demonstrated how deep RL agents can learn optimal gate-set designs under noise constraints on superconducting processors—an approach transferable to entanglement scheduling and MAC arbitration in real-time networks.
Moreover, generative QML models are being proposed to simulate entanglement propagation and predict the impact of measurement collapse across distributed sessions~\cite{Wang2023QMLComms}. These models enable simulation-accelerated stack orchestration, where LLMs make policy decisions that QML agents validate via predictive coherence metrics.
LLMs and QML form a bidirectional orchestration loop where natural-language control flows downward from the cognitive layer, and probabilistic feedback from QML simulations feeds upward, enabling intent-driven quantum networking that is reactive, self-optimizing, and cross-layer aware.

\subsection{AI/ML and Quantum Intelligence Integration}
As quantum networks become increasingly dynamic and probabilistically governed, traditional deterministic control mechanisms fall short in sustaining coherence, maintaining fidelity, and optimizing resource usage. This has led to the emergence of \textit{AI-driven orchestration}, with specific emphasis on \textit{QML}, \textit{LLMs}, and \textit{RL} for layered stack optimization.
QML techniques offer probabilistic modeling capabilities that align naturally with the intrinsic uncertainties of quantum systems. Cacciapuoti et al.~\cite{Cacciapuoti2022QMLSurvey} provided a foundational survey on how QML models can be leveraged to predict decoherence events, optimize entanglement distribution, and guide real-time path selection in entanglement routing. These capabilities are particularly relevant in environments where link stability fluctuates with quantum noise, making classical precomputed routing tables ineffective.

LLMs, particularly those trained on network policy grammars and quantum communication datasets, are increasingly used as cognitive agents in the orchestration layer (Layer 8). Chaoub and Elkotob~\cite{Chaoub2025MobileLLM6G} have demonstrated how LLMs can understand user intent, translate natural language service requests into actionable protocol configurations, and negotiate dynamic reconfiguration policies at runtime. This human-AI interface is valuable for managing stack states across heterogeneous quantum-classical hybrid infrastructures. Additionally, Prados et al.~\cite{Prados2023LLMQuantum} explored how LLMs can act as interface translators between classical orchestration APIs and quantum session managers, allowing seamless stack-level control in cross-layer models.
Furthermore, RL offers another powerful paradigm well-suited for control tasks in volatile quantum environments. Baum et al.~\cite{Baum2021DeepRLPRX} implemented deep RL on superconducting quantum hardware to train policies for error mitigation and gate fidelity maximization, concepts transferable to real-time stack adaptations in MAC scheduling and coherence-preserving transport layer policies. RL agents can adapt to noisy feedback, track fidelity variations, and suggest optimal control actions, making them critical to cross-layer stack tuning.
These AI/ML paradigms collectively signal a transition toward \textit{intent-driven}, self-optimizing quantum stacks where orchestration decisions are not only data-driven but also capable of semantic interpretation and real-time adaptation, forming the cornerstone of Layer 8 cognitive control.

\subsection{Trust and Verification Systems}
Trust establishment in quantum networks introduces multifaceted challenges due to their inherent non-determinism, probabilistic security guarantees, and the ephemeral nature of entanglement. To address these limitations, hybrid infrastructures combining \textit{blockchain}, \textit{zero-knowledge proofs (ZKPs)}, and \textit{entropy-aware verification} mechanisms are gaining traction. Blockchain offers tamper-evident logs of entanglement usage, QKD sessions, and inter-node communication metadata. Recent research by Zhou et al.~\cite{Zhou2024ZKPBlockchain} surveys how zero-knowledge proofs can reinforce decentralized identity management and cross-node trust in post-quantum environments.
In these systems, identity verification can be performed without revealing entangled session contents or private metadata, a crucial property under the constraints of the no-cloning theorem. Sezer et al.~\cite{Sezer2025PPPQB} propose a privacy-preserving blockchain that utilizes ZKPs to allow secure registration and session establishment without leaking identity credentials. These protocols are further fortified using quantum entropy sources as randomization anchors~\cite{Li2023DIQRNGZK}, ensuring non-reproducibility and verifiability of probabilistic handshake outcomes.

From a broader view, Ren et al.~\cite{Ren2025QuantumBlockchainWeb3} advocate for a Web3-aligned identity infrastructure underpinned by quantum-safe cryptographic primitives, including lattice-based ZKP circuits and entropy verifiers. In practical scenarios, Zeydan et al.~\cite{Zeydan2024QuantumID} show how combining QKD key freshness with blockchain-tracked authentication history strengthens the traceability and non-repudiation in entangled communications.
These trust systems enhance resilience against Sybil and man-in-the-middle attacks and enable compliance-aware orchestration where Layer 8 (Cognitive Plane) agents make adaptive decisions based on the integrity scores and entropy availability at lower stack levels. These mechanisms constitute a critical foundation for governance, compliance, and secure interoperability in globally distributed quantum-classical networks.

\section{Conclusion}
The rapid evolution of quantum communication technologies, driven by advances in entanglement distribution, quantum error correction, and programmable interfaces, reshapes the fundamental assumptions underpinning classical networking paradigms. This survey addressed the architectural, operational, and systemic misalignments between the classical OSI model and the requirements of quantum networking, highlighting the urgent need for a quantum-converged stack. We proposed a comprehensive architectural framework that extends the OSI model to include Layer 0 (Quantum Substrate) and Layer 8 (Cognitive Intent Plane), supported by stack-specific functionality such as quantum-aware session management, fidelity-preserving transport, and LLM-assisted orchestration.
A detailed analysis was provided across each OSI layer, identifying quantum-aware opportunities, use-case-driven stack redefinitions, enabling technologies such as QML, RIS, PQC, and SDN-Q, and an evaluation methodology centered around entanglement fidelity, entropy throughput, and coherence latency. We consolidated contemporary perspectives and simulation insights from NetSquid, QuNetSim, and QuISP to bridge theoretical models and practical validation.
We further outlined critical open challenges, including the need for stackless, real-time protocol flows; the co-evolution of LLM and QML agents for cross-layer orchestration; and the urgent demand for interoperable standards in hybrid quantum-classical infrastructures. Finally, the paper mapped out key future research directions encompassing stackable architectures for the quantum internet, programmable control via SD-Q, cross-layer QEC, and advanced benchmarking standards for quantum performance metrics like QBER, teleportation delay, and routing entropy.
The community must pursue deeper integration between AI, physics-aware networking, and secure architectural abstractions as we transition from the early deployments of quantum point-to-point systems to scalable, intelligent, and programmable global quantum internetworks. This co-design philosophy will be vital in realizing robust, policy-aware, and application-adaptive quantum networks that meet the demands of 7G and beyond.

\bibliographystyle{IEEEtran}
\bibliography{IEEEabrv,References/main}

\end{document}